\documentclass[12pt]{article}

\usepackage{CJKutf8}
\usepackage{cite}
\usepackage{graphicx}
\usepackage{amsmath}
\usepackage{amsthm}

\usepackage{geometry}
\usepackage{framed}
\usepackage{multirow}
\usepackage{rotating}
\usepackage{amsfonts}
\usepackage{longtable}
\usepackage{array} 

\usepackage{fancyhdr}
\begin{document}

\title{Handling Control System Uncertainty}
\date{}

\author{Hao Li 
\thanks{Namely \begin{CJK}{UTF8}{gbsn}李颢\end{CJK}, the same author of the works \cite{Li2026ACTPA_SJTU_2, Li2026ACTPA_SJTU_1}.} }

\maketitle

\begin{abstract}
\textbf{Control science} is a core representative of the third industrial revolution and is so important to modern civilization. \textbf{Control systems} are the main subject of control science and may involve many aspects of consideration, such as hardware consideration, software consideration, operation consideration, maintenance consideration, economy consideration, society consideration. However, besides all such aspects of consideration, one aspect that is most essential to the control system is methodology consideration in mathematical sense, knowledge on which is what we refer to as \textbf{control theory}. Besides its importance from the mathematical perspective, control theory is even more charming as it is deeply rooted in practical applications. Charms of control theory consist in both \textit{know-why} and \textit{know-how} and it is the fusion of control theory and practical applications that highlights such charms. Control theory for practical applications, especially when somewhat with so-called ``advanced'' flavour, involves several fundamental aspects. This article introduces the \textit{Handling Control System Uncertainty} aspect of \textit{Advanced Control Theory for Practical Applications} \cite{Li2026ACTPA_SJTU_2, Li2026ACTPA_SJTU_1}.
\end{abstract}

\section{Introduction}  \label{sec:PID_control}

Handling of control system uncertainty was already involved in as early as Chapter 2 \textit{Feedback Control} of the previous book \textit{Control Theory For Practical Applications} \cite{Li2024CTPA_Springer, Li2024CTPA_SJTU_1}, when the author discussed utilities of feedback control. The \textit{feedback control utility I} mentioned in Chapter 2 of the previous book is reviewed below.

\begin{framed} 
\noindent \textbf{Feedback control utility I}: \textit{Feedback control enables a control system even with uncertainty to guarantee control accuracy}.
\end{framed}

However, control system uncertainty has rarely been considered since Chapter 2 of the previous book. After discussions in Section 4.3 in Chapter 4 which may enlighten readers on importance of handling control system uncertainty, this article presents a number of representative methods for achieving this objective.

\subsection{Proportional-integral-derivative (PID) control}

\begin{figure}[h!]
\begin{center}
\includegraphics[width=0.85\columnwidth]{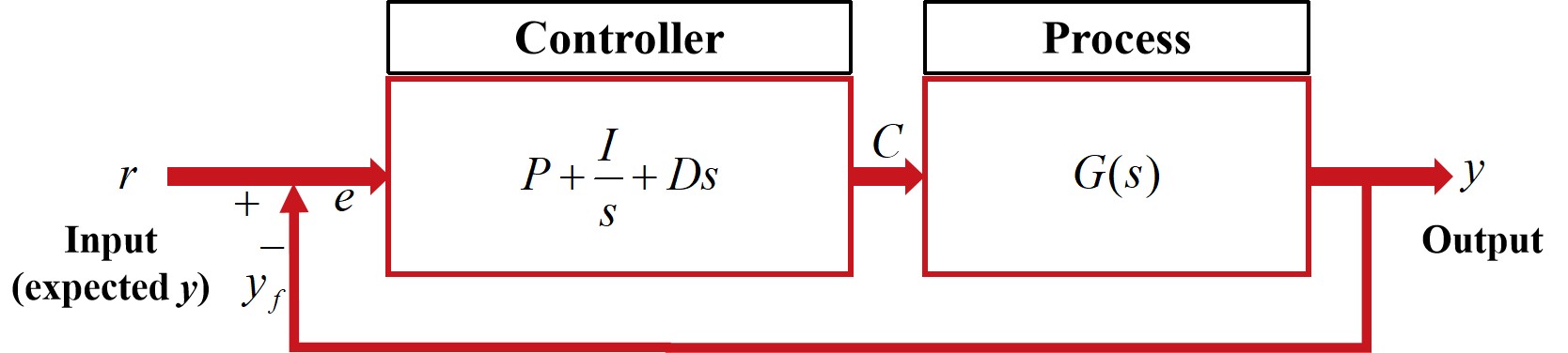}
\end{center}
\caption{Proportional-integral-derivative (PID) control}
\label{fig:PID_control}
\end{figure}

The \textbf{proportional-integral-derivative (PID)} control method or the family of PID controllers
\footnote{For expression convenience yet without causing confusion, we sometimes abuse the term \textit{control method} to mean both the general control methodology shared by a family of controllers and certain concrete controller or method with specific controller parameters in this family.},
as illustrated in Figure \ref{fig:PID_control}, is the most famous and popular control method \cite{Samad2017} and is probably the most representative control method that embodies the spirit of feedback control. The proportional-integral-derivative control method consists in generating the control law of $C$ namely the controller output by combining linearly a proportional term, an integral term, and a derivative term of the feedback error $e$ namely the controller input as
\begin{align}  \label{eq:PID_controller}
C &= P e + I \int e dt + D \frac{\mathrm{d}}{\mathrm{d} t} e,  \\
C(s) &= (P + \frac{I}{s} + D s) e(s), \nonumber
\end{align}
where the set of controller parameters for the PID control method is the set of proportional, integral, and derivative coefficients ($P$, $I$, $D$).

As explained in the previous book \textit{Control Theory For Practical Applications} \cite{Li2024CTPA_Springer, Li2024CTPA_SJTU_1}, the PID control method may also be called the \textit{proportional-integral-differential} control method where $D$ means \textit{differential}, and may also be formalized as
\begin{align}  \label{eq:PID_controller2}
C &= P e + \bar{I} \sum e + \bar{D} \mathrm{d} e,
\end{align}
where
\begin{align*}
\bar{I} \equiv I \mathrm{d} t, \qquad \bar{D} \equiv D / \mathrm{d} t.
\end{align*}
Be the proportional-integral-derivative control method or the proportional-integral-differential control method, the control method essence is the same. So we do not distinguish between the two kinds of PID terms.

The proportional-integral-derivative control method is a classical control method that is \textit{model-free}, or in other words, it does not demand any control system model. Note that a lack of any control system model might be regarded as the most severe kind of control system uncertainty, so for control tasks to which the proportional-integral-derivative control method is applicable, it naturally has ability to handle control system uncertainty.

On the other hand, the ability of the proportional-integral-derivative control method to handle control system uncertainty is so ``natural'' that the method itself seems to have nothing intentionally done to handle control system uncertainty. However, despite its popularity, the proportional-integral-derivative control method tends to be applicable only to comparatively simple control tasks. For comparatively complicated control tasks, the proportional-integral-derivative control method tends to be even not applicable, not to say any ability to handle control system uncertainty. In other words, the ability of the proportional-integral-derivative control method to handle control system uncertainty somehow relies on ``friendliness'' of control tasks. If control tasks are ``friendly'' to the proportional-integral-derivative control method, then its applicability and uncertainty handling ability are there. But if control tasks are complicated and hence ``unfriendly'' to it, then it has neither applicability nor uncertainty handling ability there.

\section{Sliding mode control}  \label{sec:sliding_mode}

\begin{figure}[h!]
\begin{center}
\includegraphics[width=0.4\columnwidth]{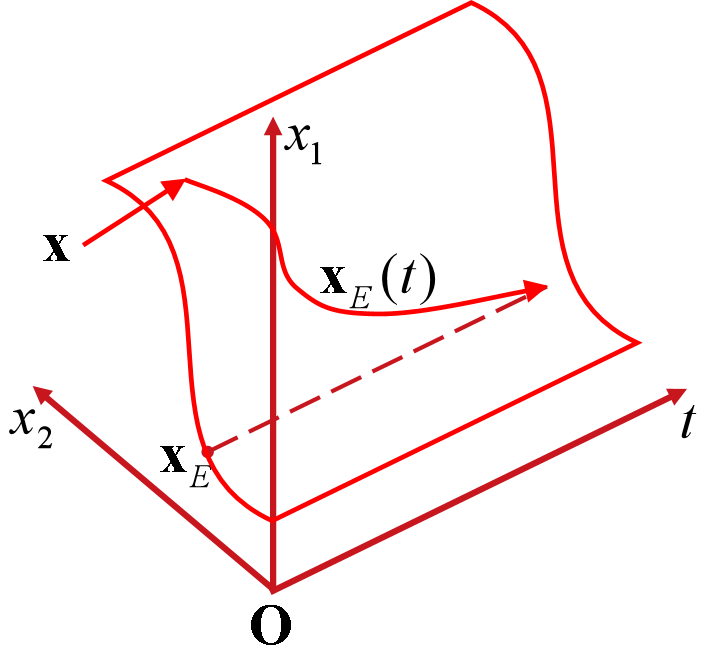}
\end{center}
\caption{Sliding mode control}
\label{fig:sliding_mode_control}
\end{figure}

Unlike proportional-integral-derivative control that seems to have nothing intentionally done to handle control system uncertainty but somehow relies on ``friendliness'' of control tasks, \textbf{sliding mode control} \cite{Utkin1977, Utkin1992, Shtessel2014} handles control system uncertainty by \textit{intentionally forcing the state to evolve onto certain designed sliding mode manifold or surface and then evolve towards the expected state in sliding mode on the manifold}. Figuratively speaking, this is like to drive the state to slide along certain sliding path on the designed manifold until the state arrives at the destination, as illustrated in Figure \ref{fig:sliding_mode_control}. It is worth noting that sliding mode control may take advantage of proportional-integral-derivative control or other kind of control to force the state to evolve onto the designed sliding mode manifold.

\begin{figure}[h!]
\begin{center}
\includegraphics[width=0.35\columnwidth]{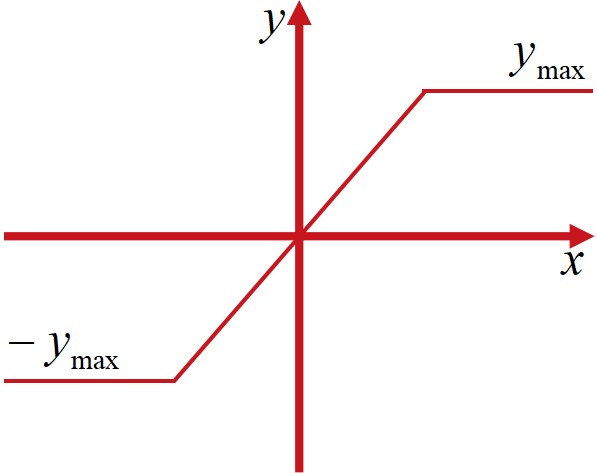}
\end{center}
\caption{Actuator saturation}
\label{fig:actuator_saturation}
\end{figure}

\begin{figure}[h!]
\begin{center}
\includegraphics[width=0.35\columnwidth]{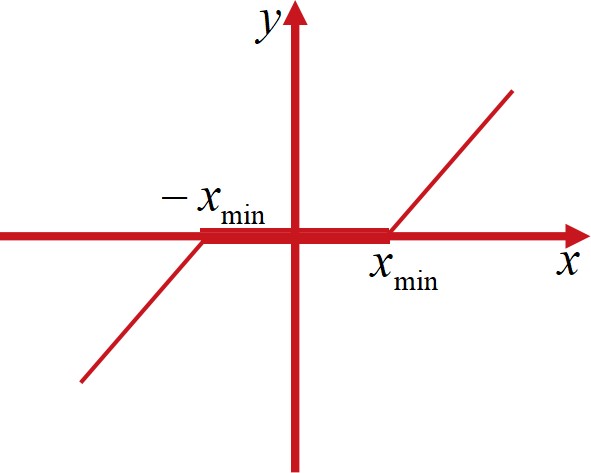}
\end{center}
\caption{Dead zone}
\label{fig:dead_zone}
\end{figure}

\begin{figure}[h!]
\begin{center}
\includegraphics[width=0.35\columnwidth]{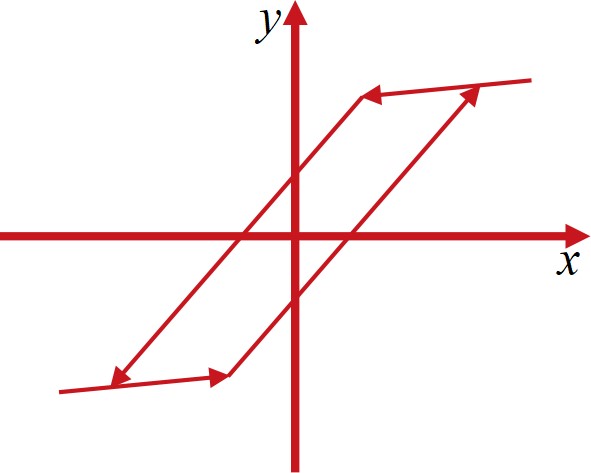}
\end{center}
\caption{Hysteresis}
\label{fig:hysteresis}
\end{figure}

Originally, sliding mode control theory aims at control systems with \textit{discontinuous dynamics} --- Existence of discontinuous dynamics may be due to a variety of factors, for example, imperfections of switching devices (time delay, dead zones, hysteresis loops), actuator saturation, model order mismatch, state feedback delay, piecewise linear approximation, etc. Some of such factors are illustrated in Figure \ref{fig:actuator_saturation}, Figure \ref{fig:dead_zone}, and Figure \ref{fig:hysteresis} --- Yet the basic spirit of sliding mode control is applicable to control systems with or without discontinuous dynamics.

\subsection{Sliding mode augmented full-state feedback control}  \label{sec:sliding_mode_augmented_FSFC}

We take concrete application examples to demonstrate how the full-state feedback control method can be augmented to more powerful versions in the spirit of sliding mode control.

\subsubsection*{Application: double inverted pendulum sliding mode control}

Consider the double inverted pendulum control method presented in Section 2.2.3 in Chapter 2. 
\footnote{Namely Chapter 2 of the author's works \cite{Li2026ACTPA_SJTU_2, Li2026ACTPA_SJTU_1}. Note that this article is Chapter 5 of the works.}
Let 
\begin{align*}  
m_1 = 1, \quad m_2 = 1, \quad L_1 = 1, \quad L_2 = 1, \quad g = 10, 
\end{align*}
and set the expected closed-loop characteristic polynomial as
\begin{align*}
C_\mathrm{E}(s) = (s + 4)^6 = s^6 + 24 s^5 + 240 s^4 + 1280 s^3 + 3840 s^2 + 6144 s + 4096
\end{align*}
for example. Then the corresponding gain matrix is
\begin{align*}
\mathbf{K} = \begin{bmatrix} -259.52 & 6.72 & 482.48 & 118.72 & 20.48 & 30.72 \end{bmatrix}^\mathrm{T}
\end{align*}
and the full-state feedback control law is
\begin{align*}
a = - \mathbf{K}^\mathrm{T} \mathbf{x} = - \begin{bmatrix} -259.52 & 6.72 & 482.48 & 118.72 & 20.48 & 30.72 \end{bmatrix} \begin{bmatrix} \theta_1 \\ \frac{\mathrm{d} \theta_1}{\mathrm{d} t} \\ \theta_2 \\ \frac{\mathrm{d} \theta_2}{\mathrm{d} t} \\ x \\ \frac{\mathrm{d} x}{\mathrm{d} t} \end{bmatrix}.
\end{align*}

The full-state feedback control law works for the double inverted pendulum control problem, if the initial deviation of the cart position $x$ is not large. However, if the initial cart position $x$ is set to a large value, the full-state feedback control law presented above no longer works. Reasons for failure of the full-state feedback control law when the initial cart position $x$ is large are two-folds: First, the state differential equation described in (1.3) that models double inverted pendulum dynamics is nonlinear and hence its linearly-approximated version described in (1.13) based on which the adopted full-state feedback control law is designed is not always valid for state space of the double inverted pendulum state. Second, the adopted full-state feedback control law only focuses on converging the final state to the expected state as soon as possible, without considering intermediate state evolution during the control process. When the initial deviation of the cart position is large, the adopted full-state feedback control law tends to generate drastic control input of cart acceleration, without considering how the double inverted pendulum state will evolve accordingly. Consequently, drastic control input of cart acceleration causes the double inverted pendulum state to evolve into state space for which the linearly-approximated model version described in (1.13) is no longer valid
\footnote{Namely (1.3) and (1.13) in the author's works \cite{Li2026ACTPA_SJTU_2, Li2026ACTPA_SJTU_1}.}.

In fact, for the double inverted pendulum control problem given arbitrary initial cart position, even above concrete full-state feedback control law can still work, not to say a better version or an optimal version of full-state feedback control. We only need to improve the original full-state feedback control method moderately by incorporating the basic spirit of sliding mode control. For the double inverted pendulum control problem, we can intuitively set the sliding mode manifold as
\begin{align*}
\begin{bmatrix} \mathbf{I}_4 & \mathbf{0} \end{bmatrix} \mathbf{x} = \mathbf{0}
\end{align*} 
namely
\begin{align*}
\begin{bmatrix} \theta_1 & \frac{\mathrm{d} \theta_1}{\mathrm{d} t} & \theta_2 & \frac{\mathrm{d} \theta_2}{\mathrm{d} t} \end{bmatrix}^\mathrm{T} = \begin{bmatrix} 0 & 0 & 0 & 0 \end{bmatrix}^\mathrm{T}
\end{align*}
and design a sliding mode as 
\begin{align}  \label{eq:DIP_sliding_mode}
\mathbf{x}_\mathrm{E} = \begin{bmatrix} 0 & 0 & 0 & 0 & \mbox{sign}(x_0) \max \{|x_0| - s_v t, 0\} & 0 \end{bmatrix}^\mathrm{T},
\end{align}
where $s_v$ is a constant parameter that tunes the sliding velocity. For example, set 
\begin{align*}
s_v = 8. 
\end{align*} 
Then the full-state feedback control law of cart acceleration $a$ is
\begin{align*}
a = -\mathbf{K}^\mathrm{T} (\mathbf{x} - \mathbf{x}_\mathrm{E}),
\end{align*}
where $\mathbf{x}_\mathrm{E}$ is no longer a constant but a time-variant function described in (\ref{eq:DIP_sliding_mode}). 

Matlab simulation code for demonstrating sliding mode control of the double inverted pendulum is given as follows. The visualization code \textbf{DisplayDIP.m} and the double inverted pendulum dynamics code \textbf{DynamicsDIP.m} are given in Section 2.2.1 in Chapter 2.

\begin{framed} 
\noindent \textbf{DoubleInvertedPendulumSMC.m} \\
\noindent \%\% Double inverted pendulum parameters \\
m1 = 1; m2 = 1; L1 = 1; L2 = 1; g = 10; \\
\%\% Simulation preliminary configuration \\
dt = 0.001; \% Numerical computation step \\
tSpan = 0:dt:8; \% Simulation time span \\
x = 20; dx = 0; \% Cart position and its velocity \\
y1 = 0.2; dy1 = 0;  \% Inverted pendulum angle theta-1 and its angular velocity \\
y2 = 0; dy2 = 0;  \% Inverted pendulum angle theta-2 and its angular velocity \\
stt = [y1; dy1; y2; dy2; x; dx]; \% Double inverted pendulum state \\
sttAll = zeros(length(stt), length(tSpan)); k = 0; \% Record states in simulation  \\
xExpected = 0; y1Expected = 0; y2Expected = 0; \% Expected equilibrium status \\
SimConfig = [m1, m2, L1, L2, g, dt]; \\
 \\
\%\% Design the gain matrix \\
A = [0, 1, 0, 0, 0, 0; ... \\
$~~~~$ (m1+m2)*g/(m1*L1), 0, -m2*g/(m1*L1), 0, 0, 0; ... \\
$~~~~$ 0, 0, 0, 1, 0, 0; ... \\
$~~~~$ -(m1+m2)*g/(m1*L2), 0, (m1+m2)*g/(m1*L2), 0, 0, 0; ... \\
$~~~~$ 0, 0, 0, 0, 0, 1; ... \\
$~~~~$ 0, 0, 0, 0, 0, 0]; \\
B = [0; -1/L1; 0; 0; 0; 1]; \\
lambdaE = [-4;-4;-4;-4;-4;-4]; \% Expected eigenvalues \\
sttK = DesignGainMatrix(A, B, lambdaE); \\
fprintf('Gain matrix K: '); sttK' \\
 \\
sttE = [0; 0; 0; 0; x; 0]; \% Sliding mode initialization \\
\%\% Simulation of double inverted pendulum control \\
for t = tSpan \\
$~~~~$ \%\% Control method \\
$~~~~$ if (sttE(5)$>$0) sttE(5) = max(sttE(5) - 8*dt, 0); \% Sliding mode design \\
$~~~~$ else sttE(5) = min(sttE(5) + 8*dt, 0); end \\
$~~~~$ acc = -sttK'*(stt-sttE); \% Full-state feedback control \\
$~~~~$  \\
$~~~~$ \%\% Double inverted pendulum dynamics \\
$~~~~$ stt = DynamicsDIP(SimConfig, stt, acc); \\
$~~~~$ sttC = num2cell(stt); [y1, dy1, y2, dy2, x, dx] = sttC\{:\}; \\
$~~~~$ if (abs(y1)$>$=pi/2 \&\& abs(y2)$>$=pi/2) fprintf('Control failure!$\backslash$n'); break; end \\
$~~~~$ k = k+1; sttAll(:,k) = stt; \\
$~~~~$ \%\% Double inverted pendulum visualization \\
$~~~~$ if (rem(k,20) == 0) \\
$~~~~$ $~~~~$ DisplayDIP(x, y1, y2, L1, L2); pause(20*dt); \\
$~~~~$ end \\
end
\end{framed}

\begin{figure}[h!]
\begin{center}
\includegraphics[width=1.0\columnwidth]{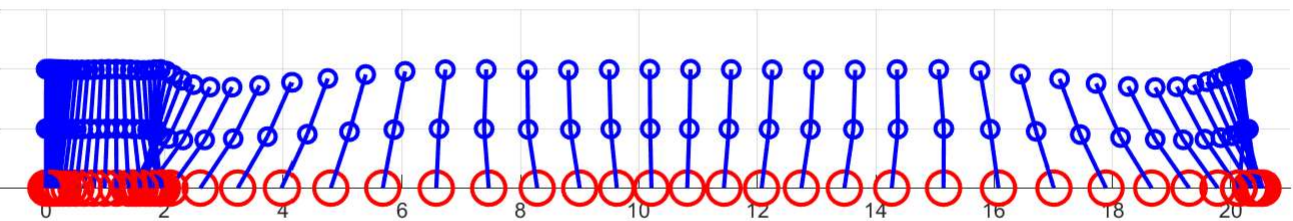}
\end{center}
\caption{Double inverted pendulum sliding mode control}
\label{fig:SlidingModeControlDIP}
\end{figure}

The performance of double inverted pendulum sliding mode control is demonstrated in Figure \ref{fig:SlidingModeControlDIP}. The concrete value $20$ to which the initial deviation of the cart position $x$ is set has no special meaning other than that purely for demonstration purpose. Readers may try the simulation code and will find that no matter how far away the initial cart position $x$ is set, the augmented version of the full-state feedback control method will always succeed in controlling the double inverted pendulum to the expected state.

\subsubsection*{Application: motorcycle sliding mode control}

The application of motorcycle lateral control for sake of motorcycle lane keeping is already demonstrated in Section 2.2.3 in Chapter 2. 
\footnote{Namely Chapter 2 of the author's works \cite{Li2026ACTPA_SJTU_2, Li2026ACTPA_SJTU_1}. Note that this article is Chapter 5 of the works.}
Now consider motorcycle control more general than lane keeping oriented motorcycle lateral control, namely to control the motorcycle to move from the initial pose to an arbitrary destination pose. Let 
\begin{align*}  
L = 1.5, \quad H = 1, \quad \tau_{\beta} = 0.02, \quad g = 10, \quad v = 10. 
\end{align*}
Set the motorcycle initial pose and the motorcycle destination pose as
\begin{align*}
\begin{bmatrix} x_I \\ y_I \\ \phi_I \end{bmatrix} = \begin{bmatrix} 0 \\ -0.2 \\ -0.1 \end{bmatrix}, \qquad \begin{bmatrix} x_D \\ y_D \\ \phi_D \end{bmatrix} = \begin{bmatrix} 32.62 \\ 21.80 \\ \frac{\pi}{4} \end{bmatrix}
\end{align*}
respectively for example. 

For the control task with above concrete configuration, set two sliding manifolds characterized by the poses
\begin{align*}
\mathbf{P}_1 = \begin{bmatrix} x_I \\ y_I \\ \frac{\phi_D}{2} \end{bmatrix},  \qquad \mathbf{P}_2 = \begin{bmatrix} x_D \\ y_D \\ \phi_D \end{bmatrix}
\end{align*}
respectively. They are actually two lines parametrized by the two poses, namely
\begin{align*}
\mbox{sliding mode (line) 1 :} \quad & \cos \frac{\phi_D}{2} (x - x_I) - \sin \frac{\phi_D}{2} (y - y_I) = 0,  \\
\mbox{sliding mode (line) 2 :} \quad & \cos \phi_D (x - x_D) - \sin \phi_D (y - y_D) = 0.
\end{align*}
The two sliding mode lines intersect at a turning point
\begin{align*}
\begin{bmatrix} x_M \\ y_M \end{bmatrix} = \begin{bmatrix} x_I \\ y_I \end{bmatrix} + \begin{bmatrix} \cos \frac{\phi_D}{2} \\ \sin \frac{\phi_D}{2} \end{bmatrix} \begin{bmatrix} 1 & 0 \end{bmatrix} \begin{bmatrix} \cos \frac{\phi_D}{2} & - \cos \phi_D \\ \sin \frac{\phi_D}{2} & - \sin \phi_D \end{bmatrix}^{-1} \begin{bmatrix} x_D - x_I \\ y_D - y_I \end{bmatrix}.
\end{align*}

First control the motorcycle to slide on the sliding mode line 1 until it approaches the sliding mode line 2. Once the motorcycle approaches within certain preview distance from the sliding mode line 2, then switch to slide on the sliding mode line 2 until it arrives at the destination pose. To force the motorcycle to slide on a target sliding mode line
\begin{align*}
\cos \phi_S (x - x_S) - \sin \phi_S (y - y_S) = 0,
\end{align*}
transform motorcycle coordinates from the global reference into the coordinates system aligned with the sliding mode line as
\begin{align*}
\begin{bmatrix} \bar{x} \\ \bar{y} \end{bmatrix} &= \begin{bmatrix} \cos \phi_S & \sin \phi_S \\ - \sin \phi_S & \cos \phi_S \end{bmatrix} \begin{bmatrix} x - x_S \\ y - y_S \end{bmatrix},  \\
\bar{\phi} &= \phi - \phi_S,
\end{align*}
then perform full-state feedback control with the transformed coordinates as
\begin{align*}
\beta_I = - \mathbf{K}^\mathrm{T} \begin{bmatrix} \bar{y} \\ \bar{\phi} \\ \theta \\ \frac{\mathrm{d} \theta}{\mathrm{d} t} \end{bmatrix},
\end{align*}
which is similar to lane keeping oriented motorcycle lateral control. In above formalism, the pose
\begin{align*}
\begin{bmatrix} x_S \\ y_S \\ \phi_S \end{bmatrix} \in \{ \begin{bmatrix} x_I \\ y_I \\ \frac{\phi_D}{2} \end{bmatrix}, \begin{bmatrix} x_D \\ y_D \\ \phi_D \end{bmatrix} \},
\end{align*}
depends on which one of the sliding lines 1 and 2 is current target sliding mode line.

Here, the gain matrix $\mathbf{K}$ can be designed in the same way as that presented in Section 2.2.3 in Chapter 2. More specifically, design the gain matrix with the simplified system model
\begin{align}  \label{eq:motorcycle_lateral_control_approximation_beta_tau=0}
\frac{\mathrm{d}}{\mathrm{d} t} \mathbf{x} = \begin{bmatrix} 0 & v & 0 & 0 \\ 0 & 0 & 0 & 0 \\ 0 & 0 & 0 & 1 \\ 0 & 0 & \frac{g}{H} & 0 \end{bmatrix} \mathbf{x} + \begin{bmatrix} 0 \\ \frac{v}{L} \\ 0 \\ -\frac{v^2}{H L} \end{bmatrix} \beta \equiv \mathbf{A} \mathbf{x} + \mathbf{B} \beta,
\end{align}
set the expected closed-loop characteristic polynomial as
\begin{align*}
C_\mathrm{E}(s) = (s + \frac{5}{2})^4 = s^4 + 10 s^3 + \frac{75}{2} s^2 + \frac{125}{2} s + \frac{625}{16},
\end{align*}
and design the gain matrix as
\begin{align*}
\mathbf{K} = \begin{bmatrix} -0.0586 & -0.9375 & -0.7711 & -0.2437 \end{bmatrix}^\mathrm{T}.
\end{align*}
Matlab simulation code for complete demonstration of motorcycle sliding mode control is given as follows.

\begin{framed} 
\noindent \textbf{MotorcycleControlSMC.m} \\
\noindent \%\% Motorcycle parameters \\
L = 1.5; \% Motorcycle wheel-base \\
H = 1; \% Motorcycle gravity center height \\
tb = 0.02; \% Steer time-constant \\
g = 10; \% Gravity coefficient \\
\%\% Simulation preliminary configuration \\
dt = 0.001; \% Numerical computation step \\
tSpan = 0:dt:10; \% Simulation time span \\
x = 0; \% Motorcycle x-position \\
y = -0.2; \% Motorcycle y-position \\
phi = -0.1; \% Motorcycle orientation (yaw angle) \\
b = 0; \% Motorcycle steering angle \\
a = 0.3; \% Motorcycle vertical angle (roll angle) \\
da = 0; \% Motorcycle vertical angular velocity \\
stt = [x; y; phi; b; a; da]; \% Motorcycle state \\
sttAll = zeros(length(stt), length(tSpan)); k = 0; \% Record states \\
SimConfig = [L, H, tb, dt, g]; \\
\%\% Design the gain matrix \\
phiI = pi/8; phiD = pi/4; Dm1 = 20; Dm2 = 20;  \\
posI = [x; y; phiI]; \% Initial sliding mode characterized by pose \\
xD = Dm1*cos(phiI)+Dm2*cos(phiD); yD = Dm1*sin(phiI)+Dm2*sin(phiD); \\
posD = [xD; yD; phiD]; \% Destination sliding mode characterized by pose \\
tID = [cos(phiI),-cos(phiD);sin(phiI),-sin(phiD)]$\backslash$(posD(1:2)-posI(1:2)); \\
xM = posI(1)+cos(phiI)*tID(1); yM = posI(2)+sin(phiI)*tID(1); \\
posS = posI; \% Sliding mode 1 \\
vC = 10; \% vC : velocity/speed control (longitudinal control) \\
A = [0, vC, 0, 0; 0, 0, 0, 0; 0, 0, 0, 1; 0, 0, g/H, 0]; \\
B = [0; vC/L; 0; -vC\^{}2/(H*L)]; \\
Kc = DesignGainMatrix(A, B, -ones(4,1)*2.5); \\
 \\
\%\% Simulation of motorcycle control \\
for t = tSpan \\
$~~~~$ \%\% Sliding mode control \\
$~~~~$ \% Switch from sliding mode 1 to sliding mode 2 \\
$~~~~$ previewSMC = 6.0; \% Preview distance for sliding mode switching \\
$~~~~$ if (sqrt((x-xM)\^{}2+(y-yM)\^{}2)$<$previewSMC) posS = posD; end \\
$~~~~$ \% Slide on manifold \\
$~~~~$ \% Rotation matrix: [cos(phiS),sin(phiS);-sin(phiS),cos(phiS)] \\
$~~~~$ posSC = num2cell(posS); [xS, yS, phiS] = posSC\{:\}; \\
$~~~~$ sC = -Kc'*[-sin(phiS)*(x-xS)+cos(phiS)*(y-yS); phi-phiS; a; da]; \\
 \\
$~~~~$ \%\% Motorcycle dynamics \\
$~~~~$ stt = DynamicsMotorcycle(SimConfig, stt, sC, vC); \\
$~~~~$ sttC = num2cell(stt); [x, y, phi, b, a, da] = sttC\{:\}; \\
$~~~~$ if (abs(a)$>$=pi/2) fprintf('Control failure!$\backslash$n'); break; end \\
$~~~~$ k = k+1; sttAll(:,k) = stt; \\
$~~~~$ if (sqrt((x-posD(1))\^{}2+(y-posD(2))\^{}2)$<$0.2) \\
$~~~~$ $~~~~$ fprintf('Arrive at destination!$\backslash$n');  \\
$~~~~$ $~~~~$ sttAll = sttAll(:,1:k); break; \\
$~~~~$ end \\
$~~~~$ \%\% Motorcycle lateral state visualization \\
$~~~~$ if (rem(k,20) == 0) \\
$~~~~$ $~~~~$ DisplayMotorcycleStateSMC(stt, SimConfig, posI, posD, posS, 3); \\
$~~~~$ $~~~~$ pause(dt); \\
$~~~~$ end \\
end
\end{framed}

The motorcycle dynamics code \textbf{DynamicsMotorcycle.m} in above code has already been given in Section 2.2.3 in Chapter 2. The gain matrix designing code \textbf{DesignGainMatrix.m} is given in Section 2.3.2 in Chapter 2. 
\footnote{Namely Chapter 2 of the author's works \cite{Li2026ACTPA_SJTU_2, Li2026ACTPA_SJTU_1}. Note that this article is Chapter 5 of the works.}
The motorcycle state visualization code \textbf{DisplayMotorcycleStateSMC.m} is given as follows.

\begin{framed} 
\noindent \textbf{DisplayMotorcycleStateSMC.m} \\
\noindent \%\% Motorcycle state visualization for sliding mode control \\
function DisplayMotorcycleStateSMC(stt, SimConfig, posI, posD, posS, W) \\
$~~~~$ sttC = num2cell(stt); [x, y, phi, b, a, da] = sttC\{:\};  \\
$~~~~$ SimConfigC = num2cell(SimConfig); [L, H, tb, dt, g] = SimConfigC\{:\}; \\
$~~~~$ phiI = posI(3); phiD = posD(3); xyDI = posD(1:2)-posI(1:2); \\
$~~~~$ tID = [cos(phiI),-cos(phiD);sin(phiI),-sin(phiD)]$\backslash$xyDI; \\
$~~~~$ xM = posI(1)+cos(phiI)*tID(1); yM = posI(2)+sin(phiI)*tID(1); \\
$~~~~$ xmin=min([posI(1),posD(1),xM])-W; xmax = max([posI(1),posD(1),xM])+W; \\
$~~~~$ ymin=min([posI(2),posD(2),yM])-W; ymax = max([posI(2),posD(2),yM])+W; \\
$~~~~$ posSC = num2cell(posS); [xS, yS, phiS] = posSC\{:\}; \\
$~~~~$ yL = -sin(phiS)*(x-xS)+cos(phiS)*(y-yS); \\
$~~~~$ figure(1), clf; set(gcf, 'Position', [100, 0, 1000, 800]); \\
$~~~~$ subplot(2,1,1), mapC = [0.8 0.8 0.8]; \\
$~~~~$ patch([xmin,xmax,xmax,xmin], [ymin,ymin,ymax,ymax], ... \\
$~~~~$ $~~~~$ mapC, 'EdgeColor', mapC); axis equal; axis off; hold on; \\
$~~~~$ plot([posI(1),xM],[posI(2),yM],'b- -','LineWidth',2); \\
$~~~~$ plot([xM,posD(1)],[yM,posD(2)],'b- -','LineWidth',2); \\
$~~~~$ x1 = x+L*cos(phi); y1 = y+L*sin(phi); wr = 0.3*H; pb = phi+b; \\
$~~~~$ line([x1-wr*cos(pb),x1+wr*cos(pb)], [y1-wr*sin(pb),y1+wr*sin(pb)], ... \\
$~~~~$ $~~~~$ 'Color', 'k', 'LineWidth', 3); \\
$~~~~$ line([x-wr*cos(phi),x+wr*cos(phi)], [y-wr*sin(phi),y+wr*sin(phi)], ... \\
$~~~~$ $~~~~$ 'Color', 'k', 'LineWidth', 3); \\
$~~~~$ line([x, x1], [y, y1], 'Color', 'r', 'LineWidth', 2); hold off; \\
$~~~~$ subplot(2,1,2), patch([-W,-W,W,W], ... \\
$~~~~$ $~~~~$ [0,-0.2,-0.2,0], mapC, 'EdgeColor', mapC); hold on;  \\
$~~~~$ line([yL,yL+2*wr*sin(a)],[0,2*wr*H*cos(a)],'Color','k','LineWidth',9); \\
$~~~~$ line([yL+H*sin(a)-(H/4)*cos(a),yL+H*sin(a)+(H/4)*cos(a)], ... \\
$~~~~$ $~~~~$ [H*cos(a)+(H/4)*sin(a),H*cos(a)-(H/4)*sin(a)], ... \\
$~~~~$ $~~~~$ 'Color', 'k', 'LineWidth', 3); \\
$~~~~$ line([yL, yL+H*sin(a)], [0, H*cos(a)], 'Color', 'r', 'LineWidth', 3);  \\
$~~~~$ plot(yL+H*sin(a),H*cos(a),'or','LineWidth',3); \\
$~~~~$ ylim([0, 1.2*H]); axis equal; axis off; hold off; \\
end
\end{framed}

\begin{figure}[h!]
\begin{center}
\includegraphics[width=0.6\columnwidth]{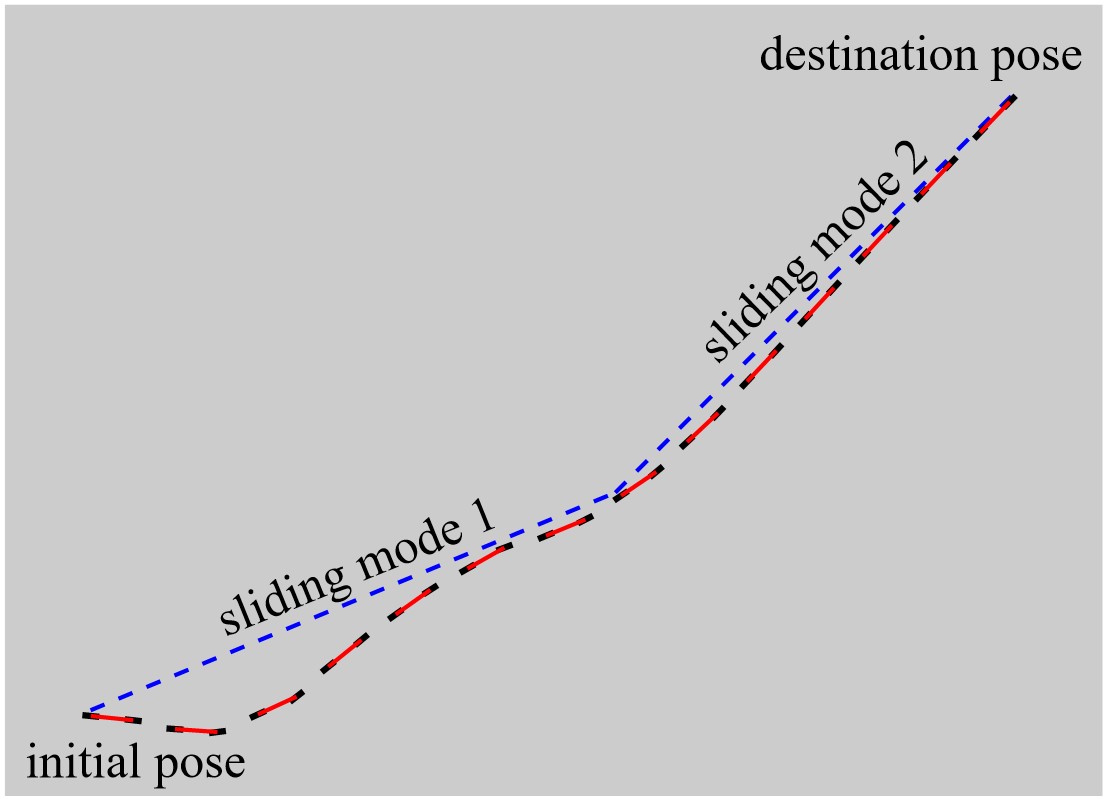}
\end{center}
\caption{Motorcycle sliding mode control}
\label{fig:motorcycle_sliding_mode_control}
\end{figure}

The performance of motorcycle sliding mode control is demonstrated in Figure \ref{fig:motorcycle_sliding_mode_control}. Readers may try the simulation code and will also find that the motorcycle can succeed in navigating automatically from the initial pose to the destination pose.

\subsection{Not best not bad}

\begin{figure}[h!]
\begin{center}
\includegraphics[width=0.6\columnwidth]{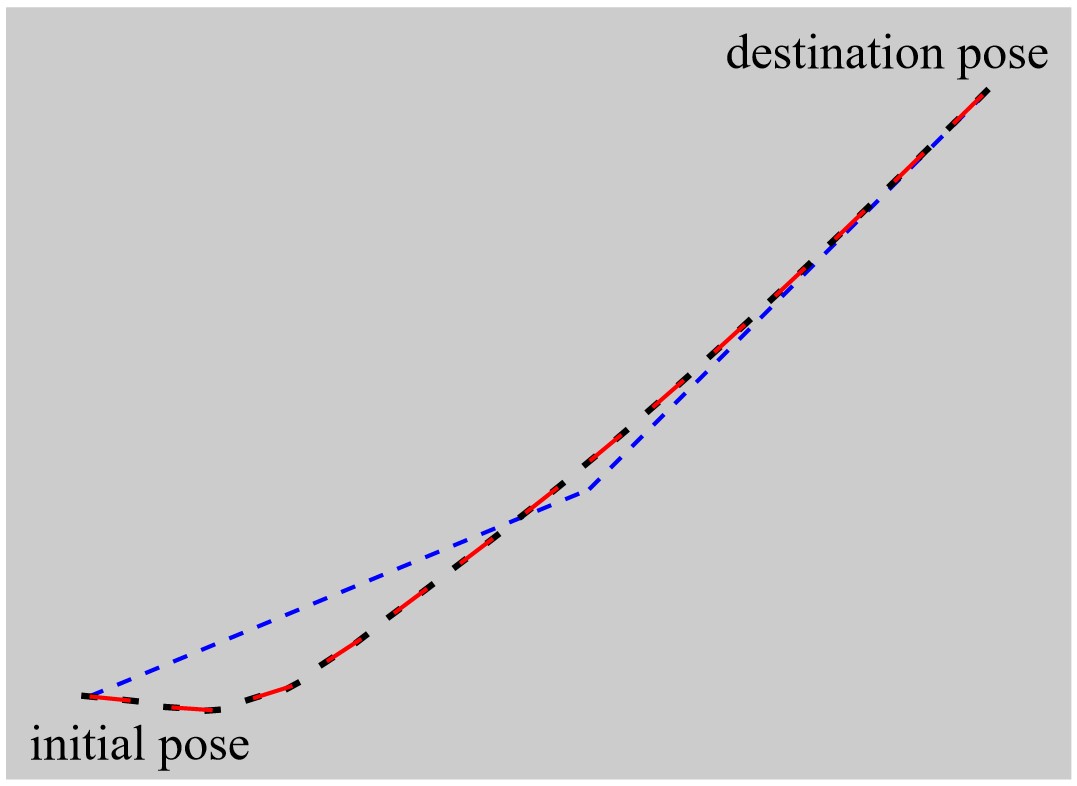}
\end{center}
\caption{Smoother motorcycle control}
\label{fig:motorcycle_smoother_than_smc}
\end{figure}

Some clarifications hover over characteristics of sliding mode control. A characteristic worth noting is that sliding mode control usually is not the optimal one among all methods or methodologies that can handle intended control tasks. Take above demonstrated motorcycle control as example, a version of control performance which is apparently better in terms of control smoothness can be achieved, as illustrated in Figure \ref{fig:motorcycle_smoother_than_smc}. Simply speaking, sliding mode control is \textit{not best}.

On the other hand, another characteristic worth noting is that sliding mode control has the merit of guaranteeing ``quality'' of state space regions into which the state may evolve. In other words, sliding mode control has the merit of forcing the state to evolve into state space regions in which guaranteed control performance can be achieved. By analogy, it is like in ancient times when people sailed on ocean, they usually chose ocean regions not too far away from continents for sailing. Although such way of sailing tends to incur a sailing distance larger than the globally optimal one, it at least guarantees that people do sail safely and arrive successfully at their destination. Such way of sailing in ancient times well reflects spirit of sliding mode control. Simply speaking, sliding mode control is \textit{not bad}, especially not bad in handling control system uncertainty because it forces the state to evolve only in state space regions without uncertainty or at least without severe uncertainty.

\begin{figure}[h!]
\begin{center}
\includegraphics[width=0.35\columnwidth]{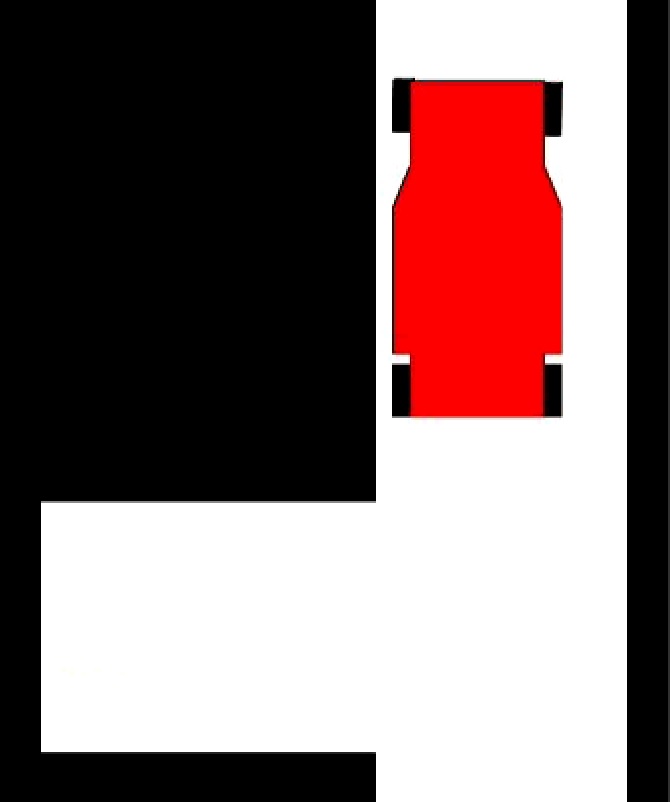}
\end{center}
\caption{Narrow space parking}
\label{fig:narrow_space_parking}
\end{figure}

Above two characteristics together convey that sliding mode control embodies the spirit of \textit{not best (but at least) not bad} control, or simply the spirit of \textit{not best not bad (NBNB)} control. It is especially useful for handling challenging control tasks such as narrow space parking 
\footnote{It is worth noting that people rarely encounter such kind of extreme scenarios in reality, yet the video of narrow space parking is just to demonstrate the ability to handle such challenging control task.}
illustrated in Figure \ref{fig:narrow_space_parking}.

\section{Robust control} \label{sec:robust_control}

It is mentioned that by sliding mode control we handle control system uncertainty in the spirit of forcing the state to evolve only in state space regions that tend to be exempt from uncertainty. On the other hand, what if the control system somewhat suffers from uncertainty definitely, be the uncertainty belonging to \textit{contingency uncertainty} or \textit{modelling uncertainty}.

\begin{figure}[h!]
\begin{center}
\includegraphics[width=0.9\columnwidth]{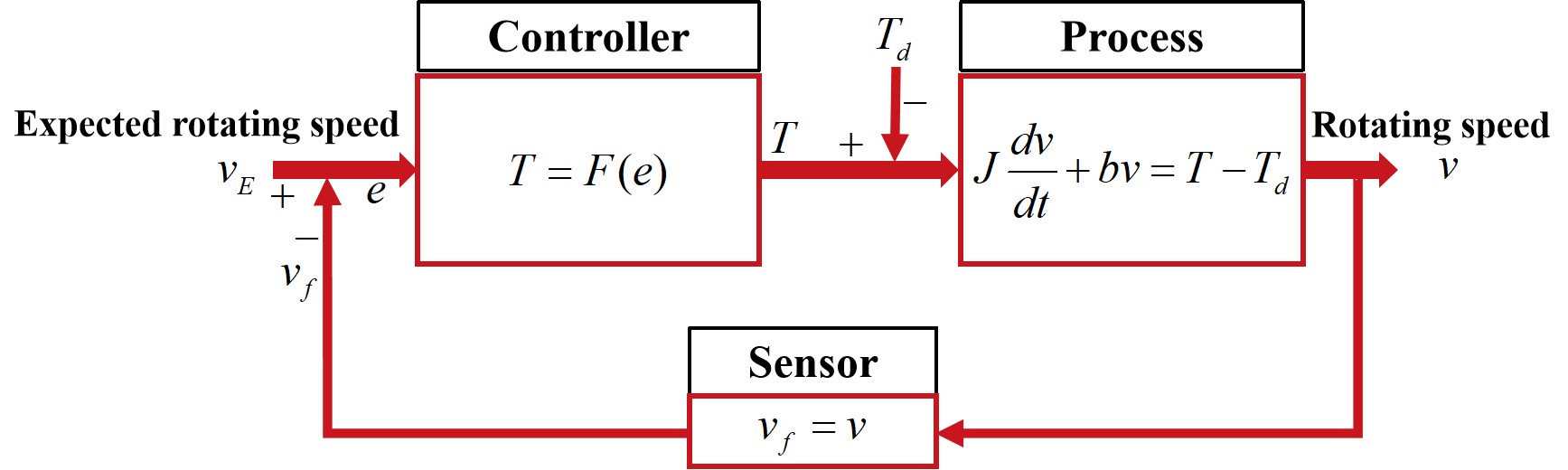}
\end{center}
\caption{Rotating disk speed control with contingency uncertainty}
\label{fig:contingency_uncertainty_RDS}
\end{figure}

Simply speaking, contingency uncertainty refers to uncertainty caused by stochastic or \textit{ad hoc} factors that people cannot know or expect \textit{a priori}. For example, consider the rotating disk speed control system illustrated in Figure \ref{fig:contingency_uncertainty_RDS}, where $T$ denotes the controller output torque and $T_d$ denotes torque disturbance. The rotating disk speed control system suffers from contingency uncertainty because the torque disturbance $T_d$ is normally unknown or cannot be expected \textit{a priori} in practical applications.

Modelling uncertainty refers to uncertainty caused by limitation of what the adopted model can describe, or even in more general terms, caused by limitation of people's knowledge of the objective world. For example, consider the single inverted pendulum control system and suppose it adopts linear state-space modelling described by
\begin{align}  \label{eq:SIP_state_DE_linear}  
\frac{\mathrm{d}}{\mathrm{d} t} \mathbf{x} \equiv \frac{\mathrm{d}}{\mathrm{d} t} \begin{bmatrix} \theta \\ \frac{\mathrm{d} \theta}{\mathrm{d} t} \\ x \\ \frac{\mathrm{d} x}{\mathrm{d} t} \end{bmatrix} = \begin{bmatrix} 0 & 1 & 0 & 0 \\ \frac{g}{L} & 0 & 0 & 0 \\ 0 & 0 & 0 & 1  \\ 0 & 0 & 0 & 0 \end{bmatrix} \begin{bmatrix} \theta \\ \frac{\mathrm{d} \theta}{\mathrm{d} t} \\ x \\ \frac{\mathrm{d} x}{\mathrm{d} t} \end{bmatrix} + \begin{bmatrix} 0 \\ -\frac{1}{L} \\ 0 \\ 1 \end{bmatrix} a \equiv \mathbf{A} \mathbf{x} + \mathbf{B} a.
\end{align}
It is worth noting that although we know that dynamics of the single inverted pendulum control system's state
\begin{align*}
\mathbf{x} \equiv \begin{bmatrix} \theta & \frac{\mathrm{d} \theta}{\mathrm{d} t} & x & \frac{\mathrm{d} x}{\mathrm{d} t} \end{bmatrix}^\mathrm{T}
\end{align*}
should be described by the state differential equation
\begin{align}  \label{eq:SIP_state_DE}
\frac{\mathrm{d}}{\mathrm{d} t} \mathbf{x} \equiv \frac{\mathrm{d}}{\mathrm{d} t} \begin{bmatrix} \theta \\ \frac{\mathrm{d} \theta}{\mathrm{d} t} \\ x \\ \frac{\mathrm{d} x}{\mathrm{d} t} \end{bmatrix} = \begin{bmatrix} \frac{\mathrm{d} \theta}{\mathrm{d} t} \\ \frac{\sin \theta}{L} g - \frac{\cos \theta}{L} a \\ \frac{\mathrm{d} x}{\mathrm{d} t} \\ a \end{bmatrix} \equiv f(\mathbf{x}, a),
\end{align}
as the linear model formalism (\ref{eq:SIP_state_DE_linear}) is actually adopted, the single inverted pendulum control system is analysed and handled as if we have limited knowledge of it. 

It is right the discrepancy between the actually adopted linear model
\begin{align*}
\frac{\mathrm{d}}{\mathrm{d} t} \mathbf{x} = \mathbf{A} \mathbf{x} + \mathbf{B} a
\end{align*}
and the nonlinear model
\begin{align*}
\frac{\mathrm{d}}{\mathrm{d} t} \mathbf{x} = f(\mathbf{x}, a)
\end{align*}
that reflects our ``limited knowledge'' and causes modelling uncertainty. In fact, above clarification of existence of modelling uncertainty can not only be applied to the single inverted pendulum control system but also be applied generally to nonlinear control systems for which linear system models are actually adopted.

If control system uncertainty is after all inevitable, then how to handle it? Section \ref{sec:robust_control} aims right at providing some solutions to the problem.

\subsection{Eigenvalue perturbation}

Any well-designed control method, even not designed with control system uncertainty taken into account, may still be able to somewhat handle control system uncertainty. To facilitate understanding of such ``natural'' ability of a control method to somewhat handle control system uncertainty, we may still take the single inverted pendulum control system as example.

Transform the state differential equation (\ref{eq:SIP_state_DE})
\begin{align*}
\frac{\mathrm{d}}{\mathrm{d} t} \mathbf{x} \equiv \frac{\mathrm{d}}{\mathrm{d} t} \begin{bmatrix} \theta \\ \frac{\mathrm{d} \theta}{\mathrm{d} t} \\ x \\ \frac{\mathrm{d} x}{\mathrm{d} t} \end{bmatrix} = \begin{bmatrix} \frac{\mathrm{d} \theta}{\mathrm{d} t} \\ \frac{\sin \theta}{L} g - \frac{\cos \theta}{L} a \\ \frac{\mathrm{d} x}{\mathrm{d} t} \\ a \end{bmatrix} \equiv f(\mathbf{x}, a)
\end{align*}
equivalently into a new model formalism for the single inverted pendulum control system as
\begin{equation}  \label{eq:uncertain_SIP_state_DE_linear}
\frac{\mathrm{d}}{\mathrm{d} t} \mathbf{x} = \begin{bmatrix} 0 & 1 & 0 & 0 \\ \frac{g}{L} \frac{\sin \theta}{\theta} & 0 & 0 & 0 \\ 0 & 0 & 0 & 1  \\ 0 & 0 & 0 & 0 \end{bmatrix} \mathbf{x} + \begin{bmatrix} 0 \\ -\frac{\cos \theta}{L} \\ 0 \\ 1 \end{bmatrix} a \equiv \mathbf{A}(\mathbf{x}) \mathbf{x} + \mathbf{B}(\mathbf{x}) a,
\end{equation}
where the state
\begin{align*}
\mathbf{x} \equiv \begin{bmatrix} \theta & \frac{\mathrm{d} \theta}{\mathrm{d} t} & x & \frac{\mathrm{d} x}{\mathrm{d} t} \end{bmatrix}^\mathrm{T}.
\end{align*}
If the inverted pendulum angle $\theta$ is close to zero, then
\begin{align*}
\frac{\sin \theta}{\theta} \approx 1, \qquad \cos \theta \approx 1.
\end{align*}
Approximate the time-variant state transition matrix $\mathbf{A}(\mathbf{x})$ and the time-variant control input matrix $\mathbf{B}(\mathbf{x})$ respectively as
\begin{align*}
\mathbf{A}(\mathbf{x}) \approx \mathbf{A}(\mathbf{0}) = \begin{bmatrix} 0 & 1 & 0 & 0 \\ \frac{g}{L} & 0 & 0 & 0 \\ 0 & 0 & 0 & 1  \\ 0 & 0 & 0 & 0 \end{bmatrix}, \quad \mathbf{B}(\mathbf{x}) \approx \mathbf{B}(\mathbf{0}) = \begin{bmatrix} 0 \\ -\frac{1}{L} \\ 0 \\ 1 \end{bmatrix}
\end{align*}
and reduce (\ref{eq:uncertain_SIP_state_DE_linear}) to the linear time-invariant formalism (\ref{eq:SIP_state_DE_linear})
\begin{align*}
\frac{\mathrm{d}}{\mathrm{d} t} \mathbf{x} = \begin{bmatrix} 0 & 1 & 0 & 0 \\ \frac{g}{L} & 0 & 0 & 0 \\ 0 & 0 & 0 & 1  \\ 0 & 0 & 0 & 0 \end{bmatrix} \mathbf{x} + \begin{bmatrix} 0 \\ -\frac{1}{L} \\ 0 \\ 1 \end{bmatrix} a = \mathbf{A}(\mathbf{0}) \mathbf{x} + \mathbf{B}(\mathbf{0}) a.
\end{align*}
We had better bear in mind that no matter how fair (\ref{eq:uncertain_SIP_state_DE_linear}) might be approximated by (\ref{eq:SIP_state_DE_linear}), there is after all some discrepancy between (\ref{eq:SIP_state_DE_linear}) and (\ref{eq:uncertain_SIP_state_DE_linear}).

Design an effective gain matrix  
\begin{align*}
\mathbf{K} \equiv \begin{bmatrix} k_1 & k_2 & k_3 & k_4 \end{bmatrix}^\mathrm{T}
\end{align*}
according to the constant state transition matrix $\mathbf{A}(\mathbf{0})$ and the constant control input matrix $\mathbf{B}(\mathbf{0})$ in (\ref{eq:SIP_state_DE_linear}) such that the constant closed-loop state transition matrix
\begin{align*}
\mathbf{A}_c(\mathbf{0}) = \mathbf{A}(\mathbf{0}) - \mathbf{B}(\mathbf{0}) \mathbf{K}^\mathrm{T} = \begin{bmatrix} 0 & 1 & 0 & 0 \\ \frac{g}{L} + \frac{k_1}{L} & \frac{k_2}{L} & \frac{k_3}{L} & \frac{k_4}{L} \\ 0 & 0 & 0 & 1  \\ - k_1 & - k_2 & - k_3 & - k_4 \end{bmatrix}
\end{align*}
is stable. Suppose $\mathbf{A}_c(\mathbf{0})$ has eigenvalues $\lambda_1$, $\lambda_2$, $\lambda_3$, $\lambda_4$ with
\begin{align}  \label{eq:SIP_eigenvalues_re_order}
\mbox{Re}(\lambda_4) \leq \mbox{Re}(\lambda_3) \leq \mbox{Re}(\lambda_2) \leq \mbox{Re}(\lambda_1) < 0.
\end{align}
Given the gain matrix $\mathbf{K}$, dynamics of the state $\mathbf{x}$ is actually described by the closed-loop feedback state differential equation
\begin{equation}  \label{eq:adaptive_SIP_closed_loop_SDE}
\frac{\mathrm{d}}{\mathrm{d} t} \mathbf{x} = \mathbf{A}_c (\mathbf{x}) \mathbf{x} \equiv (\mathbf{A}(\mathbf{x}) - \mathbf{B}(\mathbf{x}) \mathbf{K}^\mathrm{T}) \mathbf{x} = \begin{bmatrix} 0 & 1 & 0 & 0 \\ \frac{g}{L} \frac{\sin \theta}{\theta} + \frac{\cos \theta}{L} k_1 & \frac{\cos \theta}{L} k_2 & \frac{\cos \theta}{L} k_3 & \frac{\cos \theta}{L} k_4 \\ 0 & 0 & 0 & 1  \\ - k_1 & - k_2 & - k_3 & - k_4 \end{bmatrix} \mathbf{x}.
\end{equation}

Choose a matrix norm $\|| \cdot \||$ induced by an absolute vector norm and take advantage of the matrix norm $\|| \cdot \||$ to quantitatively characterize the discrepancy between $\mathbf{A}_c (\mathbf{x})$ and $\mathbf{A}_c (\mathbf{0})$ and analyse how such discrepancy influences eigenvalue locations of the closed-loop state transition matrix. Denote
\begin{subequations}  \label{eq:Delta_Ac_x}
\begin{align}
\Delta \mathbf{A}_c (\mathbf{x}) &\equiv \mathbf{A}_c (\mathbf{x}) - \mathbf{A}_c (\mathbf{0}) = \begin{bmatrix} 0 & 0 & 0 & 0 \\ (\frac{\sin \theta}{\theta} - 1) \frac{g}{L} + \frac{\cos \theta - 1}{L} k_1 & \frac{\cos \theta - 1}{L} k_2 & \frac{\cos \theta - 1}{L} k_3 & \frac{\cos \theta - 1}{L} k_4 \\ 0 & 0 & 0 & 0  \\ 0 & 0 & 0 & 0 \end{bmatrix}  \\
  &= (\frac{\sin \theta}{\theta} - 1) \frac{g}{L} \begin{bmatrix} 0 & 0 & 0 & 0 \\ 1 & 0 & 0 & 0 \\ 0 & 0 & 0 & 0  \\ 0 & 0 & 0 & 0 \end{bmatrix} + \frac{\cos \theta - 1}{L} \begin{bmatrix} 0 & 0 & 0 & 0 \\ k_1 & k_2 & k_3 & k_4 \\ 0 & 0 & 0 & 0  \\ 0 & 0 & 0 & 0 \end{bmatrix}  \\
  &= (\frac{\sin \theta}{\theta} - 1) \frac{g}{L} \mathbf{e}_2 \mathbf{e}_1^\mathrm{T} + \frac{\cos \theta - 1}{L} \mathbf{e}_2 \mathbf{K}^\mathrm{T}
\end{align}
\end{subequations}
and we have
\begin{align*}
\|| \Delta \mathbf{A}_c (\mathbf{x}) \|| &\leq \|| (\frac{\sin \theta}{\theta} - 1) \frac{g}{L} \mathbf{e}_2 \mathbf{e}_1^\mathrm{T} \|| + \|| \frac{\cos \theta - 1}{L} \mathbf{e}_2 \mathbf{K}^\mathrm{T} \||  \\
  &= | \frac{\sin \theta}{\theta} - 1 | \frac{g}{L} \cdot \|| \mathbf{e}_2 \mathbf{e}_1^\mathrm{T} \|| + | \cos \theta - 1 | \frac{1}{L} \cdot \|| \mathbf{e}_2 \mathbf{K}^\mathrm{T} \||.
\end{align*}
Note that
\begin{align*}
| \frac{\sin \theta}{\theta} - 1 | &= 1 - \frac{\sin \theta}{\theta} \leq \frac{\theta^2}{6},  \\
| \cos \theta - 1 | &= 1 - \cos \theta \leq \frac{\theta^2}{2},
\end{align*}
then we have
\begin{equation}  \label{eq:Delta_Ac_x_matrix_norm}
\|| \Delta \mathbf{A}_c (\mathbf{x}) \|| \leq \frac{\theta^2}{6} \frac{g}{L} \cdot \|| \mathbf{e}_2 \mathbf{e}_1^\mathrm{T} \|| + \frac{\theta^2}{2} \frac{1}{L} \cdot \|| \mathbf{e}_2 \mathbf{K}^\mathrm{T} \|| = \theta^2 (\frac{g}{6 L} \|| \mathbf{e}_2 \mathbf{e}_1^\mathrm{T} \|| + \frac{1}{2L} \|| \mathbf{e}_2 \mathbf{K}^\mathrm{T} \||).
\end{equation}

Suppose $\mathbf{A}_c (\mathbf{0})$ is diagonalizable and has the following similarity transformation
\footnote{Even when $\mathbf{A}_c (\mathbf{0})$ is not diagonalizable, we can always have a way to vary the gain matrix $\mathbf{K}$ by an arbitrarily infinitesimal amount such that $\mathbf{A}_c (\mathbf{0})$ has four distinct eigenvalues and hence is definitely diagonalizable. So we can fairly suppose $\mathbf{A}_c (\mathbf{0})$ is diagonalizable.}
\begin{align*}
\mathbf{A}_c (\mathbf{0}) = \mathbf{S} \begin{bmatrix} \lambda_1 & & & \\ & \lambda_2 & & \\ & & \lambda_3 & \\ & & & \lambda_4 \end{bmatrix} \mathbf{S}^{-1}.
\end{align*}
According to the \textit{Bauer-Fike theorem} \cite{Horn2012}, for an arbitrary eigenvalue $\lambda (\mathbf{x})$ of $\mathbf{A}_c (\mathbf{x})$, there is an eigenvalue $\lambda$ of $\mathbf{A}_c (\mathbf{0})$ such that
\begin{equation}  \label{eq:Bauer_Fike_theorem}
| \lambda (\mathbf{x}) - \lambda | \leq \|| \mathbf{S} \|| \cdot \|| \mathbf{S}^{-1} \|| \cdot \|| \Delta \mathbf{A}_c (\mathbf{x}) \|| = \kappa (\mathbf{S}) \|| \Delta \mathbf{A}_c (\mathbf{x}) \||,
\end{equation}
where
\begin{align*}
\kappa (\mathbf{S}) = \|| \mathbf{S} \|| \cdot \|| \mathbf{S}^{-1} \||
\end{align*}
is the \textit{condition number} of the similarity matrix $\mathbf{S}$ with respect to the matrix norm $\|| \cdot \||$.

Substitute (\ref{eq:Delta_Ac_x_matrix_norm}) into (\ref{eq:Bauer_Fike_theorem}) and obtain
\begin{equation}  \label{eq:SIP_Bauer_Fike_theorem}
| \lambda (\mathbf{x}) - \lambda | \leq \theta^2 \kappa (\mathbf{S}) (\frac{g}{6 L} \|| \mathbf{e}_2 \mathbf{e}_1^\mathrm{T} \|| + \frac{1}{2L} \|| \mathbf{e}_2 \mathbf{K}^\mathrm{T} \||).
\end{equation}
Associate (\ref{eq:SIP_eigenvalues_re_order}) with (\ref{eq:SIP_Bauer_Fike_theorem}) and conclude that if the state $\mathbf{x}$ is in the state space region satisfying
\begin{align*}
\theta < \sqrt{\frac{| \mbox{Re}(\lambda_1) |}{\kappa (\mathbf{S}) (\frac{g}{6 L} \|| \mathbf{e}_2 \mathbf{e}_1^\mathrm{T} \|| + \frac{1}{2L} \|| \mathbf{e}_2 \mathbf{K}^\mathrm{T} \||)}},
\end{align*}
then the time-variant closed-loop state transition matrix $\mathbf{A}_c(\mathbf{x})$ definitely has eigenvalues all with negative real part and hence is stable. From above analysis we can see that if the state $\mathbf{x}$ is confined within such state space region, then the full-state feedback control method with the designed gain matrix $\mathbf{K}$ has some ``natural'' ability to handle control system uncertainty that is due to uncertainty of the time-variant closed-loop state transition matrix $\mathbf{A}_c(\mathbf{x})$.

The author believes that by analogy to above analysis with the application example of single inverted pendulum control, readers would have an idea of the ``natural'' ability of a well-designed control method to somewhat handle control system uncertainty. Simply speaking, the stability characteristic of a control system does not change abruptly as the control system changes due to uncertainty.

On the other hand, in practical applications, we cannot always count on such ``natural'' ability of well-designed control methods. Instead, we need to \textit{intentionally do something} to handle control system uncertainty, and some effective methods for doing so will be presented in Section \ref{sec:robust_control_Riccati_equation_method} and Section \ref{sec:robust_control_interval_polynomial}.

\subsection{Linear state-space modelling with uncertainty}  \label{sec:LSS_model_uncertainty}

Revisit (\ref{eq:uncertain_SIP_state_DE_linear}) and generalize it to the generic time-variant state differential equation
\begin{equation}  \label{eq:adaptive_state_DE_linear}
\frac{\mathrm{d}}{\mathrm{d} t} \mathbf{x} = \mathbf{A}(\mathbf{x}) \mathbf{x} + \mathbf{B}(\mathbf{x}) \mathbf{u},
\end{equation}
which can also be formalized as
\begin{equation}  \label{eq:uncertain_state_DE_linear}
\frac{\mathrm{d}}{\mathrm{d} t} \mathbf{x} = [\mathbf{A}(\mathbf{0}) + \Delta \mathbf{A}(\mathbf{x})] \mathbf{x} + [\mathbf{B}(\mathbf{0}) + \Delta \mathbf{B}(\mathbf{x})] \mathbf{u},
\end{equation}
where
\begin{align*}
\Delta \mathbf{A}(\mathbf{x}) \equiv \mathbf{A}(\mathbf{x}) - \mathbf{A}(\mathbf{0}), \qquad \Delta \mathbf{B}(\mathbf{x}) \equiv \mathbf{B}(\mathbf{x}) - \mathbf{B}(\mathbf{0}).
\end{align*}

The state transition matrix difference $\Delta \mathbf{A}(\mathbf{x})$ and the control input matrix difference $\Delta \mathbf{B}(\mathbf{x})$ in (\ref{eq:uncertain_state_DE_linear}) can be interpreted as uncertainty factors superposed on the control system modelled by the linear time-invariant formalism
\begin{align}  \label{eq:state_differential_equation_linear2}
\frac{\mathrm{d}}{\mathrm{d} t} \mathbf{x} = \mathbf{A}(\mathbf{0}) \mathbf{x} + \mathbf{B}(\mathbf{0}) \mathbf{u}.
\end{align} 
With interpretation of $\Delta \mathbf{A}(\mathbf{x})$ and $\Delta \mathbf{B}(\mathbf{x})$ as uncertainty factors and without causing confusion, simplify the formalism (\ref{eq:uncertain_state_DE_linear}) into
\begin{equation}  \label{eq:uncertain_state_DE_linear2}
\frac{\mathrm{d}}{\mathrm{d} t} \mathbf{x} = (\mathbf{A} + \Delta \mathbf{A}) \mathbf{x} + (\mathbf{B} + \Delta \mathbf{B}) \mathbf{u},
\end{equation}
where
\begin{align*}
\mathbf{A} \equiv \mathbf{A}(\mathbf{0}), \qquad \mathbf{B} \equiv \mathbf{B}(\mathbf{0}), \qquad \Delta \mathbf{A} \equiv \Delta \mathbf{A}(\mathbf{x}), \qquad \Delta \mathbf{B} \equiv \Delta \mathbf{B}(\mathbf{x}).
\end{align*}
(\ref{eq:uncertain_state_DE_linear2}) gives a generic formalism of \textbf{linear state-space modelling with uncertainty}.

The full-state feedback control method presented in Chapter 2, 
\footnote{Namely Chapter 2 of the author's works \cite{Li2026ACTPA_SJTU_2, Li2026ACTPA_SJTU_1}. Note that this article is Chapter 5 of the works.}
which is general enough to handle a large variety of multiple-input-multiple-output control problems, relies on its natural ability to somewhat handle control system uncertainty. However, given some control system, a gain matrix designed according to $\mathbf{A}$ and $\mathbf{B}$ may not be applicable to all 
\begin{align*}
\Omega_{\Delta \mathbf{A}, \Delta \mathbf{B}} \equiv \{ \mathbf{A} + \Delta \mathbf{A}, \quad \mathbf{B} + \Delta \mathbf{B} \}
\end{align*}
that can potentially be encountered during operation of the control system.

For example, consider single inverted pendulum control. As explained in Section 4.1.4 in Chapter 4, 
\footnote{Namely Chapter 4 of the author's works \cite{Li2026ACTPA_SJTU_2, Li2026ACTPA_SJTU_1}.}
since the cart position $x$ is purely a linear factor for dynamics of the state $\mathbf{x}$ and causes no uncertainty to the linear system model, we focus on handling the partial or reduced state
\begin{align*}
\mathbf{x}_P \equiv \begin{bmatrix} \theta & \frac{\mathrm{d} \theta}{\mathrm{d} t} & \frac{\mathrm{d} x}{\mathrm{d} t} \end{bmatrix}^\mathrm{T}
\end{align*}
that consists of the inverted pendulum angle, the inverted pendulum angular velocity, and the cart velocity. Extract the sub-model associated with the partial state $\mathbf{x}_P$ from (\ref{eq:uncertain_SIP_state_DE_linear}) and obtain
\begin{equation}  \label{eq:uncertain_SIP_state_DE_linear_partial}
\frac{\mathrm{d}}{\mathrm{d} t} \mathbf{x}_P = \mathbf{A}_P(\mathbf{x}_P) \mathbf{x}_P + \mathbf{B}_P(\mathbf{x}_P) a,
\end{equation}
where
\begin{align*}
\mathbf{A}_P(\mathbf{x}_P) \equiv \begin{bmatrix} 0 & 1 & 0 \\ \frac{g}{L} \frac{\sin \theta}{\theta} & 0 & 0 \\ 0 & 0 & 0 \end{bmatrix}, \quad \mathbf{B}_P(\mathbf{x}_P) \equiv \begin{bmatrix} 0 \\ -\frac{\cos \theta}{L} \\ 1 \end{bmatrix}.
\end{align*}

Apply the full-state feedback control method and design a gain matrix according to the approximated linear time-invariant version of (\ref{eq:uncertain_SIP_state_DE_linear_partial}) namely
\begin{equation}  \label{eq:SIP_state_DE_linear_partial}
\frac{\mathrm{d}}{\mathrm{d} t} \mathbf{x}_P =\begin{bmatrix} 0 & 1 & 0 \\ \frac{g}{L} & 0 & 0 \\ 0 & 0 & 0 \end{bmatrix} \mathbf{x}_P + \begin{bmatrix} 0 \\ -\frac{1}{L} \\ 1 \end{bmatrix} a.
\end{equation}
Set the expected closed-loop characteristic polynomial as
\begin{align*}
C_\mathrm{E}(s) = (s + 4)^3 = s^3 + 12 s^2 + 48 s + 64
\end{align*}
and compute the corresponding gain matrix as
\begin{align*}
\mathbf{K}_P = \begin{bmatrix} -58 & -18.4 & -6.4 \end{bmatrix}^\mathrm{T}.
\end{align*}
The full-state feedback control law is
\begin{align*}
a = - \mathbf{K}_P^\mathrm{T} \mathbf{x}_P = 58 \theta + 18.4 \frac{\mathrm{d} \theta}{\mathrm{d} t} + 6.4 \frac{\mathrm{d} x}{\mathrm{d} t}.
\end{align*}

Except for the equilibrium state, there is always a discrepancy between (\ref{eq:SIP_state_DE_linear_partial}) and (\ref{eq:uncertain_SIP_state_DE_linear_partial}) and hence there is always some control system uncertainty. On one hand, the specific full-state feedback control method can work for the state space region where the inverted pendulum angle $\theta$ is not far away from zero. This reflects its natural ability to somewhat handle control system uncertainty. On the other hand, when the control system operates in the state space region where the inverted pendulum angle $\theta$ is far away from zero, the specific full-state feedback control method tends to fail. Matlab simulation code for relevant demonstration is given as follows.

\begin{framed} 
\noindent \textbf{SingleInvertedPendulumNonRobustFailure.m} \\
\noindent \%\% Single inverted pendulum parameters \\
m1 = 1; L1 = 1; g = 10; \\
\%\% Design the gain matrix via the general method \\
A = [0, 1, 0, 0; g/L1, 0, 0, 0; 0, 0, 0, 1; 0, 0, 0, 0]; \\
B = [0; -1/L1; 0; 1]; \\
Ap = A([1,2,4], [1,2,4]); Bp = B([1,2,4]); \\
sttKp = DesignGainMatrix(Ap, Bp, [-4;-4;-4]); \% Obtain the gain matrix \\
fprintf('Gain matrix Kp: '); sttKp' \\
 \\
\%\% Simulation preliminary configuration \\
dt = 0.001; \% Numerical computation step \\
tSpan = 0:dt:5; \% Simulation time span \\
x = 0.2; \% Cart position  \\
dx = 0; \% Cart velocity \\
y = 0.2; \% Inverted pendulum angle theta  \\
dy = 0;  \% Inverted pendulum angular velocity \\
stt = [y; dy; x; dx]; \% Single inverted pendulum state \\
sttAll = zeros(length(stt), length(tSpan)); k = 0; \% Record states \\
xExpected = 0; yExpected = 0; \% Expected equilibrium status \\
SimConfig = [m1, L1, g, dt]; \\
\%\% Simulation of single inverted pendulum control \\
for t = tSpan \\
$~~~~$ \%\% Control method \\
$~~~~$ acc = -sttKp'*stt([1,2,4]); \% FSFC of the cart acceleration \\
 \\
$~~~~$ \%\% Single inverted pendulum dynamics \\
$~~~~$ stt = DynamicsSIP(SimConfig, stt, acc); \\
$~~~~$ sttC = num2cell(stt); [y, dy, x, dx] = sttC\{:\}; \\
$~~~~$ if (y\^{}2+dy\^{}2+dx\^{}2$<$0.0001) fprintf('Control success!$\backslash$n'); break; end \\
$~~~~$ if (abs(y)$>$=pi/2) fprintf('Control failure!$\backslash$n'); break; end \\
$~~~~$ k = k+1; sttAll(:,k) = stt; \\
$~~~~$ \%\% Single inverted pendulum visualization \\
$~~~~$ if (rem(k,20) == 0) \\
$~~~~$ $~~~~$ DisplaySIP(x, y, L1); pause(dt); \\
$~~~~$ end \\
end \\
 \\
\%\% Simulation of SIP control under large uncertainty \\
x = 0.2; \% Cart position  \\
dx = 0; \% Cart velocity \\
y = 0.4*pi; \% Inverted pendulum angle theta  \\
dy = 0;  \% Inverted pendulum angular velocity \\
stt = [y; dy; x; dx]; \% Single inverted pendulum state \\
sttAll2 = zeros(length(stt), length(tSpan)); k = 0; \% Record states \\
for t = tSpan \\
$~~~~$ \%\% Control method \\
$~~~~$ acc = -sttKp'*stt([1,2,4]); \% FSFC of the cart acceleration \\
 \\
$~~~~$ \%\% Single inverted pendulum dynamics \\
$~~~~$ stt = DynamicsSIP(SimConfig, stt, acc); \\
$~~~~$ sttC = num2cell(stt); [y, dy, x, dx] = sttC\{:\}; \\
$~~~~$ if (abs(y)$>$=pi/2) fprintf('Control failure!$\backslash$n'); break; end \\
$~~~~$ k = k+1; sttAll2(:,k) = stt; \\
$~~~~$ \%\% Single inverted pendulum visualization \\
$~~~~$ if (rem(k,20) == 0) \\
$~~~~$ $~~~~$ DisplaySIP(x, y, L1); pause(dt); \\
$~~~~$ end \\
end
\end{framed}

As clarified above, a full-state feedback control method with the gain matrix designed according to $\mathbf{A}$ and $\mathbf{B}$ may not work for all $\Omega_{\Delta \mathbf{A}, \Delta \mathbf{B}}$ that can potentially be encountered during operation of the control system. On the other hand, a question arises: Is there a full-state feedback control method (or more generally a control method) that can work for all potential $\Omega_{\Delta \mathbf{A}, \Delta \mathbf{B}}$ in concerned practical applications? If so, how to design such kind of full-state feedback control method (or such kind of control method)? In other words, how to find a ``magic'' gain matrix that is applicable to all potential $\Omega_{\Delta \mathbf{A}, \Delta \mathbf{B}}$?

Providing answers to questions like these is right the objective of control methods coined commonly as \textbf{robust control} \cite{Zames1981, ZhouK1998, Dullerud2000}, representative methods of which will be presented next in Section \ref{sec:robust_control_Riccati_equation_method} and Section \ref{sec:robust_control_interval_polynomial}. Simply speaking, \textit{the core spirit of robust control is to design control methods that can always work no matter how control system uncertainty is}.

\subsection{Interval matrices}

Before we move to Section \ref{sec:robust_control_Riccati_equation_method}, some matrix-related concepts need to be clarified.

Given a generic $m$-by-$n$ time-variant matrix 
\begin{align*}
\begin{bmatrix} x_{ij}(t) \end{bmatrix} \equiv \begin{bmatrix} x_{11}(t) & x_{12}(t) & \cdots & x_{1n}(t) \\ x_{21}(t) & x_{22}(t) & \cdots & x_{2n}(t) \\ \vdots & \vdots & \ddots & \vdots \\ x_{m1}(t) & x_{m2}(t) & \cdots & x_{mn}(t) \end{bmatrix},
\end{align*}
define its \textit{element-wise minimum value matrix} as
\begin{equation}  \label{eq:element_min_matrix}
\underline{\begin{bmatrix} x_{ij} \end{bmatrix}} \equiv \begin{bmatrix} \min x_{ij}(t) \end{bmatrix} = \begin{bmatrix} \min x_{11}(t) & \min x_{12}(t) & \cdots & \min x_{1n}(t) \\ \min x_{21}(t) & \min x_{22}(t) & \cdots & \min x_{2n}(t) \\ \vdots & \vdots & \ddots & \vdots \\ \min x_{m1}(t) & \min x_{m2}(t) & \cdots & \min x_{mn}(t) \end{bmatrix},
\end{equation}
its \textit{element-wise maximum value matrix} as
\begin{equation}  \label{eq:element_max_matrix}
\overline{\begin{bmatrix} x_{ij} \end{bmatrix}} \equiv \begin{bmatrix} \max x_{ij}(t) \end{bmatrix} = \begin{bmatrix} \max x_{11}(t) & \max x_{12}(t) & \cdots & \max x_{1n}(t) \\ \max x_{21}(t) & \max x_{22}(t) & \cdots & \max x_{2n}(t) \\ \vdots & \vdots & \ddots & \vdots \\ \max x_{m1}(t) & \max x_{m2}(t) & \cdots & \max x_{mn}(t) \end{bmatrix},
\end{equation}
and its \textit{element-wise absolute value matrix} as
\begin{equation}  \label{eq:element_abs_matrix}
| \begin{bmatrix} x_{ij} \end{bmatrix} | \equiv \begin{bmatrix} | x_{ij}(t) | \end{bmatrix} = \begin{bmatrix} | x_{11}(t) | & | x_{12}(t) | & \cdots & | x_{1n}(t) | \\ | x_{21}(t) | & | x_{22}(t) | & \cdots & | x_{2n}(t) | \\ \vdots & \vdots & \ddots & \vdots \\ | x_{m1}(t) | & | x_{m2}(t) | & \cdots & | x_{mn}(t) | \end{bmatrix}.
\end{equation}

Given two generic $m$-by-$n$ time-variant matrices
\begin{align*}
\mathbf{L} \equiv \begin{bmatrix} l_{ij}(t) \end{bmatrix}, \qquad \mathbf{H} \equiv \begin{bmatrix} h_{ij}(t) \end{bmatrix},
\end{align*}
define the \textit{element-wise larger} notation $\succ$ (i.e. the bent $>$) as
\begin{equation}  \label{eq:element_matrix_larger}
\mathbf{H} \succ \mathbf{L} \iff h_{ij}(t) > l_{ij}(t) \quad \forall i \in \{1, \cdots, m\}, j \in \{1, \cdots, n\},
\end{equation}
the \textit{element-wise smaller} notation $\prec$ (i.e. the bent $<$) as
\begin{equation}  \label{eq:element_matrix_smaller}
\mathbf{L} \prec \mathbf{H} \iff l_{ij}(t) < h_{ij}(t) \quad \forall i \in \{1, \cdots, m\}, j \in \{1, \cdots, n\},
\end{equation}
the \textit{element-wise larger and equal} notation $\succeq$ (i.e. the bent $\geq$) as
\begin{equation}  \label{eq:element_matrix_larger+equal}
\mathbf{H} \succeq \mathbf{L} \iff h_{ij}(t) \geq l_{ij}(t) \quad \forall i \in \{1, \cdots, m\}, j \in \{1, \cdots, n\},
\end{equation}
the \textit{element-wise smaller and equal} notation $\preceq$ (i.e. the bent $\leq$) as
\begin{equation}  \label{eq:element_matrix_smaller+equal}
\mathbf{L} \preceq \mathbf{H} \iff l_{ij}(t) \leq h_{ij}(t) \quad \forall i \in \{1, \cdots, m\}, j \in \{1, \cdots, n\}.
\end{equation}

Given two generic $m$-by-$n$ time-variant matrices
\begin{align*}
\mathbf{L} \equiv \begin{bmatrix} l_{ij}(t) \end{bmatrix}, \qquad \mathbf{H} \equiv \begin{bmatrix} h_{ij}(t) \end{bmatrix}
\end{align*}
that satisfies
\begin{align*}
\mathbf{L} \preceq \mathbf{H} \iff \mathbf{H} \succeq \mathbf{L},
\end{align*}
define the \textbf{interval matrix set} or for short the \textbf{interval matrix}
\begin{equation}  \label{eq:interval_matrix}
[\mathbf{L}, \mathbf{H}] \equiv \{ \mathbf{X} \mbox{ } | \mbox{ } \mathbf{L} \preceq \mathbf{X} \preceq \mathbf{H} \}.
\end{equation}

In practical applications, control system uncertainty may be unknown, yet it is usually somehow bounded. For example, given the generic formalism of linear state-space modelling with uncertainty described in (\ref{eq:uncertain_state_DE_linear2})
\begin{align*}
\frac{\mathrm{d}}{\mathrm{d} t} \mathbf{x} = (\mathbf{A} + \Delta \mathbf{A}) \mathbf{x} + (\mathbf{B} + \Delta \mathbf{B}) \mathbf{u},
\end{align*}
the uncertainty factors $\Delta \mathbf{A}$ and $\Delta \mathbf{B}$ are at least bounded between their respective element-wise minimum value matrix and element-wise maximum value matrix, namely
\begin{subequations}  \label{eq:A+DA_B+DB_interval_matrix1}
\begin{align}
&\underline{\Delta \mathbf{A}} \preceq \Delta \mathbf{A} \preceq \overline{\Delta \mathbf{A}},  \\
&\underline{\Delta \mathbf{B}} \preceq \Delta \mathbf{B} \preceq \overline{\Delta \mathbf{B}}
\end{align}
\end{subequations}
or equivalently
\begin{subequations}  \label{eq:A+DA_B+DB_interval_matrix2}
\begin{align}
&\mathbf{A} + \underline{\Delta \mathbf{A}} \preceq \mathbf{A} + \Delta \mathbf{A} \preceq \mathbf{A} + \overline{\Delta \mathbf{A}}, \\
&\mathbf{B} + \underline{\Delta \mathbf{B}} \preceq \mathbf{B} + \Delta \mathbf{B} \preceq \mathbf{B} + \overline{\Delta \mathbf{B}}.
\end{align}
\end{subequations}

In the context of full-state feedback control, as mentioned previously, the objective of robust control is to find a ``magic'' gain matrix that is applicable to all potential 
\begin{align*}
\Omega_{\Delta \mathbf{A}, \Delta \mathbf{B}} \equiv \{ \mathbf{A} + \Delta \mathbf{A}, \quad \mathbf{B} + \Delta \mathbf{B} \}.
\end{align*}
This requirement can be formalized as follows: How to find the gain matrix $\mathbf{K}$ such that
\footnote{The function \textit{Eig} means getting the vector of eigenvalues of the given matrix.}
\begin{equation}  \label{eq:robust_gain_matrix_design}
\forall \mathbf{A}^* \in [\mathbf{A} + \underline{\Delta \mathbf{A}}, \mathbf{A} + \overline{\Delta \mathbf{A}}], \mathbf{B}^* \in [\mathbf{B} + \underline{\Delta \mathbf{B}}, \mathbf{B} + \overline{\Delta \mathbf{B}}], \quad \mbox{Re}(\mbox{Eig}(\mathbf{A}^* - \mathbf{B}^* \mathbf{K}^\mathrm{T})) \prec \mathbf{0}
\end{equation}
namely the interval matrix 
\begin{align*}
\{ \mathbf{A}^* - \mathbf{B}^* \mathbf{K}^\mathrm{T} \mbox{ } | \mbox{ } \mathbf{A}^* \in [\mathbf{A} + \underline{\Delta \mathbf{A}}, \mathbf{A} + \overline{\Delta \mathbf{A}}], \mathbf{B}^* \in [\mathbf{B} + \underline{\Delta \mathbf{B}}, \mathbf{B} + \overline{\Delta \mathbf{B}}] \}
\end{align*}
is always stable.

\subsection{Riccati equation method}  \label{sec:robust_control_Riccati_equation_method}

How to provide a solution to the problem formalized in (\ref{eq:robust_gain_matrix_design})? For such purpose, we first need a more refined way of characterizing control system uncertainty than that described in (\ref{eq:A+DA_B+DB_interval_matrix1}) or (\ref{eq:A+DA_B+DB_interval_matrix2}). The \textbf{rank-one matrix decomposition} method \cite{Petersen1986} provides a representative refined way of characterizing control system uncertainty.

The rank-one matrix decomposition method consists in conceiving summations of non-negative rank-one matrices bounded within which the interval matrices corresponding to the uncertainty factors $\Delta \mathbf{A}$ and $\Delta \mathbf{B}$ are. More specifically, decompose the uncertainty factor $\Delta \mathbf{A}$ into a time-variant weighted combination of non-negative rank-one matrices
\begin{equation}  \label{eq:DA_decompose_rank_one}
\Delta \mathbf{A} = \sum_{k=1}^p a_k (t) \mathbf{a}_k \mathbf{x}_k^\mathrm{T} \equiv \sum_{k=1}^p a_k \mathbf{a}_k \mathbf{x}_k^\mathrm{T}
\end{equation}
and constrain the time-variant weights 
\begin{align*}
a_k \equiv a_k (t)
\end{align*}
by certain common bound $\bar{a}$ as
\begin{equation}  \label{eq:DA_weight_bound}
| a_k | = | a_k (t) | \leq \bar{a}	\qquad \forall k \in \{1, \cdots, p\}.
\end{equation}
In (\ref{eq:DA_decompose_rank_one}), $p$ is certain number and
\begin{align*}
\mathbf{a}_k &\succeq \mathbf{0} \qquad \forall k \in \{1, \cdots, p\},  \\
\mathbf{x}_k &\succeq \mathbf{0} \qquad \forall k \in \{1, \cdots, p\}.
\end{align*}
In fact, each $\mathbf{a}_k \mathbf{x}_k^\mathrm{T}$ represents a \textit{non-negative matrix factorization} 
\footnote{The Matlab built-in function \textit{nnmf} can perform non-negative matrix factorization.}
of the corresponding non-negative rank-one matrix. Define a summation of the non-negative rank-one matrices as
\begin{equation}  \label{eq:DA_sum_rank_one}
\mathbf{\Sigma}_{\Delta \mathbf{A}} \equiv \bar{a} \sum_{k=1}^p \mathbf{a}_k \mathbf{x}_k^\mathrm{T}.
\end{equation}
Then we have
\begin{equation}  \label{eq:DA_sum_rank_one_bound}
- \mathbf{\Sigma}_{\Delta \mathbf{A}} \preceq \Delta \mathbf{A} \preceq \mathbf{\Sigma}_{\Delta \mathbf{A}}.
\end{equation}

Similarly, decompose the uncertainty factor $\Delta \mathbf{B}$ into a time-variant weighted combination of non-negative rank-one matrices
\begin{equation}  \label{eq:DB_decompose_rank_one}
\Delta \mathbf{B} = \sum_{k=1}^q b_k (t) \mathbf{b}_k \mathbf{y}_k^\mathrm{T} \equiv \sum_{k=1}^q b_k \mathbf{b}_k \mathbf{y}_k^\mathrm{T}
\end{equation}
and constrain the time-variant weights 
\begin{align*}
b_k \equiv b_k (t)
\end{align*}
by certain common bound $\bar{b}$ as
\begin{equation}  \label{eq:DB_weight_bound}
| b_k | = | b_k (t) | \leq \bar{b}	\qquad \forall k \in \{1, \cdots, q\}.
\end{equation}
In (\ref{eq:DB_decompose_rank_one}), $q$ is certain number and
\begin{align*}
\mathbf{b}_k &\succeq \mathbf{0} \qquad \forall k \in \{1, \cdots, q\},  \\
\mathbf{y}_k &\succeq \mathbf{0} \qquad \forall k \in \{1, \cdots, q\}.
\end{align*}
Each $\mathbf{b}_k \mathbf{y}_k^\mathrm{T}$ also represents a non-negative matrix factorization of the corresponding non-negative rank-one matrix. Define a summation of the non-negative rank-one matrices as
\begin{equation}  \label{eq:DB_sum_rank_one}
\mathbf{\Sigma}_{\Delta \mathbf{B}} \equiv \bar{b} \sum_{k=1}^q \mathbf{b}_k \mathbf{y}_k^\mathrm{T}
\end{equation}
Then we have
\begin{equation}  \label{eq:DB_sum_rank_one_bound}
- \mathbf{\Sigma}_{\Delta \mathbf{B}} \preceq \Delta \mathbf{B} \preceq \mathbf{\Sigma}_{\Delta \mathbf{B}}.
\end{equation}

To find a solution of the gain matrix $\mathbf{K}$ that satisfies (\ref{eq:robust_gain_matrix_design}), \textit{Petersen} and \textit{Hollot} \cite{Petersen1986} proposed a representative method namely the \textbf{Riccati equation method}. Here, an improved formalism of the Riccati equation method \cite{YangY2004} is adopted
\footnote{Originally in \cite{YangY2004}, there is error in the presented formalism, which is corrected here. Besides, a simplified proof of the conclusion about (\ref{eq:robust_gain_matrix_design_solution}) is provided below.}. 
Define four auxiliary matrices
\begin{subequations}  \label{eq:Sigma_a_x_b_y}
\begin{align}
\mathbf{\Sigma_a} &\equiv \bar{a} \sum_{k=1}^p \mathbf{a}_k \mathbf{a}_k^\mathrm{T},  \\
\mathbf{\Sigma_x} &\equiv \bar{a} \sum_{k=1}^p \mathbf{x}_k \mathbf{x}_k^\mathrm{T},  \\
\mathbf{\Sigma_b} &\equiv \bar{b} \sum_{k=1}^q \mathbf{b}_k \mathbf{b}_k^\mathrm{T},  \\
\mathbf{\Sigma_y} &\equiv \bar{b} \sum_{k=1}^q \mathbf{y}_k \mathbf{y}_k^\mathrm{T}
\end{align}
\end{subequations}
and establish a \textit{Riccati equation} of the third formalism (1.55) as
\footnote{Namely (1.55) in the author's works \cite{Li2026ACTPA_SJTU_2, Li2026ACTPA_SJTU_1}. Note that this article is Chapter 5 of the works.}
\begin{equation}  \label{eq:Petersen_YangY_Riccati}
\mathbf{P} \mathbf{A} + \mathbf{A}^\mathrm{T} \mathbf{P} - \mathbf{P} \mathbf{M} \mathbf{P} + \mathbf{Q_\Sigma} = \mathbf{0},
\end{equation}
where
\begin{align*}
\mathbf{M} &= \mathbf{B} (\mathbf{R} + \epsilon \mathbf{\Sigma_y})^{-1} (2 \mathbf{R} + \epsilon \mathbf{\Sigma_y}) (\mathbf{R} + \epsilon \mathbf{\Sigma_y})^{-1} \mathbf{B}^\mathrm{T} - \mathbf{\Sigma_a} - \frac{1}{\epsilon} \mathbf{\Sigma_b},  \\
\mathbf{Q_\Sigma} &= \mathbf{\Sigma_x} + \mathbf{Q}.
\end{align*}
In (\ref{eq:Petersen_YangY_Riccati}), $\mathbf{Q}$ and $\mathbf{R}$ are user-defined positive definite matrices, and $\epsilon$ is a user-defined positive scalar value.

If the Riccati equation (\ref{eq:Petersen_YangY_Riccati}) has a positive definite solution $\mathbf{P}$, then the gain matrix 
\begin{equation}  \label{eq:robust_gain_matrix_design_solution}
\mathbf{K} = [(\mathbf{R} + \epsilon \mathbf{\Sigma_y})^{-1} \mathbf{B}^\mathrm{T} \mathbf{P}]^\mathrm{T} = \mathbf{P} \mathbf{B} (\mathbf{R} + \epsilon \mathbf{\Sigma_y})^{-1}
\end{equation}
is a solution to the problem formalized in (\ref{eq:robust_gain_matrix_design}). Note that $\mathbf{P}$, $\mathbf{R}$, and $\mathbf{\Sigma_y}$ in (\ref{eq:robust_gain_matrix_design_solution}) are all symmetric matrices.

\begin{proof}
Consider the gain matrix specified in (\ref{eq:robust_gain_matrix_design_solution}), substitute the full-state feedback control law
\begin{align*}
\mathbf{u} = - \mathbf{K}^\mathrm{T} \mathbf{x} = - (\mathbf{R} + \epsilon \mathbf{\Sigma_y})^{-1} \mathbf{B}^\mathrm{T} \mathbf{P} \mathbf{x}
\end{align*}
into (\ref{eq:uncertain_state_DE_linear2}) and obtain
\begin{equation}  \label{eq:uncertain_state_DE_linear2_closed}
\frac{\mathrm{d}}{\mathrm{d} t} \mathbf{x} = [(\mathbf{A} + \Delta \mathbf{A}) - (\mathbf{B} + \Delta \mathbf{B}) \mathbf{K}^\mathrm{T}] \mathbf{x}.
\end{equation}
Define a Lyapunov equation style matrix as
\begin{equation}  \label{eq:uncertain_state_DE_linear2_closed_L}
\mathbf{L} = \mathbf{P} [(\mathbf{A} + \Delta \mathbf{A}) - (\mathbf{B} + \Delta \mathbf{B}) \mathbf{K}^\mathrm{T}] + [(\mathbf{A} + \Delta \mathbf{A}) - (\mathbf{B} + \Delta \mathbf{B}) \mathbf{K}^\mathrm{T}]^\mathrm{T} \mathbf{P}.
\end{equation}
Expand (\ref{eq:uncertain_state_DE_linear2_closed_L}) and obtain
\begin{align*}
\mathbf{L} &= \mathbf{P} \mathbf{A} + \mathbf{A}^\mathrm{T} \mathbf{P} - \mathbf{P} \mathbf{B} \mathbf{K}^\mathrm{T} - \mathbf{K} \mathbf{B}^\mathrm{T} \mathbf{P} + (\mathbf{P} \Delta \mathbf{A} + \Delta \mathbf{A}^\mathrm{T} \mathbf{P}) - (\mathbf{P} \Delta \mathbf{B} \mathbf{K}^\mathrm{T} + \mathbf{K} \Delta \mathbf{B}^\mathrm{T} \mathbf{P})  \\
  &= \mathbf{P} \mathbf{A} + \mathbf{A}^\mathrm{T} \mathbf{P} - \mathbf{P} \mathbf{B} (\mathbf{R} + \epsilon \mathbf{\Sigma_y})^{-1} \mathbf{B}^\mathrm{T} \mathbf{P} - \mathbf{P} \mathbf{B} (\mathbf{R} + \epsilon \mathbf{\Sigma_y})^{-1} \mathbf{B}^\mathrm{T} \mathbf{P}  \\
  &\qquad + (\mathbf{P} \Delta \mathbf{A} + \Delta \mathbf{A}^\mathrm{T} \mathbf{P}) - (\mathbf{P} \Delta \mathbf{B} \mathbf{K}^\mathrm{T} + \mathbf{K} \Delta \mathbf{B}^\mathrm{T} \mathbf{P}),
\end{align*}
i.e.
\begin{equation}  \label{eq:uncertain_state_DE_linear2_closed_L2}
\mathbf{L} = \mathbf{P} \mathbf{A} + \mathbf{A}^\mathrm{T} \mathbf{P} - 2 \mathbf{P} \mathbf{B} (\mathbf{R} + \epsilon \mathbf{\Sigma_y})^{-1} \mathbf{B}^\mathrm{T} \mathbf{P} + \mathbf{L}_1 - \mathbf{L}_2,
\end{equation}
where
\begin{align*}
\mathbf{L}_1 &= \mathbf{P} \Delta \mathbf{A} + \Delta \mathbf{A}^\mathrm{T} \mathbf{P} = \sum_{k=1}^p a_k (\mathbf{P} \mathbf{a}_k \mathbf{x}_k^\mathrm{T} + \mathbf{x}_k \mathbf{a}_k^\mathrm{T} \mathbf{P}),  \\
\mathbf{L}_2 &= \mathbf{P} \Delta \mathbf{B} \mathbf{K}^\mathrm{T} + \mathbf{K} \Delta \mathbf{B}^\mathrm{T} \mathbf{P} = \sum_{k=1}^q b_k (\mathbf{P} \mathbf{b}_k \mathbf{y}_k^\mathrm{T} \mathbf{K}^\mathrm{T} + \mathbf{K} \mathbf{y}_k \mathbf{b}_k^\mathrm{T} \mathbf{P})
\end{align*}
with (\ref{eq:DA_decompose_rank_one}) and (\ref{eq:DB_decompose_rank_one}) applied.

Since
\begin{align*}
\mathbf{P} \mathbf{a}_k \mathbf{a}_k^\mathrm{T} \mathbf{P} + \mathbf{x}_k \mathbf{x}_k^\mathrm{T} \geq 0
\end{align*}
and
\begin{equation}  \label{eq:robust_GM_L1_sub}
\mathbf{P} \mathbf{a}_k \mathbf{a}_k^\mathrm{T} \mathbf{P} + \mathbf{x}_k \mathbf{x}_k^\mathrm{T} - (\mathbf{P} \mathbf{a}_k \mathbf{x}_k^\mathrm{T} + \mathbf{x}_k \mathbf{a}_k^\mathrm{T} \mathbf{P}) = (\mathbf{P} \mathbf{a}_k - \mathbf{x}_k) (\mathbf{P} \mathbf{a}_k - \mathbf{x}_k)^\mathrm{T} \geq 0,
\end{equation}
we have
\begin{align*}
a_k (\mathbf{P} \mathbf{a}_k \mathbf{x}_k^\mathrm{T} + \mathbf{x}_k \mathbf{a}_k^\mathrm{T} \mathbf{P}) \leq | a_k | \cdot (\mathbf{P} \mathbf{a}_k \mathbf{a}_k^\mathrm{T} \mathbf{P} + \mathbf{x}_k \mathbf{x}_k^\mathrm{T}) \leq \bar{a} (\mathbf{P} \mathbf{a}_k \mathbf{a}_k^\mathrm{T} \mathbf{P} + \mathbf{x}_k \mathbf{x}_k^\mathrm{T})
\end{align*}
and hence
\begin{equation}  \label{eq:robust_GM_L1}
\mathbf{L}_1 \leq \sum_{k=1}^p \bar{a} (\mathbf{P} \mathbf{a}_k \mathbf{a}_k^\mathrm{T} \mathbf{P} + \mathbf{x}_k \mathbf{x}_k^\mathrm{T}) = \mathbf{P} \mathbf{\Sigma_a} \mathbf{P} + \mathbf{\Sigma_x}
\end{equation}
with the first and second equations of (\ref{eq:Sigma_a_x_b_y}) applied.

Since
\begin{align*}
\epsilon \mathbf{K} \mathbf{y}_k \mathbf{y}_k^\mathrm{T} \mathbf{K}^\mathrm{T} + \frac{1}{\epsilon} \mathbf{P} \mathbf{b}_k \mathbf{b}_k^\mathrm{T} \mathbf{P} \geq 0
\end{align*}
and
\begin{align}  \label{eq:robust_GM_L2_sub}
&\epsilon \mathbf{K} \mathbf{y}_k \mathbf{y}_k^\mathrm{T} \mathbf{K}^\mathrm{T} + \frac{1}{\epsilon} \mathbf{P} \mathbf{b}_k \mathbf{b}_k^\mathrm{T} \mathbf{P} - (\mathbf{P} \mathbf{b}_k \mathbf{y}_k^\mathrm{T} \mathbf{K}^\mathrm{T} + \mathbf{K} \mathbf{y}_k \mathbf{b}_k^\mathrm{T} \mathbf{P})  \nonumber \\ 
=& (\sqrt{\epsilon} \mathbf{K} \mathbf{y}_k - \frac{1}{\sqrt{\epsilon}} \mathbf{P} \mathbf{b}_k) (\sqrt{\epsilon} \mathbf{K} \mathbf{y}_k - \frac{1}{\sqrt{\epsilon}} \mathbf{P} \mathbf{b}_k)^\mathrm{T} \geq 0,
\end{align}
we have
\begin{align*}
- b_k (\mathbf{P} \mathbf{b}_k \mathbf{y}_k^\mathrm{T} \mathbf{K}^\mathrm{T} + \mathbf{K} \mathbf{y}_k \mathbf{b}_k^\mathrm{T} \mathbf{P}) &\leq | - b_k | \cdot (\epsilon \mathbf{K} \mathbf{y}_k \mathbf{y}_k^\mathrm{T} \mathbf{K}^\mathrm{T} + \frac{1}{\epsilon} \mathbf{P} \mathbf{b}_k \mathbf{b}_k^\mathrm{T} \mathbf{P})  \\ 
 &\leq \bar{b} (\epsilon \mathbf{K} \mathbf{y}_k \mathbf{y}_k^\mathrm{T} \mathbf{K}^\mathrm{T} + \frac{1}{\epsilon} \mathbf{P} \mathbf{b}_k \mathbf{b}_k^\mathrm{T} \mathbf{P})
\end{align*}
and hence
\begin{equation}  \label{eq:robust_GM_L2}
- \mathbf{L}_2 \leq \sum_{k=1}^q \bar{b} (\epsilon \mathbf{K} \mathbf{y}_k \mathbf{y}_k^\mathrm{T} \mathbf{K}^\mathrm{T} + \frac{1}{\epsilon} \mathbf{P} \mathbf{b}_k \mathbf{b}_k^\mathrm{T} \mathbf{P}) = \epsilon \mathbf{K} \mathbf{\Sigma_y} \mathbf{K}^\mathrm{T} + \frac{1}{\epsilon} \mathbf{P} \mathbf{\Sigma_b} \mathbf{P}
\end{equation}
with the third and fourth equations of (\ref{eq:Sigma_a_x_b_y}) applied.

Substitute (\ref{eq:robust_gain_matrix_design_solution}), (\ref{eq:robust_GM_L1}), and (\ref{eq:robust_GM_L2}) into (\ref{eq:uncertain_state_DE_linear2_closed_L2}) and obtain
\begin{align*}
\mathbf{L} &\leq \mathbf{P} \mathbf{A} + \mathbf{A}^\mathrm{T} \mathbf{P} - 2 \mathbf{P} \mathbf{B} (\mathbf{R} + \epsilon \mathbf{\Sigma_y})^{-1} \mathbf{B}^\mathrm{T} \mathbf{P} + (\mathbf{P} \mathbf{\Sigma_a} \mathbf{P} + \mathbf{\Sigma_x}) + (\epsilon \mathbf{K} \mathbf{\Sigma_y} \mathbf{K}^\mathrm{T} + \frac{1}{\epsilon} \mathbf{P} \mathbf{\Sigma_b} \mathbf{P})  \\
  &= \mathbf{P} \mathbf{A} + \mathbf{A}^\mathrm{T} \mathbf{P} + \mathbf{\Sigma_x}  \\
  &\qquad - \mathbf{P} [2 \mathbf{B} (\mathbf{R} + \epsilon \mathbf{\Sigma_y})^{-1} \mathbf{B}^\mathrm{T} - \epsilon \mathbf{B} (\mathbf{R} + \epsilon \mathbf{\Sigma_y})^{-1} \mathbf{\Sigma_y} (\mathbf{R} + \epsilon \mathbf{\Sigma_y})^{-1} \mathbf{B}^\mathrm{T} - \mathbf{\Sigma_a} - \frac{1}{\epsilon} \mathbf{\Sigma_b}] \mathbf{P}  \\
  &= \mathbf{P} \mathbf{A} + \mathbf{A}^\mathrm{T} \mathbf{P} - \mathbf{P} \mathbf{M} \mathbf{P} + \mathbf{Q_\Sigma} - \mathbf{Q} = - \mathbf{Q} < 0.
\end{align*}

Note that $\Delta \mathbf{A}$ and $\Delta \mathbf{B}$ in the Lyapunov equation style matrix $\mathbf{L}$ defined in (\ref{eq:uncertain_state_DE_linear2_closed_L}) can be arbitrary, so above analysis conveys that $\mathbf{L}$ is guaranteed to be negative definite given the gain matrix $\mathbf{K}$ specified in (\ref{eq:robust_gain_matrix_design_solution}) and the positive definite solution $\mathbf{P}$ satisfying (\ref{eq:Petersen_YangY_Riccati}). In other words, no matter for what $\Delta \mathbf{A}$ and $\Delta \mathbf{B}$, there is always a positive definite matrix $\mathbf{P}$ such that
\begin{align*}
\mathbf{P} [(\mathbf{A} + \Delta \mathbf{A}) - (\mathbf{B} + \Delta \mathbf{B}) \mathbf{K}^\mathrm{T}] + [(\mathbf{A} + \Delta \mathbf{A}) - (\mathbf{B} + \Delta \mathbf{B}) \mathbf{K}^\mathrm{T}]^\mathrm{T} \mathbf{P} < 0.
\end{align*}
Then according to the \textit{Lyapunov criterion III-B} presented in Section 1.4.1 in Chapter 1, 
\footnote{Namely Chapter 1 of the author's works \cite{Li2026ACTPA_SJTU_2, Li2026ACTPA_SJTU_1}.}
the closed-loop state transition matrix
\begin{align*}
(\mathbf{A} + \Delta \mathbf{A}) - (\mathbf{B} + \Delta \mathbf{B}) \mathbf{K}^\mathrm{T}
\end{align*}
is guaranteed to be stable. The proof is done.
\end{proof}

\subsubsection*{Application: single inverted pendulum robust control}

Consider single inverted pendulum control. As in Section 4.1.4 in Chapter 4 and in Section \ref{sec:LSS_model_uncertainty}, we focus on handling the partial or reduced state
\begin{align*}
\mathbf{x}_P \equiv \begin{bmatrix} \theta & \frac{\mathrm{d} \theta}{\mathrm{d} t} & \frac{\mathrm{d} x}{\mathrm{d} t} \end{bmatrix}^\mathrm{T}
\end{align*}
that consists of the inverted pendulum angle, the inverted pendulum angular velocity, and the cart velocity. Extract the sub-model associated with the partial state $\mathbf{x}_P$ from (\ref{eq:uncertain_SIP_state_DE_linear}) and obtain (\ref{eq:uncertain_SIP_state_DE_linear_partial})
\begin{align*}
\frac{\mathrm{d}}{\mathrm{d} t} \mathbf{x}_P = \mathbf{A}_P(\mathbf{x}_P) \mathbf{x}_P + \mathbf{B}_P(\mathbf{x}_P) a,
\end{align*}
where
\begin{align*}
\mathbf{A}_P(\mathbf{x}_P) \equiv \begin{bmatrix} 0 & 1 & 0 \\ \frac{g}{L} \frac{\sin \theta}{\theta} & 0 & 0 \\ 0 & 0 & 0 \end{bmatrix}, \quad \mathbf{B}_P(\mathbf{x}_P) \equiv \begin{bmatrix} 0 \\ -\frac{\cos \theta}{L} \\ 1 \end{bmatrix}.
\end{align*}

Suppose the operation range of the inverted pendulum angle $\theta$ is
\begin{align}  \label{eq:SIP_theta_operation_range}
- \theta_{\max} \leq \theta \leq \theta_{\max}.
\end{align}
Then we can transform (\ref{eq:uncertain_SIP_state_DE_linear_partial}) into an instantiation of the generic formalism (\ref{eq:uncertain_state_DE_linear2}) as
\begin{align*}
\frac{\mathrm{d}}{\mathrm{d} t} \mathbf{x} = (\mathbf{A} + \Delta \mathbf{A}) \mathbf{x} + (\mathbf{B} + \Delta \mathbf{B}) \mathbf{u},
\end{align*}
where
\begin{align*}
\mathbf{A} \equiv \begin{bmatrix} 0 & 1 & 0 \\ \frac{g}{L} & 0 & 0 \\ 0 & 0 & 0 \end{bmatrix}, \qquad \mathbf{B} \equiv \begin{bmatrix} 0 \\ -\frac{1}{L} \\ 1 \end{bmatrix}
\end{align*}
and
\begin{align*}
| \Delta \mathbf{A} | &\preceq \Delta \mathbf{A}_{\max} \equiv \begin{bmatrix} 0 & 0 & 0 \\ \frac{g}{L} (1 - \frac{\sin \theta_{\max}}{\theta_{\max}}) & 0 & 0 \\ 0 & 0 & 0 \end{bmatrix},  \\ 
| \Delta \mathbf{B} | &\preceq \Delta \mathbf{B}_{\max} \equiv \begin{bmatrix} 0 \\ \frac{1 - \cos \theta_{\max}}{L} \\ 0 \end{bmatrix}.
\end{align*}

For concrete configuration of parameters, let 
\begin{align*}  
L = 1, \quad g = 10, \quad \theta_{\max} = 0.4 \pi, 
\end{align*}
then we have
\begin{align*}
\mathbf{A} &= \begin{bmatrix} 0 & 1 & 0 \\ 10 & 0 & 0 \\ 0 & 0 & 0 \end{bmatrix}, \qquad \mathbf{B} = \begin{bmatrix} 0 \\ -1 \\ 1 \end{bmatrix},  \\
\Delta \mathbf{A}_{\max} &= \begin{bmatrix} 0 & 0 & 0 \\ 2.43 & 0 & 0 \\ 0 & 0 & 0 \end{bmatrix}, \qquad \Delta \mathbf{B}_{\max} = \begin{bmatrix} 0 \\ 0.69 \\ 0 \end{bmatrix}.
\end{align*}
Decompose the uncertainty factor $\Delta \mathbf{A}$ into
\begin{align*}
\Delta \mathbf{A} = a_1 (t) \begin{bmatrix} 0 \\ 0.0081 \\ 0 \end{bmatrix} \begin{bmatrix} 1 & 0 & 0 \end{bmatrix} \equiv a_1 \mathbf{a}_1 \mathbf{x}_1^\mathrm{T},
\end{align*}
where
\begin{align*}
| a_1 | = | a_1 (t) | \leq \bar{a} \equiv 300.
\end{align*}
Decompose the uncertainty factor $\Delta \mathbf{B}$ into 
\begin{align*}
\Delta \mathbf{B} = b_1 (t) \begin{bmatrix} 0 \\ 0.0023 \\ 0 \end{bmatrix} \begin{bmatrix} 1 \end{bmatrix} \equiv b_1 \mathbf{b}_1 \mathbf{y}_1^\mathrm{T},
\end{align*}
where
\begin{align*}
| b_1 | = | b_1 (t) | \leq \bar{b} \equiv 300.
\end{align*}
We also have
\begin{align*}
\Delta \mathbf{A}_{\max} = \bar{a} \mathbf{a}_1 \mathbf{x}_1^\mathrm{T}, \qquad \Delta \mathbf{B}_{\max} = \bar{b} \mathbf{b}_1 \mathbf{y}_1^\mathrm{T}.
\end{align*}

Compute the four auxiliary matrices defined in (\ref{eq:Sigma_a_x_b_y})
\begin{align*}
\mathbf{\Sigma_a} &= \bar{a} \mathbf{a}_1 \mathbf{a}_1^\mathrm{T} = 300 \begin{bmatrix} 0 \\ 0.0081 \\ 0 \end{bmatrix} \begin{bmatrix} 0 & 0.0081 & 0 \end{bmatrix} = \begin{bmatrix} 0 & 0 & 0 \\ 0 & 0.0197 & 0 \\ 0 & 0 & 0 \end{bmatrix},  \\
\mathbf{\Sigma_x} &= \bar{a} \mathbf{x}_1 \mathbf{x}_1^\mathrm{T} = 300 \begin{bmatrix} 1 \\ 0 \\ 0 \end{bmatrix} \begin{bmatrix} 1 & 0 & 0 \end{bmatrix} = \begin{bmatrix} 300 & 0 & 0 \\ 0 & 0 & 0 \\ 0 & 0 & 0 \end{bmatrix},  \\
\mathbf{\Sigma_b} &= \bar{b} \mathbf{b}_1 \mathbf{b}_1^\mathrm{T} = 300 \begin{bmatrix} 0 \\ 0.0023 \\ 0 \end{bmatrix} \begin{bmatrix} 0 & 0.0023 & 0 \end{bmatrix} = \begin{bmatrix} 0 & 0 & 0 \\ 0 & 0.0016 & 0 \\ 0 & 0 & 0 \end{bmatrix},  \\
\mathbf{\Sigma_y} &= \bar{b} \mathbf{y}_1 \mathbf{y}_1^\mathrm{T} = 300 \begin{bmatrix} 1 \end{bmatrix} \begin{bmatrix} 1 \end{bmatrix} = 300.
\end{align*}
Set
\begin{align*}
\mathbf{Q} = \begin{bmatrix} 1 & & \\ & 1 & \\ & & 1 \end{bmatrix}, \qquad \mathbf{R} = 0.01, \qquad \epsilon = 0.01
\end{align*}
and compute
\begin{align*}
\mathbf{M} &= \mathbf{B} (\mathbf{R} + \epsilon \mathbf{\Sigma_y})^{-1} (2 \mathbf{R} + \epsilon \mathbf{\Sigma_y}) (\mathbf{R} + \epsilon \mathbf{\Sigma_y})^{-1} \mathbf{B}^\mathrm{T} - \mathbf{\Sigma_a} - \frac{1}{\epsilon} \mathbf{\Sigma_b} = 
\begin{bmatrix} 0 & 0 & 0 \\ 0 & 0.15 & -0.33 \\ 0 & -0.33 & 0.33 \end{bmatrix},  \\
\mathbf{Q_\Sigma} &= \mathbf{\Sigma_x} + \mathbf{Q} = \begin{bmatrix} 301 & & \\ & 1 & \\ & & 1 \end{bmatrix}.
\end{align*}

Solve the Riccati equation (\ref{eq:Petersen_YangY_Riccati})
\begin{align*}
\mathbf{P} \mathbf{A} + \mathbf{A}^\mathrm{T} \mathbf{P} - \mathbf{P} \mathbf{M} \mathbf{P} + \mathbf{Q_\Sigma} = \mathbf{0}
\end{align*}
and obtain a positive definite solution
\begin{align*}
\mathbf{P} = \begin{bmatrix} 2042.9 & 644.3 & 132.1 \\ 644.3 & 208.3 & 43.5 \\ 132.1 & 43.5 & 11.6 \end{bmatrix}.
\end{align*}
Then compute the robust gain matrix via (\ref{eq:robust_gain_matrix_design_solution}) as
\begin{equation}  \label{eq:SIP_robust_gain_matrix}
\mathbf{K}_P = \mathbf{P} \mathbf{B} (\mathbf{R} + \epsilon \mathbf{\Sigma_y})^{-1} = \begin{bmatrix} -170.2 & -54.7 & -10.6 \end{bmatrix}^\mathrm{T}.
\end{equation}
The corresponding robust full-state feedback control law is
\begin{equation}  \label{eq:SIP_robust_FSFC}
a = - \mathbf{K}_P^\mathrm{T} \mathbf{x}_P = 170.2 \theta + 54.7 \frac{\mathrm{d} \theta}{\mathrm{d} t} + 10.6  \frac{\mathrm{d} x}{\mathrm{d} t}.
\end{equation}

\subsubsection*{Verification of control robustness}

To verify that the robust gain matrix $\mathbf{K}_P$ given in (\ref{eq:SIP_robust_gain_matrix}) is indeed a solution to the problem formalized in (\ref{eq:robust_gain_matrix_design}), or in other words, to verify that the robust full-state feedback control law (\ref{eq:SIP_robust_FSFC}) can work under all conditions of $\Delta \mathbf{A}$ and $\Delta \mathbf{B}$ for single inverted pendulum control, we traverse inverted pendulum angles densely in the $\theta$ operation range described in (\ref{eq:SIP_theta_operation_range}) and compute eigenvalue real parts of their associated closed-loop state transition matrices
\footnote{In the context of single inverted pendulum control, variation of $\mathbf{A}_P(\mathbf{x}_P)$ and $\mathbf{B}_P(\mathbf{x}_P)$ actually depends on the inverted pendulum angle $\theta$ only.}
namely
\begin{align*}
\{ \mbox{Re}(\mbox{Eig}(\mathbf{A}_P(\mathbf{x}_P) - \mathbf{B}_P(\mathbf{x}_P) \mathbf{K}_P^\mathrm{T})) \mbox{ } | \mbox{ } - \theta_{\max} \leq \theta \leq \theta_{\max} \},
\end{align*}
which are listed in Table \ref{tab:SIP_eig_re_traversed}. Here, ``Re. Eig.'' means ``real part of eigenvalue''. As we can see in Table \ref{tab:SIP_eig_re_traversed}, all the eigenvalues have negative real parts, which implies that all the closed-loop state transition matrices are stable.

\begin{longtable}{|c c c c|}
\caption{Re. Eig. associated with densely traversed $\theta$\label{tab:SIP_eig_re_traversed}} \\
\hline
$\theta$ (degree) & 1st Re. Eig. & 2nd Re. Eig. & 3rd Re. Eig. \\ \hline
-72 & -2.0366 & -2.0366 & -2.2358 \\
-71 & -2.5055 & -2.5055 & -2.2039 \\
-70 & -2.9730 & -2.9730 & -2.1695 \\
-69 & -3.4390 & -3.4390 & -2.1326 \\
-68 & -3.9032 & -3.9032 & -2.0931 \\
-67 & -4.3654 & -4.3654 & -2.0513 \\
-66 & -4.8254 & -4.8254 & -2.0075 \\
-65 & -5.2829 & -5.2829 & -1.9618 \\
-64 & -5.7376 & -5.7376 & -1.9149 \\
-63 & -6.1890 & -6.1890 & -1.8671 \\
-62 & -6.6368 & -6.6368 & -1.8191 \\
-61 & -8.0478 & -6.1131 & -1.7713 \\
-60 & -9.9006 & -5.1391 & -1.7241 \\
-59 & -11.2187 & -4.6903 & -1.6781 \\
-58 & -12.3703 & -4.3980 & -1.6334 \\
-57 & -13.4305 & -4.1866 & -1.5904 \\
-56 & -14.4301 & -4.0250 & -1.5491 \\
-55 & -15.3848 & -3.8971 & -1.5098 \\
-54 & -16.3039 & -3.7933 & -1.4723 \\
-53 & -17.1933 & -3.7074 & -1.4368 \\
-52 & -18.0571 & -3.6353 & -1.4031 \\
-51 & -18.8980 & -3.5740 & -1.3712 \\
-50 & -19.7181 & -3.5213 & -1.3411 \\
-49 & -20.5187 & -3.4756 & -1.3126 \\
-48 & -21.3011 & -3.4357 & -1.2858 \\
-47 & -22.0660 & -3.4006 & -1.2604 \\
-46 & -22.8142 & -3.3694 & -1.2364 \\
-45 & -23.5460 & -3.3417 & -1.2137 \\
-44 & -24.2619 & -3.3169 & -1.1923 \\
-43 & -24.9621 & -3.2946 & -1.1721 \\
-42 & -25.6468 & -3.2745 & -1.1530 \\
-41 & -26.3162 & -3.2562 & -1.1349 \\
-40 & -26.9704 & -3.2397 & -1.1178 \\
-39 & -27.6095 & -3.2245 & -1.1016 \\
-38 & -28.2334 & -3.2107 & -1.0863 \\
-37 & -28.8422 & -3.1980 & -1.0717 \\
-36 & -29.4360 & -3.1864 & -1.0580 \\
-35 & -30.0145 & -3.1757 & -1.0450 \\
-34 & -30.5779 & -3.1658 & -1.0326 \\
-33 & -31.1261 & -3.1566 & -1.0209 \\
-32 & -31.6590 & -3.1481 & -1.0098 \\
-31 & -32.1766 & -3.1403 & -0.9993 \\
-30 & -32.6787 & -3.1330 & -0.9894 \\
-29 & -33.1654 & -3.1263 & -0.9799 \\
-28 & -33.6365 & -3.1200 & -0.9710 \\
-27 & -34.0920 & -3.1141 & -0.9626 \\
-26 & -34.5317 & -3.1087 & -0.9546 \\
-25 & -34.9557 & -3.1036 & -0.9470 \\
-24 & -35.3637 & -3.0989 & -0.9399 \\
-23 & -35.7558 & -3.0945 & -0.9331 \\
-22 & -36.1319 & -3.0904 & -0.9268 \\
-21 & -36.4918 & -3.0866 & -0.9208 \\
-20 & -36.8356 & -3.0831 & -0.9152 \\
-19 & -37.1631 & -3.0798 & -0.9099 \\
-18 & -37.4743 & -3.0768 & -0.9049 \\
-17 & -37.7690 & -3.0740 & -0.9003 \\
-16 & -38.0473 & -3.0714 & -0.8960 \\
-15 & -38.3090 & -3.0689 & -0.8920 \\
-14 & -38.5542 & -3.0667 & -0.8883 \\
-13 & -38.7827 & -3.0647 & -0.8848 \\
-12 & -38.9945 & -3.0629 & -0.8817 \\
-11 & -39.1895 & -3.0612 & -0.8788 \\
-10 & -39.3678 & -3.0597 & -0.8762 \\
-9 & -39.5292 & -3.0583 & -0.8738 \\
-8 & -39.6737 & -3.0571 & -0.8718 \\
-7 & -39.8013 & -3.0561 & -0.8699 \\
-6 & -39.9119 & -3.0552 & -0.8683 \\
-5 & -40.0056 & -3.0544 & -0.8670 \\
-4 & -40.0822 & -3.0538 & -0.8659 \\
-3 & -40.1419 & -3.0533 & -0.8651 \\
-2 & -40.1845 & -3.0530 & -0.8645 \\
-1 & -40.2101 & -3.0528 & -0.8641 \\
0 & -40.2186 & -3.0527 & -0.8640 \\
1 & -40.2101 & -3.0528 & -0.8641 \\
2 & -40.1845 & -3.0530 & -0.8645 \\
3 & -40.1419 & -3.0533 & -0.8651 \\
4 & -40.0822 & -3.0538 & -0.8659 \\
5 & -40.0056 & -3.0544 & -0.8670 \\
6 & -39.9119 & -3.0552 & -0.8683 \\
7 & -39.8013 & -3.0561 & -0.8699 \\
8 & -39.6737 & -3.0571 & -0.8718 \\
9 & -39.5292 & -3.0583 & -0.8738 \\
10 & -39.3678 & -3.0597 & -0.8762 \\
11 & -39.1895 & -3.0612 & -0.8788 \\
12 & -38.9945 & -3.0629 & -0.8817 \\
13 & -38.7827 & -3.0647 & -0.8848 \\
14 & -38.5542 & -3.0667 & -0.8883 \\
15 & -38.3090 & -3.0689 & -0.8920 \\
16 & -38.0473 & -3.0714 & -0.8960 \\
17 & -37.7690 & -3.0740 & -0.9003 \\
18 & -37.4743 & -3.0768 & -0.9049 \\
19 & -37.1631 & -3.0798 & -0.9099 \\
20 & -36.8356 & -3.0831 & -0.9152 \\
21 & -36.4918 & -3.0866 & -0.9208 \\
22 & -36.1319 & -3.0904 & -0.9268 \\
23 & -35.7558 & -3.0945 & -0.9331 \\
24 & -35.3637 & -3.0989 & -0.9399 \\
25 & -34.9557 & -3.1036 & -0.9470 \\
26 & -34.5317 & -3.1087 & -0.9546 \\
27 & -34.0920 & -3.1141 & -0.9626 \\
28 & -33.6365 & -3.1200 & -0.9710 \\
29 & -33.1654 & -3.1263 & -0.9799 \\
30 & -32.6787 & -3.1330 & -0.9894 \\
31 & -32.1766 & -3.1403 & -0.9993 \\
32 & -31.6590 & -3.1481 & -1.0098 \\
33 & -31.1261 & -3.1566 & -1.0209 \\
34 & -30.5779 & -3.1658 & -1.0326 \\
35 & -30.0145 & -3.1757 & -1.0450 \\
36 & -29.4360 & -3.1864 & -1.0580 \\
37 & -28.8422 & -3.1980 & -1.0717 \\
38 & -28.2334 & -3.2107 & -1.0863 \\
39 & -27.6095 & -3.2245 & -1.1016 \\
40 & -26.9704 & -3.2397 & -1.1178 \\
41 & -26.3162 & -3.2562 & -1.1349 \\
42 & -25.6468 & -3.2745 & -1.1530 \\
43 & -24.9621 & -3.2946 & -1.1721 \\
44 & -24.2619 & -3.3169 & -1.1923 \\
45 & -23.5460 & -3.3417 & -1.2137 \\
46 & -22.8142 & -3.3694 & -1.2364 \\
47 & -22.0660 & -3.4006 & -1.2604 \\
48 & -21.3011 & -3.4357 & -1.2858 \\
49 & -20.5187 & -3.4756 & -1.3126 \\
50 & -19.7181 & -3.5213 & -1.3411 \\
51 & -18.8980 & -3.5740 & -1.3712 \\
52 & -18.0571 & -3.6353 & -1.4031 \\
53 & -17.1933 & -3.7074 & -1.4368 \\
54 & -16.3039 & -3.7933 & -1.4723 \\
55 & -15.3848 & -3.8971 & -1.5098 \\
56 & -14.4301 & -4.0250 & -1.5491 \\
57 & -13.4305 & -4.1866 & -1.5904 \\
58 & -12.3703 & -4.3980 & -1.6334 \\
59 & -11.2187 & -4.6903 & -1.6781 \\
60 & -9.9006 & -5.1391 & -1.7241 \\
61 & -8.0478 & -6.1131 & -1.7713 \\
62 & -6.6368 & -6.6368 & -1.8191 \\
63 & -6.1890 & -6.1890 & -1.8671 \\
64 & -5.7376 & -5.7376 & -1.9149 \\
65 & -5.2829 & -5.2829 & -1.9618 \\
66 & -4.8254 & -4.8254 & -2.0075 \\
67 & -4.3654 & -4.3654 & -2.0513 \\
68 & -3.9032 & -3.9032 & -2.0931 \\
69 & -3.4390 & -3.4390 & -2.1326 \\
70 & -2.9730 & -2.9730 & -2.1695 \\
71 & -2.5055 & -2.5055 & -2.2039 \\
72 & -2.0366 & -2.0366 & -2.2358 \\ \hline
\end{longtable}

Matlab simulation code for complete demonstration of single inverted pendulum robust control is given as follows. The visualization code \textbf{DisplaySIP.m} and the single inverted pendulum dynamics code \textbf{DynamicsSIP.m} are given in Section 2.2.3 in Chapter 2. The gain matrix designing code \textbf{DesignGainMatrix.m} is given in Section 2.3.2 in Chapter 2. The Riccati equation solving code \textbf{SolveRiccatiEquation.m} is given in Section 1.4.4 in Chapter 1.
\footnote{Namely Chapters 1 and 2 of the author's works \cite{Li2026ACTPA_SJTU_2, Li2026ACTPA_SJTU_1}.}

\begin{framed} 
\noindent \textbf{SingleInvertedPendulumRobustFSFC.m} \\
\noindent \%\% Single inverted pendulum parameters \\
m1 = 1; L1 = 1; g = 10; \\
\%\% Simulation preliminary configuration \\
dt = 0.001; \% Numerical computation step \\
tSpan = 0:dt:20; \% Simulation time span \\
x = 0.2; \% Cart position  \\
dx = 0; \% Cart velocity \\
y = 0.4*pi; \% Inverted pendulum angle theta  \\
dy = 0;  \% Inverted pendulum angular velocity \\
stt = [y; dy; x; dx]; \% Single inverted pendulum state \\
sttAll = zeros(length(stt), length(tSpan)); k = 0; \% Record states \\
xExpected = 0; yExpected = 0; \% Expected equilibrium status \\
SimConfig = [m1, L1, g, dt]; \\
 \\
\%\% Design robust gain matrix \\
A = [0, 1, 0, 0; g/L1, 0, 0, 0; 0, 0, 0, 1; 0, 0, 0, 0]; \\
B = [0; -1/L1; 0; 1]; \\
sttK = DesignGainMatrix(A, B, [-4;-4+2i;-4-2i;-4]); \% For SMC \\
fprintf('SMC gain matrix K: '); disp(sttK'); \\
A1 = A([1,2,4], [1,2,4]); B1 = B([1,2,4]); \\
a = 0.4*pi; A2 = [0, 1, 0; (g/L1)*(sin(a)/a), 0, 0; 0, 0, 0];  \\
B2 = [0; (-1/L1)*cos(a); 1]; \\
Ap = A1; dA = abs(A2-Ap); Bp = B1; dB = abs(B2-Bp); \\
\textit{} [ak, xk] = nnmf(dA,1); xk = xk'; am = 300; ak = ak/am; \\
\textit{} [bk, yk] = nnmf(dB,1); yk = yk'; bm = 300; bk = bk/bm; \\
Sa = am*ak*ak'; Sx = am*xk*xk'; Sb = bm*bk*bk'; Sy = bm*yk*yk'; \\
Q = eye(3); R = 0.01; ep = 0.01; \\
M = Bp*inv(R+ep*Sy)*(2*R+ep*Sy)*inv(R+ep*Sy)*Bp'-Sa-(1/ep)*Sb;  \\
QS = Sx+Q; \\
P = SolveRiccatiEquation(Ap, M, QS, 'hamilton'); \\
sttKp = (inv(R+ep*Sy)*Bp'*P)'; \\
fprintf('Riccati equation solution P: '); disp(P); \\
fprintf('Robust gain matrix K: '); disp(sttKp'); \\
Ac1=A1-B1*sttKp'; Ac2=A1-B2*sttKp'; Ac3=A2-B1*sttKp'; Ac4=A2-B2*sttKp'; \\
fprintf('Eig. Re. of bounding closed-loop state transition matrices:$\backslash$n'); \\
disp(real([eig(Ac1),eig(Ac2),eig(Ac3),eig(Ac4)])); \\
cRobust=1; fprintf('Start robust control$\backslash$n');  \\
 \\
\%\% Simulation of single inverted pendulum control \\
for t = tSpan \\
$~~~~$ \%\% Control method \\
$~~~~$ if (y\^{}2+dy\^{}2+dx\^{}2 $>$ 1 \&\& cRobust==1) \\
$~~~~$ $~~~~$ acc = -sttKp'*stt([1,2,4]); \% Robust FSFC  \\
$~~~~$ else \\
$~~~~$ $~~~~$ if (1==cRobust) \\
$~~~~$ $~~~~$ $~~~~$ cRobust = 0; sttE = [0; 0; x; 0]; \\
$~~~~$ $~~~~$ $~~~~$ fprintf('Switch to sliding mode control$\backslash$n'); \\
$~~~~$ $~~~~$ end \\
$~~~~$ $~~~~$ if (sttE(3)$>$0) sttE(3) = max(sttE(3) - 8*dt, 0); \\
$~~~~$ $~~~~$ else sttE(3) = min(sttE(3) + 8*dt, 0); end \\
$~~~~$ $~~~~$ acc = -sttK'*(stt-sttE); \% Sliding mode FSFC \\
$~~~~$ end \\
 \\
$~~~~$ \%\% Single inverted pendulum dynamics \\
$~~~~$ stt = DynamicsSIP(SimConfig, stt, acc); \\
$~~~~$ sttC = num2cell(stt); [y, dy, x, dx] = sttC\{:\}; \\
$~~~~$ if (x\^{}2+y\^{}2+dy\^{}2+dx\^{}2$<$0.001) fprintf('Control success!$\backslash$n'); break; end \\
$~~~~$ if (abs(y)$>$=pi/2) fprintf('Control failure!$\backslash$n'); break; end \\
$~~~~$ k = k+1; sttAll(:,k) = stt; \\
$~~~~$ \%\% Single inverted pendulum visualization \\
$~~~~$ if (rem(k,20) == 0) \\
$~~~~$ $~~~~$ DisplaySIP(x, y, L1); pause(dt); \\
$~~~~$ end \\
end
\end{framed}

Note that
\begin{align*}
&\mathbf{A}_{\min} \preceq \mathbf{A}_P(\mathbf{x}_P) \preceq \mathbf{A}_{\max},  \\
&\mathbf{B}_{\min} \preceq \mathbf{B}_P(\mathbf{x}_P) \preceq \mathbf{B}_{\max},
\end{align*}
where
\begin{align*}
\mathbf{A}_{\min} = \begin{bmatrix} 0 & 1 & 0 \\ \frac{g}{L} \frac{\sin \theta_{\max}}{\theta_{\max}} & 0 & 0 \\ 0 & 0 & 0 \end{bmatrix}, \qquad \mathbf{A}_{\max} = \begin{bmatrix} 0 & 1 & 0 \\ \frac{g}{L} & 0 & 0 \\ 0 & 0 & 0 \end{bmatrix},
\end{align*}
\begin{align*}
\mathbf{B}_{\min} = \begin{bmatrix} 0 \\ -\frac{1}{L} \\ 1 \end{bmatrix}, \qquad \mathbf{B}_{\max} = \begin{bmatrix} 0 \\ -\frac{\cos \theta_{\max}}{L} \\ 1 \end{bmatrix}.
\end{align*}
For modelling via (\ref{eq:uncertain_state_DE_linear2})
\begin{align*}
\frac{\mathrm{d}}{\mathrm{d} t} \mathbf{x} = (\mathbf{A} + \Delta \mathbf{A}) \mathbf{x} + (\mathbf{B} + \Delta \mathbf{B}) \mathbf{u},
\end{align*}
we may also set
\begin{align*}
\mathbf{A} &\equiv \frac{1}{2} (\mathbf{A}_{\min} + \mathbf{A}_{\max}),  \\
\mathbf{B} &\equiv \frac{1}{2} (\mathbf{B}_{\min} + \mathbf{B}_{\max}),  \\
\Delta \mathbf{A}_{\max} &\equiv \frac{1}{2} (\mathbf{A}_{\max} - \mathbf{A}_{\min}),  \\
\Delta \mathbf{B}_{\max} &\equiv \frac{1}{2} (\mathbf{B}_{\max} - \mathbf{B}_{\min})
\end{align*}
and then apply the Riccati equation method. Matlab simulation code for complete demonstration of single inverted pendulum robust control in such way is given as follows.

\begin{framed} 
\noindent \textbf{SingleInvertedPendulumRobustFSFC2.m} \\
\noindent \%\% Single inverted pendulum parameters \\
m1 = 1; L1 = 1; g = 10; \\
\%\% Simulation preliminary configuration \\
dt = 0.001; \% Numerical computation step \\
tSpan = 0:dt:20; \% Simulation time span \\
x = 0.2; \% Cart position  \\
dx = 0; \% Cart velocity \\
y = 0.4*pi; \% Inverted pendulum angle theta  \\
dy = 0;  \% Inverted pendulum angular velocity \\
stt = [y; dy; x; dx]; \% Single inverted pendulum state \\
sttAll = zeros(length(stt), length(tSpan)); k = 0; \% Record states \\
xExpected = 0; yExpected = 0; \% Expected equilibrium status \\
SimConfig = [m1, L1, g, dt]; \\
 \\
\%\% Design robust gain matrix \\
A = [0, 1, 0, 0; g/L1, 0, 0, 0; 0, 0, 0, 1; 0, 0, 0, 0]; \\
B = [0; -1/L1; 0; 1]; \\
sttK = DesignGainMatrix(A, B, [-4;-4+2i;-4-2i;-4]); \% For SMC \\
fprintf('SMC gain matrix K: '); disp(sttK'); \\
A1 = A([1,2,4], [1,2,4]); B1 = B([1,2,4]); \\
a = 0.4*pi; A2 = [0, 1, 0; (g/L1)*(sin(a)/a), 0, 0; 0, 0, 0];  \\
B2 = [0; (-1/L1)*cos(a); 1]; \\
Ap = (A1+A2)/2; dA = abs(A2-Ap); Bp = (B1+B2)/2; dB = abs(B2-Bp); \\
\textit{} [ak, xk] = nnmf(dA,1); xk = xk'; am = 50; ak = ak/am; \\
\textit{} [bk, yk] = nnmf(dB,1); yk = yk'; bm = 50; bk = bk/bm; \\
Sa = am*ak*ak'; Sx = am*xk*xk'; Sb = bm*bk*bk'; Sy = bm*yk*yk'; \\
Q = eye(3); R = 0.01; ep = 0.01; \\
M = Bp*inv(R+ep*Sy)*(2*R+ep*Sy)*inv(R+ep*Sy)*Bp'-Sa-(1/ep)*Sb;  \\
QS = Sx+Q; \\
P = SolveRiccatiEquation(Ap, M, QS, 'hamilton'); \\
sttKp = (inv(R+ep*Sy)*Bp'*P)'; \\
fprintf('Riccati equation solution P: '); disp(P); \\
fprintf('Robust gain matrix K: '); disp(sttKp'); \\
Ac1=A1-B1*sttKp'; Ac2=A1-B2*sttKp'; Ac3=A2-B1*sttKp'; Ac4=A2-B2*sttKp'; \\
fprintf('Eig. Re. of bounding closed-loop state transition matrices:$\backslash$n'); \\
disp(real([eig(Ac1),eig(Ac2),eig(Ac3),eig(Ac4)])); \\
cRobust=1; fprintf('Start robust control$\backslash$n');  \\
 \\
\%\% Simulation of single inverted pendulum control \\
for t = tSpan \\
$~~~~$ \%\% Control method \\
$~~~~$ if (y\^{}2+dy\^{}2+dx\^{}2 $>$ 1 \&\& cRobust==1) \\
$~~~~$ $~~~~$ acc = -sttKp'*stt([1,2,4]); \% Robust FSFC  \\
$~~~~$ else \\
$~~~~$ $~~~~$ if (1==cRobust) \\
$~~~~$ $~~~~$ $~~~~$ cRobust = 0; sttE = [0; 0; x; 0]; \\
$~~~~$ $~~~~$ $~~~~$ fprintf('Switch to sliding mode control$\backslash$n'); \\
$~~~~$ $~~~~$ end \\
$~~~~$ $~~~~$ if (sttE(3)$>$0) sttE(3) = max(sttE(3) - 8*dt, 0); \\
$~~~~$ $~~~~$ else sttE(3) = min(sttE(3) + 8*dt, 0); end \\
$~~~~$ $~~~~$ acc = -sttK'*(stt-sttE); \% Sliding mode FSFC \\
$~~~~$ end \\
 \\
$~~~~$ \%\% Single inverted pendulum dynamics \\
$~~~~$ stt = DynamicsSIP(SimConfig, stt, acc); \\
$~~~~$ sttC = num2cell(stt); [y, dy, x, dx] = sttC\{:\}; \\
$~~~~$ if (x\^{}2+y\^{}2+dy\^{}2+dx\^{}2$<$0.001) fprintf('Control success!$\backslash$n'); break; end \\
$~~~~$ if (abs(y)$>$=pi/2) fprintf('Control failure!$\backslash$n'); break; end \\
$~~~~$ k = k+1; sttAll(:,k) = stt; \\
$~~~~$ \%\% Single inverted pendulum visualization \\
$~~~~$ if (rem(k,20) == 0) \\
$~~~~$ $~~~~$ DisplaySIP(x, y, L1); pause(dt); \\
$~~~~$ end \\
end
\end{framed}

\subsubsection*{Tuning of relevant parameters for Riccati equation solving}

The Riccati equation (\ref{eq:Petersen_YangY_Riccati})
\begin{align*}
\mathbf{P} \mathbf{A} + \mathbf{A}^\mathrm{T} \mathbf{P} - \mathbf{P} \mathbf{M} \mathbf{P} + \mathbf{Q_\Sigma} = \mathbf{0},
\end{align*}
where
\begin{align*}
\mathbf{M} &= \mathbf{B} (\mathbf{R} + \epsilon \mathbf{\Sigma_y})^{-1} (2 \mathbf{R} + \epsilon \mathbf{\Sigma_y}) (\mathbf{R} + \epsilon \mathbf{\Sigma_y})^{-1} \mathbf{B}^\mathrm{T} - \mathbf{\Sigma_a} - \frac{1}{\epsilon} \mathbf{\Sigma_b},  \\
\mathbf{Q_\Sigma} &= \mathbf{\Sigma_x} + \mathbf{Q},
\end{align*}
is not guaranteed to have any positive definite solution $\mathbf{P}$, if relevant parameters especially $\bar{a}$, $\bar{b}$, and $\epsilon$ are tuned casually.

As a rule of thumb, when the Riccati equation (\ref{eq:Petersen_YangY_Riccati}) has no positive definite solution, one may somewhat increase $\bar{a}$ and $\bar{b}$ and somewhat decrease $\epsilon$ until (\ref{eq:Petersen_YangY_Riccati}) has a positive definite solution $\mathbf{P}$. It is worth noting that increasing $\bar{a}$ means not only increasing $\bar{a}$ but also shrinking $\mathbf{a}_k$ in the uncertainty factor $\Delta \mathbf{A}$ decomposition (\ref{eq:DA_decompose_rank_one})
\begin{align*}
\Delta \mathbf{A} = \sum_{k=1}^p a_k (t) \mathbf{a}_k \mathbf{x}_k^\mathrm{T} \equiv \sum_{k=1}^p a_k \mathbf{a}_k \mathbf{x}_k^\mathrm{T}
\end{align*}
accordingly to the maximum extent, only if the bounding condition (\ref{eq:DA_sum_rank_one_bound})
\begin{align*}
- \mathbf{\Sigma}_{\Delta \mathbf{A}} \preceq \Delta \mathbf{A} \preceq \mathbf{\Sigma}_{\Delta \mathbf{A}}
\end{align*}
is not violated. Similarly, increasing $\bar{b}$ means not only increasing $\bar{b}$ but also shrinking $\mathbf{b}_k$ in the uncertainty factor $\Delta \mathbf{B}$ decomposition (\ref{eq:DB_decompose_rank_one})
\begin{align*}
\Delta \mathbf{B} = \sum_{k=1}^q b_k (t) \mathbf{b}_k \mathbf{y}_k^\mathrm{T} \equiv \sum_{k=1}^q b_k \mathbf{b}_k \mathbf{y}_k^\mathrm{T}
\end{align*}
accordingly to the maximum extent, only if the bounding condition (\ref{eq:DB_sum_rank_one_bound})
\begin{align*}
- \mathbf{\Sigma}_{\Delta \mathbf{B}} \preceq \Delta \mathbf{B} \preceq \mathbf{\Sigma}_{\Delta \mathbf{B}}
\end{align*}
is not violated.

However, if the Riccati equation (\ref{eq:Petersen_YangY_Riccati}) still has no positive definite solution $\mathbf{P}$ even when $\bar{a}$ and $\bar{b}$ are increased to very large values and $\epsilon$ is decreased to a very small value, then we may empirically venture a conclusion that no solution to the problem formalized in (\ref{eq:robust_gain_matrix_design}) exists and no corresponding robust control law exists either. 

\subsection{Interval polynomial method}  \label{sec:robust_control_interval_polynomial}

In 1978, \textit{Kharitonov} published a milestone article on interval polynomial stability analysis \cite{Kharitonov1978}. Define the \textbf{interval polynomial set} or for short the \textbf{interval polynomial}
\begin{equation}  \label{eq:interval_polynomial}
f_{[\underline{\mathbf{a}}, \overline{\mathbf{a}}]} (s) \equiv \{ f(s) = a_n s^n + a_{n-1} s^{n-1} + \cdots + a_1 s + a_0 \mbox{ } | \mbox{ } \underline{\mathbf{a}} \preceq \mathbf{a} \preceq \overline{\mathbf{a}} \}
\end{equation}
where
\begin{align*}
\mathbf{a} &\equiv \begin{bmatrix} a_n & a_{n-1} & \cdots & a_1 & a_0 \end{bmatrix}^\mathrm{T},  \\
\underline{\mathbf{a}} &\equiv \begin{bmatrix} \underline{a_n} & \underline{a_{n-1}} & \cdots & \underline{a_1} & \underline{a_0} \end{bmatrix}^\mathrm{T}, \\
\overline{\mathbf{a}} &\equiv \begin{bmatrix} \overline{a_n} & \overline{a_{n-1}} & \cdots & \overline{a_1} & \overline{a_0} \end{bmatrix}^\mathrm{T}.
\end{align*}

Consider four special polynomials in the interval polynomial $f_{[\underline{\mathbf{a}}, \overline{\mathbf{a}}]} (s)$ namely
\begin{subequations}  \label{eq:Kharitonov_bounding_polynomial}
\begin{align}
f_{K_1 [\underline{\mathbf{a}}, \overline{\mathbf{a}}]} (s) \equiv \underline{a_0} + \underline{a_1} s + \overline{a_2} s^2 + \overline{a_3} s^3 + \underline{a_4} s^4 + \underline{a_5} s^5 + \overline{a_6} s^6 + \overline{a_7} s^7 + \cdots  \\
f_{K_2 [\underline{\mathbf{a}}, \overline{\mathbf{a}}]} (s) \equiv \underline{a_0} + \overline{a_1} s + \overline{a_2} s^2 + \underline{a_3} s^3 + \underline{a_4} s^4 + \overline{a_5} s^5 + \overline{a_6} s^6 + \underline{a_7} s^7 + \cdots  \\
f_{K_3 [\underline{\mathbf{a}}, \overline{\mathbf{a}}]} (s) \equiv \overline{a_0} + \underline{a_1} s + \underline{a_2} s^2 + \overline{a_3} s^3 + \overline{a_4} s^4 + \underline{a_5} s^5 + \underline{a_6} s^6 + \overline{a_7} s^7 + \cdots  \\
f_{K_4 [\underline{\mathbf{a}}, \overline{\mathbf{a}}]} (s) \equiv \overline{a_0} + \overline{a_1} s + \underline{a_2} s^2 + \underline{a_3} s^3 + \overline{a_4} s^4 + \overline{a_5} s^5 + \underline{a_6} s^6 + \underline{a_7} s^7 + \cdots
\end{align}
\end{subequations}
which are called \textbf{Kharitonov bounding polynomials} or \textbf{Kharitonov polynomials}.

\textit{Kharitonov} proves that \textit{the interval polynomial $f_{[\underline{\mathbf{a}}, \overline{\mathbf{a}}]} (s)$ defined in (\ref{eq:interval_polynomial}) is stable (namely the polynomials of the interval polynomial $f_{[\underline{\mathbf{a}}, \overline{\mathbf{a}}]} (s)$ are all stable) if and only if the four Kharitonov bounding polynomials 
\begin{align*}
f_{K_1 [\underline{\mathbf{a}}, \overline{\mathbf{a}}]} (s), \quad f_{K_2 [\underline{\mathbf{a}}, \overline{\mathbf{a}}]} (s), \quad f_{K_3 [\underline{\mathbf{a}}, \overline{\mathbf{a}}]} (s), \quad f_{K_4 [\underline{\mathbf{a}}, \overline{\mathbf{a}}]} (s) 
\end{align*}
defined in (\ref{eq:Kharitonov_bounding_polynomial}) are stable.}

\begin{framed} 
\noindent \textbf{Kharitonov stability criterion}: \textit{the interval polynomial $f_{[\underline{\mathbf{a}}, \overline{\mathbf{a}}]} (s)$ is stable if and only if the four Kharitonov bounding polynomials $f_{K_1 [\underline{\mathbf{a}}, \overline{\mathbf{a}}]} (s)$, $f_{K_2 [\underline{\mathbf{a}}, \overline{\mathbf{a}}]} (s)$, $f_{K_3 [\underline{\mathbf{a}}, \overline{\mathbf{a}}]} (s)$, and $f_{K_4 [\underline{\mathbf{a}}, \overline{\mathbf{a}}]} (s)$ are stable.}
\end{framed}

\subsubsection*{Symbolic operation based method}

Similar to the symbolic operation based method presented in Section 2.2.1 in Chapter 2, 
\footnote{Namely Chapter 2 of the author's works \cite{Li2026ACTPA_SJTU_2, Li2026ACTPA_SJTU_1}.}
for control problems of small scales, we may take advantage of symbolic operation as well for designing of a robust gain matrix.

Consider the generic formalism of linear state-space modelling with uncertainty described in (\ref{eq:uncertain_state_DE_linear2})
\begin{align*}
\frac{\mathrm{d}}{\mathrm{d} t} \mathbf{x} = (\mathbf{A} + \Delta \mathbf{A}) \mathbf{x} + (\mathbf{B} + \Delta \mathbf{B}) \mathbf{u},
\end{align*}
where the uncertainty factors $\Delta \mathbf{A}$ and $\Delta \mathbf{B}$ are bounded between their respective element-wise minimum value matrix and element-wise maximum value matrix, as described in (\ref{eq:A+DA_B+DB_interval_matrix1})
\begin{align*}
&\underline{\Delta \mathbf{A}} \preceq \Delta \mathbf{A} \preceq \overline{\Delta \mathbf{A}},  \\
&\underline{\Delta \mathbf{B}} \preceq \Delta \mathbf{B} \preceq \overline{\Delta \mathbf{B}}.
\end{align*}
Compute the closed-loop characteristic polynomial 
\begin{align*}
\det (s \mathbf{I} - (\mathbf{A}^* - \mathbf{B}^* \mathbf{K}^\mathrm{T}))
\end{align*}
via symbolic operation and express it as a characteristic polynomial with parametrized coefficients in terms of $\mathbf{A}^*$, $\mathbf{B}^*$, and the gain matrix
\begin{align*}
\mathbf{K} \equiv \begin{bmatrix} k_1 & k_2 & \cdots & k_n \end{bmatrix}^\mathrm{T}, 
\end{align*}
where
\begin{equation}  \label{eq:uncertain_A+B_interval}
\mathbf{A}^* \in [\mathbf{A} + \underline{\Delta \mathbf{A}}, \mathbf{A} + \overline{\Delta \mathbf{A}}], \qquad \mathbf{B}^* \in [\mathbf{B} + \underline{\Delta \mathbf{B}}, \mathbf{B} + \overline{\Delta \mathbf{B}}].
\end{equation}
Denote the parametrized characteristic polynomial
\begin{align}  \label{eq:GM_CP_parametrized_K+A+B}
\det (s \mathbf{I} - (\mathbf{A}^* - \mathbf{B}^* \mathbf{K}^\mathrm{T})) = s^n + a_{n-1}(\mathbf{A}^*, \mathbf{B}^*, \mathbf{K}) s^{n-1} + \cdots + a_0(\mathbf{A}^*, \mathbf{B}^*, \mathbf{K}).
\end{align}

As explained in Section 2.2.1 in Chapter 2, each parametrized coefficient $a_k(\mathbf{A}^*, \mathbf{B}^*, \mathbf{K})$ ($k \in \{0, \cdots, n-1\}$) in (\ref{eq:GM_CP_parametrized_K+A+B}) is linear in terms of the gain matrix $\mathbf{K}$. So we can express each parametrized coefficient $a_k(\mathbf{A}^*, \mathbf{B}^*, \mathbf{K})$ as
\begin{equation}  \label{eq:GM_CP_parametrized_K_linear}
a_k(\mathbf{A}^*, \mathbf{B}^*, \mathbf{K}) = \mathbf{a}_k(\mathbf{A}^*, \mathbf{B}^*)^\mathrm{T} \begin{bmatrix} \mathbf{K} \\ 1 \end{bmatrix} \equiv a_{k,0}(\mathbf{A}^*, \mathbf{B}^*) + \sum_{i=1}^n a_{k,i}(\mathbf{A}^*, \mathbf{B}^*) k_i,
\end{equation}
where
\begin{align*}
\mathbf{a}_k(\mathbf{A}^*, \mathbf{B}^*) \equiv \begin{bmatrix} a_{k,0}(\mathbf{A}^*, \mathbf{B}^*) \\ a_{k,1}(\mathbf{A}^*, \mathbf{B}^*) \\ \vdots \\ a_{k,n}(\mathbf{A}^*, \mathbf{B}^*) \end{bmatrix}.
\end{align*}
Since $\mathbf{A}^*$ and $\mathbf{B}^*$ are bounded as specified in (\ref{eq:uncertain_A+B_interval}), each $a_{k,i}(\mathbf{A}^*, \mathbf{B}^*)$ ($i \in \{0, 1, \cdots, n\}$) in (\ref{eq:GM_CP_parametrized_K_linear}), which is a polynomial in terms of $\mathbf{A}^*$ and $\mathbf{B}^*$ elements, must also be bounded. In other words, each parametrized vector $\mathbf{a}_k(\mathbf{A}^*, \mathbf{B}^*)$ is also bounded, i.e.
\begin{equation}  \label{eq:GM_ak_A+B_bound}
\underline{\mathbf{a}_k(\mathbf{A}^*, \mathbf{B}^*)} \preceq \mathbf{a}_k(\mathbf{A}^*, \mathbf{B}^*) \preceq \overline{\mathbf{a}_k(\mathbf{A}^*, \mathbf{B}^*)}.
\end{equation} 
(\ref{eq:GM_CP_parametrized_K_linear}) and (\ref{eq:GM_ak_A+B_bound}) imply that each parametrized coefficient $a_k(\mathbf{A}^*, \mathbf{B}^*, \mathbf{K})$ is also bounded, i.e.
\begin{equation}  \label{eq:GM_ak_A+B+K_bound}
\underline{\mathbf{a}(\mathbf{A}^*, \mathbf{B}^*, \mathbf{K})} \preceq \mathbf{a}(\mathbf{A}^*, \mathbf{B}^*, \mathbf{K}) \preceq \overline{\mathbf{a}(\mathbf{A}^*, \mathbf{B}^*, \mathbf{K})},
\end{equation}
where
\begin{align*}
\mathbf{a}(\mathbf{A}^*, \mathbf{B}^*, \mathbf{K}) \equiv \begin{bmatrix} a_0(\mathbf{A}^*, \mathbf{B}^*, \mathbf{K}) \\ a_1(\mathbf{A}^*, \mathbf{B}^*, \mathbf{K}) \\ \vdots \\ a_{n-1}(\mathbf{A}^*, \mathbf{B}^*, \mathbf{K}) \end{bmatrix} = \begin{bmatrix} \mathbf{a}_0(\mathbf{A}^*, \mathbf{B}^*)^\mathrm{T} \\ \mathbf{a}_1(\mathbf{A}^*, \mathbf{B}^*)^\mathrm{T} \\ \vdots \\ \mathbf{a}_{n-1}(\mathbf{A}^*, \mathbf{B}^*)^\mathrm{T} \end{bmatrix} \begin{bmatrix} \mathbf{K} \\ 1 \end{bmatrix}.
\end{align*}

So the parametrized characteristic polynomial described in (\ref{eq:GM_CP_parametrized_K+A+B}) is an parametrized interval polynomial 
\begin{align*}
\det (s \mathbf{I} - (\mathbf{A}^* - \mathbf{B}^* \mathbf{K}^\mathrm{T})) |_{\mathbf{A}^* \in [\mathbf{A} + \underline{\Delta \mathbf{A}}, \mathbf{A} + \overline{\Delta \mathbf{A}}], \mathbf{B}^* \in [\mathbf{B} + \underline{\Delta \mathbf{B}}, \mathbf{B} + \overline{\Delta \mathbf{B}}]} = f_{[\underline{\mathbf{a}(\mathbf{A}^*, \mathbf{B}^*, \mathbf{K})}, \overline{\mathbf{a}(\mathbf{A}^*, \mathbf{B}^*, \mathbf{K})}]} (s)
\end{align*}
in terms of the gain matrix $\mathbf{K}$. According to the \textit{Kharitonov stability criterion}, for the parametrized interval polynomial to be stable, we only need to consider stability of the four Kharitonov bounding polynomials which are among the finite number of all bounding polynomials of the parametrized interval polynomial. Since each parametrized coefficient $a_k(\mathbf{A}^*, \mathbf{B}^*, \mathbf{K})$ is linear in terms of the gain matrix $\mathbf{K}$ as formalized in (\ref{eq:GM_CP_parametrized_K_linear}), each bounding polynomial of the parametrized interval polynomial must be obtained with each $a_{k,i}(\mathbf{A}^*, \mathbf{B}^*)$ being certain one of the two corresponding bounds described in (\ref{eq:GM_ak_A+B_bound}).

We can enumerate all relevant bounding polynomials all with their coefficients parametrized in terms of the gain matrix $\mathbf{K}$. Apply the \textit{Routh-Hurwitz criterion} method to each bounding polynomial and establish a group of inequalities in terms of $\mathbf{K}$. Solve the group of inequalities to find a solution of $\mathbf{K}$ which then is a robust gain matrix.

\subsubsection*{Application: single inverted pendulum robust control}

To clarify the symbolic operation based interval polynomial method that takes advantage of the \textit{Kharitonov stability criterion}, still consider single inverted pendulum control as presented in Section \ref{sec:robust_control_Riccati_equation_method}. Extract the sub-model associated with the partial state
\begin{align*}
\mathbf{x}_P \equiv \begin{bmatrix} \theta & \frac{\mathrm{d} \theta}{\mathrm{d} t} & \frac{\mathrm{d} x}{\mathrm{d} t} \end{bmatrix}^\mathrm{T}
\end{align*}
from (\ref{eq:uncertain_SIP_state_DE_linear}) and obtain (\ref{eq:uncertain_SIP_state_DE_linear_partial})
\begin{align*}
\frac{\mathrm{d}}{\mathrm{d} t} \mathbf{x}_P = \mathbf{A}_P(\mathbf{x}_P) \mathbf{x}_P + \mathbf{B}_P(\mathbf{x}_P) a,
\end{align*}
where
\begin{align*}
\mathbf{A}_P(\mathbf{x}_P) \equiv \begin{bmatrix} 0 & 1 & 0 \\ \frac{g}{L} \frac{\sin \theta}{\theta} & 0 & 0 \\ 0 & 0 & 0 \end{bmatrix}, \quad \mathbf{B}_P(\mathbf{x}_P) \equiv \begin{bmatrix} 0 \\ -\frac{\cos \theta}{L} \\ 1 \end{bmatrix}.
\end{align*}

Suppose the inverted pendulum angle $\theta$ is in the operation range (\ref{eq:SIP_theta_operation_range})
\begin{align*}
- \theta_{\max} \leq \theta \leq \theta_{\max}.
\end{align*}
The time-variant state transition matrix $\mathbf{A}_P(\mathbf{x}_P)$ and control input matrix $\mathbf{B}_P(\mathbf{x}_P)$ in (\ref{eq:uncertain_SIP_state_DE_linear_partial}) can also be expressed concisely as
\begin{equation}  \label{eq:SIP_time_variant_A+B_concise}
\mathbf{A}_P(\mathbf{x}_P) \equiv \mathbf{A}^* \equiv \begin{bmatrix} 0 & 1 & 0 \\ a & 0 & 0 \\ 0 & 0 & 0 \end{bmatrix}, \quad \mathbf{B}_P(\mathbf{x}_P) \equiv \mathbf{B}^* \equiv \begin{bmatrix} 0 \\ -b \\ 1 \end{bmatrix},
\end{equation}
where
\begin{subequations}  \label{eq:SIP_time_variant_a+b_bound}
\begin{align}
\underline{a} \equiv \frac{g}{L} \frac{\sin \theta_{\max}}{\theta_{\max}} \leq a &\leq \overline{a} \equiv \frac{g}{L},  \\
\underline{b} \equiv \frac{\cos \theta_{\max}}{L} \leq b &\leq \overline{b} \equiv \frac{1}{L},
\end{align}
\end{subequations}
and (\ref{eq:uncertain_SIP_state_DE_linear_partial}) can be expressed accordingly as
\begin{align*}
\frac{\mathrm{d}}{\mathrm{d} t} \mathbf{x}_P = \mathbf{A}^* \mathbf{x}_P + \mathbf{B}^* a.
\end{align*}
For concrete configuration of parameters, let 
\begin{align*}  
L = 1, \quad g = 10, \quad \theta_{\max} = 0.4 \pi, 
\end{align*}
then we have
\begin{align*}
\underline{a} = 7.57, \quad \overline{a} = 10, \quad \underline{b} = 0.31, \quad \overline{b} = 1
\end{align*}
and
\begin{align*}
\mathbf{A} + \underline{\Delta \mathbf{A}} \equiv \begin{bmatrix} 0 & 1 & 0 \\ 7.57 & 0 & 0 \\ 0 & 0 & 0 \end{bmatrix} \preceq \mathbf{A}^* &\preceq \mathbf{A} + \overline{\Delta \mathbf{A}} \equiv \begin{bmatrix} 0 & 1 & 0 \\ 10 & 0 & 0 \\ 0 & 0 & 0 \end{bmatrix},  \\
\mathbf{B} + \underline{\Delta \mathbf{B}} \equiv \begin{bmatrix} 0 \\ -1 \\ 1 \end{bmatrix} \preceq \mathbf{B}^* &\preceq \mathbf{B} + \overline{\Delta \mathbf{B}} \equiv \begin{bmatrix} 0 \\ -0.31 \\ 1 \end{bmatrix}.
\end{align*}

Denote the gain matrix
\begin{align*}
\mathbf{K}_P \equiv \begin{bmatrix} k_1 & k_2 & k_3 \end{bmatrix}^\mathrm{T} 
\end{align*}
and compute the parametrized characteristic polynomial via symbolic operation
\footnote{Of course we may also do the computation manually, which just incurs more time.}
\begin{align}  \label{eq:SIP_GM_CP_parametrized_K+A+B}
\det (s \mathbf{I} - (\mathbf{A}^* - \mathbf{B}^* \mathbf{K}_P^\mathrm{T})) = s^3 + a_2(\mathbf{A}^*, \mathbf{B}^*, \mathbf{K}_P) s^2 + a_1(\mathbf{A}^*, \mathbf{B}^*, \mathbf{K}_P) s + a_0(\mathbf{A}^*, \mathbf{B}^*, \mathbf{K}_P),
\end{align}
where
\begin{align*}
a_2 \equiv a_2(\mathbf{A}^*, \mathbf{B}^*, \mathbf{K}_P) &= - b k_2 + k3,  \\
a_1 \equiv a_1(\mathbf{A}^*, \mathbf{B}^*, \mathbf{K}_P) &= - a - b k_1,  \\
a_0 \equiv a_0(\mathbf{A}^*, \mathbf{B}^*, \mathbf{K}_P) &= - a k_3.
\end{align*}
The bounds of the parametrized coefficients can be obtained when $a$ and $b$ are at their respective bounds, namely
\begin{align*}
&\underline{a_2(\mathbf{A}^*, \mathbf{B}^*, \mathbf{K}_P)} = - b k_2 + k3 |_{\exists b \in \{\underline{b}, \overline{b}\}}, \quad 
&\overline{a_2(\mathbf{A}^*, \mathbf{B}^*, \mathbf{K}_P)} = - b k_2 + k3 |_{\exists b \in \{\underline{b}, \overline{b}\}},  \\
&\underline{a_1(\mathbf{A}^*, \mathbf{B}^*, \mathbf{K}_P)} = - a - b k_1 |_{\exists a \in \{\underline{a}, \overline{a}\}, b \in \{\underline{b}, \overline{b}\}}, \quad 
&\overline{a_1(\mathbf{A}^*, \mathbf{B}^*, \mathbf{K}_P)} = - a - b k_1 |_{\exists a \in \{\underline{a}, \overline{a}\}, b \in \{\underline{b}, \overline{b}\}},  \\
&\underline{a_0(\mathbf{A}^*, \mathbf{B}^*, \mathbf{K}_P)} = - a k_3 |_{\exists a \in \{\underline{a}, \overline{a}\}}, \quad 
&\overline{a_0(\mathbf{A}^*, \mathbf{B}^*, \mathbf{K}_P)} = - a k_3 |_{\exists a \in \{\underline{a}, \overline{a}\}}.
\end{align*}

Establish the Routh array for the parametrized characteristic polynomial
\begin{align*}
\begin{matrix}
s^3 & | & 1 & a_1  \\
s^2 & | & a_2 = - b k_2 + k3 & a_0  \\
s^1 & | & a_1 - \frac{a_0}{a_2} = - b k_1 + \frac{a b k_2}{- b k_2 + k3} &  \\
s^0 & | & a_0 = - a k_3 & 
\end{matrix}
\end{align*}
elements in the first column of which always need to be positive. Consider only bounds of $a$ and $b$ and establish a group of inequalities in terms of the gain matrix elements
\begin{subequations}  \label{eq:SIP_interval_polynomial_ineq_group}
\begin{align}
- \underline{a} k_3 &> 0,  \\
- \overline{a} k_3 &> 0,  \\
- \underline{b} k_2 + k3 &> 0,  \\
- \overline{b} k_2 + k3 &> 0,  \\
- \underline{b} k_1 + \frac{\underline{a} \underline{b} k_2}{- \underline{b} k_2 + k3} &> 0,  \\
- \underline{b} k_1 + \frac{\overline{a} \underline{b} k_2}{- \underline{b} k_2 + k3} &> 0,  \\
- \overline{b} k_1 + \frac{\underline{a} \overline{b} k_2}{- \overline{b} k_2 + k3} &> 0,  \\
- \overline{b} k_1 + \frac{\overline{a} \overline{b} k_2}{- \overline{b} k_2 + k3} &> 0.
\end{align}
\end{subequations}
The first and second inequalities of (\ref{eq:SIP_interval_polynomial_ineq_group}) leads to
\begin{equation}  \label{eq:SIP_IP_ineq_k3}
k_3 < 0.
\end{equation}
Following (\ref{eq:SIP_IP_ineq_k3}), the third and fourth inequalities of (\ref{eq:SIP_interval_polynomial_ineq_group}) leads to
\begin{equation}  \label{eq:SIP_IP_ineq_k2}
k_2 < \frac{1}{\underline{b}} k_3.
\end{equation}
Substitute (\ref{eq:SIP_IP_ineq_k3}) and (\ref{eq:SIP_IP_ineq_k2}) into the fifth to eighth inequalities of (\ref{eq:SIP_interval_polynomial_ineq_group}) and obtain
\begin{equation}  \label{eq:SIP_IP_ineq_k1}
k_1 < \frac{\overline{a}}{- \underline{b} k_2 + k_3} k_2.
\end{equation}

So for the concrete configuration of parameters, we have
\begin{subequations}  \label{eq:SIP_IP_concrete_ineq_K}
\begin{align}
k_3 &< 0,  \\
k_2 &< \frac{1}{0.31} k_3 = 3.23 k_3,  \\
k_1 &< \frac{10}{- 0.31 k_2 + k_3} k_2.
\end{align}
\end{subequations}
Any gain matrix
\begin{align*}
\mathbf{K}_P \equiv \begin{bmatrix} k_1 & k_2 & k_3 \end{bmatrix}^\mathrm{T}
\end{align*}
that satisfies (\ref{eq:SIP_IP_concrete_ineq_K}) is a robust gain matrix for single inverted pendulum control given the concrete configuration of parameters.

For example, it can be verified that the already known robust gain matrix described in (\ref{eq:SIP_robust_gain_matrix}) namely
\begin{align*}
\mathbf{K}_P \equiv \begin{bmatrix} k_1 & k_2 & k_3 \end{bmatrix}^\mathrm{T} = \begin{bmatrix} -170.2 & -54.7 & -10.6 \end{bmatrix}^\mathrm{T}
\end{align*}
indeed satisfies (\ref{eq:SIP_IP_concrete_ineq_K})
\begin{align*}
-10.6 &< 0,  \\
-54.7 &< 3.23 \cdot (-10.6) = -34.24,  \\
-170.2 &< \frac{10}{- 0.31 \cdot (-54.7) + (-10.6)} \cdot (-54.7) = -86.05.
\end{align*}
For another example, it can also be verified that the gain matrix
\begin{equation}  \label{eq:SIP_robust_gain_matrix2}
\mathbf{K}_P \equiv \begin{bmatrix} k_1 & k_2 & k_3 \end{bmatrix}^\mathrm{T} = \begin{bmatrix} -110 & -50 & -10 \end{bmatrix}^\mathrm{T}
\end{equation}
satisfies (\ref{eq:SIP_IP_concrete_ineq_K}) and hence is also a robust gain matrix. 

To verify robustness of the gain matrix described in (\ref{eq:SIP_robust_gain_matrix2}), traverse inverted pendulum angles densely in the $\theta$ operation range described in (\ref{eq:SIP_theta_operation_range}), compute and list eigenvalue real parts of their associated closed-loop state transition matrices namely
\begin{align*}
\{ \mbox{Re}(\mbox{Eig}(\mathbf{A}_P(\mathbf{x}_P) - \mathbf{B}_P(\mathbf{x}_P) \mathbf{K}_P^\mathrm{T})) \mbox{ } | \mbox{ } - \theta_{\max} \leq \theta \leq \theta_{\max} \}
\end{align*}
in Table \ref{tab:SIP_eig_re_traversed2}, like in Table \ref{tab:SIP_eig_re_traversed}. As we can see in Table \ref{tab:SIP_eig_re_traversed2}, all the eigenvalues have negative real parts, which implies that all the closed-loop state transition matrices are stable.

\begin{longtable}{|c c c c|}
\caption{Re. Eig. associated with densely traversed $\theta$\label{tab:SIP_eig_re_traversed2}} \\
\hline
$\theta$ (degree) & 1st Re. Eig. & 2nd Re. Eig. & 3rd Re. Eig. \\ \hline
-72 & -0.8412 & -0.8412 & -3.7684 \\
-71 & -1.1387 & -1.1387 & -4.0010 \\
-70 & -1.4004 & -1.4004 & -4.3001 \\
-69 & -1.6140 & -1.6140 & -4.6904 \\
-68 & -5.1971 & -1.7666 & -1.7666 \\
-67 & -5.8306 & -1.8530 & -1.8530 \\
-66 & -6.5719 & -1.8825 & -1.8825 \\
-65 & -7.3815 & -1.8747 & -1.8747 \\
-64 & -8.2229 & -1.8478 & -1.8478 \\
-63 & -9.0729 & -1.8133 & -1.8133 \\
-62 & -9.9192 & -1.7772 & -1.7772 \\
-61 & -10.7563 & -1.7421 & -1.7421 \\
-60 & -11.5817 & -1.7091 & -1.7091 \\
-59 & -12.3944 & -1.6787 & -1.6787 \\
-58 & -13.1942 & -1.6509 & -1.6509 \\
-57 & -13.9811 & -1.6254 & -1.6254 \\
-56 & -14.7554 & -1.6021 & -1.6021 \\
-55 & -15.5171 & -1.5809 & -1.5809 \\
-54 & -16.2666 & -1.5613 & -1.5613 \\
-53 & -17.0039 & -1.5434 & -1.5434 \\
-52 & -17.7292 & -1.5269 & -1.5269 \\
-51 & -18.4426 & -1.5117 & -1.5117 \\
-50 & -19.1440 & -1.4977 & -1.4977 \\
-49 & -19.8336 & -1.4847 & -1.4847 \\
-48 & -20.5114 & -1.4726 & -1.4726 \\
-47 & -21.1772 & -1.4613 & -1.4613 \\
-46 & -21.8312 & -1.4509 & -1.4509 \\
-45 & -22.4731 & -1.4411 & -1.4411 \\
-44 & -23.1030 & -1.4320 & -1.4320 \\
-43 & -23.7208 & -1.4234 & -1.4234 \\
-42 & -24.3263 & -1.4154 & -1.4154 \\
-41 & -24.9196 & -1.4079 & -1.4079 \\
-40 & -25.5004 & -1.4009 & -1.4009 \\
-39 & -26.0687 & -1.3943 & -1.3943 \\
-38 & -26.6245 & -1.3880 & -1.3880 \\
-37 & -27.1674 & -1.3822 & -1.3822 \\
-36 & -27.6976 & -1.3766 & -1.3766 \\
-35 & -28.2147 & -1.3714 & -1.3714 \\
-34 & -28.7188 & -1.3665 & -1.3665 \\
-33 & -29.2098 & -1.3619 & -1.3619 \\
-32 & -29.6874 & -1.3575 & -1.3575 \\
-31 & -30.1516 & -1.3534 & -1.3534 \\
-30 & -30.6024 & -1.3495 & -1.3495 \\
-29 & -31.0394 & -1.3458 & -1.3458 \\
-28 & -31.4628 & -1.3423 & -1.3423 \\
-27 & -31.8723 & -1.3390 & -1.3390 \\
-26 & -32.2679 & -1.3359 & -1.3359 \\
-25 & -32.6494 & -1.3330 & -1.3330 \\
-24 & -33.0168 & -1.3302 & -1.3302 \\
-23 & -33.3700 & -1.3276 & -1.3276 \\
-22 & -33.7089 & -1.3252 & -1.3252 \\
-21 & -34.0333 & -1.3229 & -1.3229 \\
-20 & -34.3432 & -1.3207 & -1.3207 \\
-19 & -34.6385 & -1.3187 & -1.3187 \\
-18 & -34.9192 & -1.3168 & -1.3168 \\
-17 & -35.1852 & -1.3150 & -1.3150 \\
-16 & -35.4363 & -1.3134 & -1.3134 \\
-15 & -35.6726 & -1.3119 & -1.3119 \\
-14 & -35.8939 & -1.3104 & -1.3104 \\
-13 & -36.1002 & -1.3091 & -1.3091 \\
-12 & -36.2915 & -1.3079 & -1.3079 \\
-11 & -36.4677 & -1.3068 & -1.3068 \\
-10 & -36.6287 & -1.3059 & -1.3059 \\
-9 & -36.7745 & -1.3050 & -1.3050 \\
-8 & -36.9051 & -1.3042 & -1.3042 \\
-7 & -37.0204 & -1.3035 & -1.3035 \\
-6 & -37.1203 & -1.3029 & -1.3029 \\
-5 & -37.2050 & -1.3024 & -1.3024 \\
-4 & -37.2743 & -1.3020 & -1.3020 \\
-3 & -37.3282 & -1.3016 & -1.3016 \\
-2 & -37.3667 & -1.3014 & -1.3014 \\
-1 & -37.3898 & -1.3013 & -1.3013 \\
0 & -37.3975 & -1.3012 & -1.3012 \\
1 & -37.3898 & -1.3013 & -1.3013 \\
2 & -37.3667 & -1.3014 & -1.3014 \\
3 & -37.3282 & -1.3016 & -1.3016 \\
4 & -37.2743 & -1.3020 & -1.3020 \\
5 & -37.2050 & -1.3024 & -1.3024 \\
6 & -37.1203 & -1.3029 & -1.3029 \\
7 & -37.0204 & -1.3035 & -1.3035 \\
8 & -36.9051 & -1.3042 & -1.3042 \\
9 & -36.7745 & -1.3050 & -1.3050 \\
10 & -36.6287 & -1.3059 & -1.3059 \\
11 & -36.4677 & -1.3068 & -1.3068 \\
12 & -36.2915 & -1.3079 & -1.3079 \\
13 & -36.1002 & -1.3091 & -1.3091 \\
14 & -35.8939 & -1.3104 & -1.3104 \\
15 & -35.6726 & -1.3119 & -1.3119 \\
16 & -35.4363 & -1.3134 & -1.3134 \\
17 & -35.1852 & -1.3150 & -1.3150 \\
18 & -34.9192 & -1.3168 & -1.3168 \\
19 & -34.6385 & -1.3187 & -1.3187 \\
20 & -34.3432 & -1.3207 & -1.3207 \\
21 & -34.0333 & -1.3229 & -1.3229 \\
22 & -33.7089 & -1.3252 & -1.3252 \\
23 & -33.3700 & -1.3276 & -1.3276 \\
24 & -33.0168 & -1.3302 & -1.3302 \\
25 & -32.6494 & -1.3330 & -1.3330 \\
26 & -32.2679 & -1.3359 & -1.3359 \\
27 & -31.8723 & -1.3390 & -1.3390 \\
28 & -31.4628 & -1.3423 & -1.3423 \\
29 & -31.0394 & -1.3458 & -1.3458 \\
30 & -30.6024 & -1.3495 & -1.3495 \\
31 & -30.1516 & -1.3534 & -1.3534 \\
32 & -29.6874 & -1.3575 & -1.3575 \\
33 & -29.2098 & -1.3619 & -1.3619 \\
34 & -28.7188 & -1.3665 & -1.3665 \\
35 & -28.2147 & -1.3714 & -1.3714 \\
36 & -27.6976 & -1.3766 & -1.3766 \\
37 & -27.1674 & -1.3822 & -1.3822 \\
38 & -26.6245 & -1.3880 & -1.3880 \\
39 & -26.0687 & -1.3943 & -1.3943 \\
40 & -25.5004 & -1.4009 & -1.4009 \\
41 & -24.9196 & -1.4079 & -1.4079 \\
42 & -24.3263 & -1.4154 & -1.4154 \\
43 & -23.7208 & -1.4234 & -1.4234 \\
44 & -23.1030 & -1.4320 & -1.4320 \\
45 & -22.4731 & -1.4411 & -1.4411 \\
46 & -21.8312 & -1.4509 & -1.4509 \\
47 & -21.1772 & -1.4613 & -1.4613 \\
48 & -20.5114 & -1.4726 & -1.4726 \\
49 & -19.8336 & -1.4847 & -1.4847 \\
50 & -19.1440 & -1.4977 & -1.4977 \\
51 & -18.4426 & -1.5117 & -1.5117 \\
52 & -17.7292 & -1.5269 & -1.5269 \\
53 & -17.0039 & -1.5434 & -1.5434 \\
54 & -16.2666 & -1.5613 & -1.5613 \\
55 & -15.5171 & -1.5809 & -1.5809 \\
56 & -14.7554 & -1.6021 & -1.6021 \\
57 & -13.9811 & -1.6254 & -1.6254 \\
58 & -13.1942 & -1.6509 & -1.6509 \\
59 & -12.3944 & -1.6787 & -1.6787 \\
60 & -11.5817 & -1.7091 & -1.7091 \\
61 & -10.7563 & -1.7421 & -1.7421 \\
62 & -9.9192 & -1.7772 & -1.7772 \\
63 & -9.0729 & -1.8133 & -1.8133 \\
64 & -8.2229 & -1.8478 & -1.8478 \\
65 & -7.3815 & -1.8747 & -1.8747 \\
66 & -6.5719 & -1.8825 & -1.8825 \\
67 & -5.8306 & -1.8530 & -1.8530 \\
68 & -5.1971 & -1.7666 & -1.7666 \\
69 & -1.6140 & -1.6140 & -4.6904 \\
70 & -1.4004 & -1.4004 & -4.3001 \\
71 & -1.1387 & -1.1387 & -4.0010 \\
72 & -0.8412 & -0.8412 & -3.7684 \\ \hline
\end{longtable}

Matlab simulation code for complete demonstration of single inverted pendulum control using the robust gain matrix described in (\ref{eq:SIP_robust_gain_matrix2}) is given as follows. The visualization code \textbf{DisplaySIP.m} and the single inverted pendulum dynamics code \textbf{DynamicsSIP.m} are given in Section 2.2.3 in Chapter 2. The gain matrix designing code \textbf{DesignGainMatrix.m} is given in Section 2.3.2 in Chapter 2.

\begin{framed} 
\noindent \textbf{SingleInvertedPendulumIntervalPolynomial.m} \\
\noindent \%\% Single inverted pendulum parameters \\
m1 = 1; L1 = 1; g = 10; \\
\%\% Simulation preliminary configuration \\
dt = 0.001; \% Numerical computation step \\
tSpan = 0:dt:20; \% Simulation time span \\
x = 0.2; \% Cart position  \\
dx = 0; \% Cart velocity \\
y = 0.4*pi; \% Inverted pendulum angle theta  \\
dy = 0;  \% Inverted pendulum angular velocity \\
stt = [y; dy; x; dx]; \% Single inverted pendulum state \\
sttAll = zeros(length(stt), length(tSpan)); k = 0; \% Record states \\
xExpected = 0; yExpected = 0; \% Expected equilibrium status \\
SimConfig = [m1, L1, g, dt]; \\
 \\
\%\% Design robust gain matrix \\
A = [0, 1, 0, 0; g/L1, 0, 0, 0; 0, 0, 0, 1; 0, 0, 0, 0]; \\
B = [0; -1/L1; 0; 1]; \\
sttK = DesignGainMatrix(A, B, [-4;-4+2i;-4-2i;-4]); \% For SMC \\
fprintf('SMC gain matrix K: '); disp(sttK'); \\
syms a b k1 k2 k3 s \\
Ap = [0, 1, 0; a, 0, 0; 0, 0, 0]; Bp = [0; -b; 1]; Kp = [k1;k2;k3]; \\
Cs = collect(det(s*eye(3)-(Ap-Bp*transpose(Kp)))); \\
fprintf('Parametrized characteristic polynomial: '); disp(Cs); \\
amin = (g/L1)*(sin(y)/y); amax = (g/L1);  \\
bmin = (1/L1)*cos(y); bmax = (1/L1); \\
k3 = -10; \\
k2 = floor((k3/bmin)/10)*10 - 10; \\
k1 = floor((amax*k2/(-bmin*k2+k3))/10)*10 - 10; \\
sttKp = [k1;k2;k3]; \\
fprintf('Robust gain matrix K: '); disp(sttKp'); \\
cRobust=1; fprintf('Start robust control$\backslash$n');  \\
 \\
\%\% Simulation of single inverted pendulum control \\
for t = tSpan \\
$~~~~$ \%\% Control method \\
$~~~~$ if (y\^{}2+dy\^{}2+dx\^{}2 $>$ 1 \&\& cRobust==1) \\
$~~~~$ $~~~~$ acc = -sttKp'*stt([1,2,4]); \% Robust FSFC  \\
$~~~~$ else \\
$~~~~$ $~~~~$ if (1==cRobust) \\
$~~~~$ $~~~~$ $~~~~$ cRobust = 0; sttE = [0; 0; x; 0]; \\
$~~~~$ $~~~~$ $~~~~$ fprintf('Switch to sliding mode control$\backslash$n'); \\
$~~~~$ $~~~~$ end \\
$~~~~$ $~~~~$ if (sttE(3)$>$0) sttE(3) = max(sttE(3) - 8*dt, 0); \\
$~~~~$ $~~~~$ else sttE(3) = min(sttE(3) + 8*dt, 0); end \\
$~~~~$ $~~~~$ acc = -sttK'*(stt-sttE); \% Sliding mode FSFC \\
$~~~~$ end \\
 \\
$~~~~$ \%\% Single inverted pendulum dynamics \\
$~~~~$ stt = DynamicsSIP(SimConfig, stt, acc); \\
$~~~~$ sttC = num2cell(stt); [y, dy, x, dx] = sttC\{:\}; \\
$~~~~$ if (x\^{}2+y\^{}2+dy\^{}2+dx\^{}2$<$0.001) fprintf('Control success!$\backslash$n'); break; end \\
$~~~~$ if (abs(y)$>$=pi/2) fprintf('Control failure!$\backslash$n'); break; end \\
$~~~~$ k = k+1; sttAll(:,k) = stt; \\
$~~~~$ \%\% Single inverted pendulum visualization \\
$~~~~$ if (rem(k,20) == 0) \\
$~~~~$ $~~~~$ DisplaySIP(x, y, L1); pause(dt); \\
$~~~~$ end \\
end
\end{framed}

\section{Adaptive Control}  \label{sec:adaptive_control}

\subsection{Adaptive modelling}

In practical applications, we usually adopt a basic assumption that the control system is \textit{time-invariant}. However, time-invariance is ideal whereas \textit{time-variance} is ubiquitous. Normally, what we refer to as time-invariant control systems are those for which time-invariant system modelling can be fairly adopted in their operation space. However, approximating a \textit{time-variant} control system by a corresponding time-invariant version inevitably incurs a discrepancy between the time-variant control system that exists objectively and its time-invariant counterpart that we actually handle. The discrepancy can be regarded as a kind of control system uncertainty. In many circumstances, such control system uncertainty may be negligible, but in many circumstances as well, it can by no means be neglected.

\begin{figure}[h!]
\begin{center}
\includegraphics[width=0.9\columnwidth]{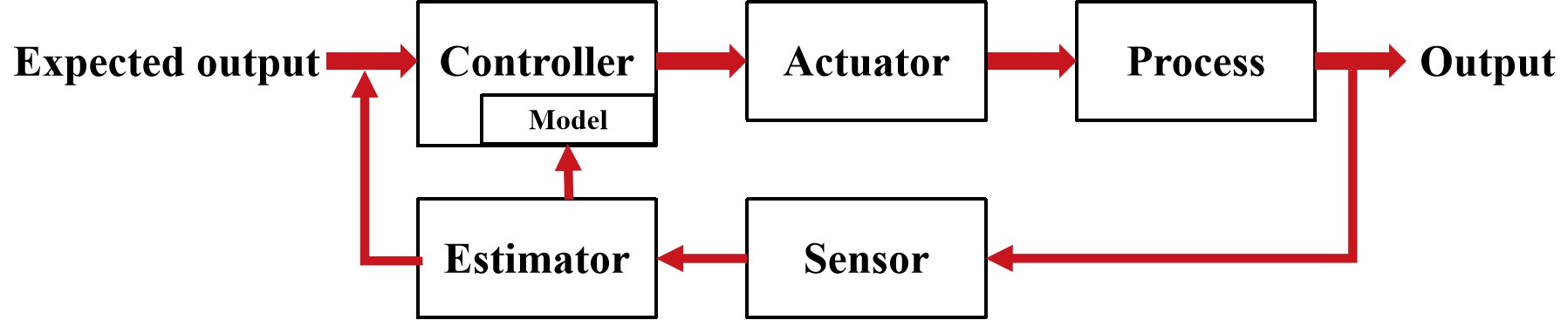}
\end{center}
\caption{Adaptive closed-loop feedback control system}
\label{fig:adaptive_control_system}
\end{figure}

Since any control system is after all time-variant by nature, theoretically speaking, instead of adopting an ideal time-invariant system model established off-line, a control system had better possess a mechanism to adapt its system model via on-line estimation or identification so that the adaptive system model would better match its dynamics during its real-time operation, and had better also generate its control law with the adaptive system model taken into account in real time. This is the basic spirit of \textbf{adaptive control} \cite{Bellman1959, Astrom2008, ChaiT2016}, which is illustrated in Figure \ref{fig:adaptive_control_system}.

Adaptive control does not neglect control system uncertainty. In fact, robust control already presented in Section \ref{sec:robust_control} does not neglect control system uncertainty either. Yet unlike robust control which aims at providing control methods intended to work even when facing the maximum degrees of control system uncertainty (though without any idea of actual situation of control system uncertainty), adaptive control handles control system uncertainty directly via certain on-line estimation mechanism that adjusts the adopted system model adaptively for sake of making the control system be exempt from uncertainty. In other words, adaptive control takes advantage of certain on-line estimation mechanism to turn original control system uncertainty (i.e. things that we do not know) into certainty (i.e. things that we know) or at least into considerably less uncertainty (i.e. things that we are not so uncertain but could know to considerable extent). The basic spirit of adaptive control illustrated in Figure \ref{fig:adaptive_control_system} is rather generally applicable and can be naturally incorporated into other control methods. 

\subsection{Adaptive linear state-space modelling based control}  \label{sec:adaptive_linear_SSM_control}

Given a nonlinear control system, even suppose it is time-invariant, we may sometimes still follow the basic spirit of adaptive control and handle the nonlinear control system as a linear control system with time-variant system model parameters. Given a nonlinear control system that adopts generic state-space modelling described by
\begin{align}  \label{eq:state_differential_equation}
\frac{\mathrm{d}}{\mathrm{d} t} \mathbf{x} = f(\mathbf{x}, \mathbf{u}).
\end{align}
The idea is to approximate the system model function $f(\mathbf{x}, \mathbf{u})$, if possible, by an adaptive linear form with time-variant parameters (usually in terms of the state $\mathbf{x}$)
\begin{align*}
f(\mathbf{x}, \mathbf{u}) \approx \mathbf{A}(\mathbf{x}) \mathbf{x} + \mathbf{B}(\mathbf{x}) \mathbf{u}.
\end{align*} 
Then the nonlinear control system is approximately modelled by (\ref{eq:adaptive_state_DE_linear})
\begin{align*}  
\frac{\mathrm{d}}{\mathrm{d} t} \mathbf{x} = \mathbf{A}(\mathbf{x}) \mathbf{x} + \mathbf{B}(\mathbf{x}) \mathbf{u},
\end{align*}
which gives a generic formalism of \textbf{adaptive linear state-space modelling}. The linear system model described by (1.11) can be regarded as a special case of the adaptive linear system model described by (\ref{eq:adaptive_state_DE_linear}). In other words, if we set the state transition matrix $\mathbf{A}(\mathbf{x})$ and the control input matrix $\mathbf{B}(\mathbf{x})$ described in (\ref{eq:adaptive_state_DE_linear}) to constant matrices $\mathbf{A}$ and $\mathbf{B}$ respectively, then (\ref{eq:adaptive_state_DE_linear}) will be reduced to (1.11).
\footnote{Namely (1.11) in the author's works \cite{Li2026ACTPA_SJTU_2, Li2026ACTPA_SJTU_1}.}

In (\ref{eq:adaptive_state_DE_linear}), both the state transition matrix $\mathbf{A}(\mathbf{x})$ and the control input matrix $\mathbf{B}(\mathbf{x})$ are time-variant but \textit{can be known in real time during on-line operation of the control system}. More specifically, here we intentionally take advantage of the fact that both $\mathbf{A}(\mathbf{x})$ and $\mathbf{B}(\mathbf{x})$ can be known or estimated on-line and design control methods according to on-line estimated values of $\mathbf{A}(\mathbf{x})$ and $\mathbf{B}(\mathbf{x})$. The practice here is totally different from how we treat the state transition matrix $\mathbf{A}(\mathbf{x})$ and the control input matrix $\mathbf{B}(\mathbf{x})$ in (\ref{eq:uncertain_state_DE_linear})
\begin{align*}
\frac{\mathrm{d}}{\mathrm{d} t} \mathbf{x} = [\mathbf{A}(\mathbf{0}) + \Delta \mathbf{A}(\mathbf{x})] \mathbf{x} + [\mathbf{B}(\mathbf{0}) + \Delta \mathbf{B}(\mathbf{x})] \mathbf{u},
\end{align*}
where $\mathbf{A}(\mathbf{x})$ and $\mathbf{B}(\mathbf{x})$ are treated as their linear time-invariant counterparts superposed with \textit{unknown} control system uncertainty.

Compared with the linear system model described by (1.11), the adaptive linear system model described by (\ref{eq:adaptive_state_DE_linear}) enjoys the advantage of being able to better match dynamics of the control system's state during real-time operation. Compared with the original nonlinear system model described by (\ref{eq:state_differential_equation}), the adaptive linear system model described by (\ref{eq:adaptive_state_DE_linear}) enjoys the obvious advantage of control convenience, especially note that a variety of mature control methods exist for linear control systems.

Based on the adaptive linear system model described by (\ref{eq:adaptive_state_DE_linear}), adaptive control can be naturally performed. For example, consider the full-state feedback control method described by
\begin{align}  \label{eq:full_state_feedback_control}
\mathbf{u} = -\mathbf{K}^\mathrm{T} \mathbf{x}.
\end{align}
In the spirit of adaptive control, we no longer fix the gain matrix $\mathbf{K}$, but allow the gain matrix $\mathbf{K}$ to be adapted according to the time-variant state transition matrix $\mathbf{A}(\mathbf{x})$ and the time-variant control input matrix $\mathbf{B}(\mathbf{x})$. The adaptive full-state feedback control method can be described by
\begin{equation}  \label{eq:adaptive_full_state_feedback_control}
\mathbf{u} = -\mathbf{K}_{\mathbf{A}(\mathbf{x}), \mathbf{B}(\mathbf{x})}^\mathrm{T} \mathbf{x},
\end{equation}
where the gain matrix $\mathbf{K}_{\mathbf{A}(\mathbf{x}), \mathbf{B}(\mathbf{x})}$ is recomputed according to $\mathbf{A}(\mathbf{x})$ and $\mathbf{B}(\mathbf{x})$ regularly.

Sometimes, such kind of adaptive linear state-space modelling based control strategy may facilitate handling of nonlinear and time-variant control problems. For demonstration of this point, we still resort to the example of single inverted pendulum control presented in Section \ref{sec:robust_control_Riccati_equation_method} and Section \ref{sec:robust_control_interval_polynomial}.

\subsubsection*{Application: single inverted pendulum adaptive control}

Extract the sub-model associated with the partial state
\begin{align*}
\mathbf{x}_P \equiv \begin{bmatrix} \theta & \frac{\mathrm{d} \theta}{\mathrm{d} t} & \frac{\mathrm{d} x}{\mathrm{d} t} \end{bmatrix}^\mathrm{T}
\end{align*}
from (\ref{eq:uncertain_SIP_state_DE_linear}) and obtain (\ref{eq:uncertain_SIP_state_DE_linear_partial})
\begin{align*}
\frac{\mathrm{d}}{\mathrm{d} t} \mathbf{x}_P = \mathbf{A}_P(\mathbf{x}_P) \mathbf{x}_P + \mathbf{B}_P(\mathbf{x}_P) a,
\end{align*}
where
\begin{align*}
\mathbf{A}_P(\mathbf{x}_P) \equiv \begin{bmatrix} 0 & 1 & 0 \\ \frac{g}{L} \frac{\sin \theta}{\theta} & 0 & 0 \\ 0 & 0 & 0 \end{bmatrix}, \quad \mathbf{B}_P(\mathbf{x}_P) \equiv \begin{bmatrix} 0 \\ -\frac{\cos \theta}{L} \\ 1 \end{bmatrix}.
\end{align*}
Suppose the inverted pendulum angle $\theta$ is in the operation range (\ref{eq:SIP_theta_operation_range})
\begin{align*}
- \theta_{\max} \leq \theta \leq \theta_{\max}.
\end{align*}

Apply the adaptive full-state feedback control method as
\begin{align}  \label{eq:adaptive_SIP_FSFC_a}
a = -\mathbf{K}_{\mathbf{A}_P(\mathbf{x}_P), \mathbf{B}_P(\mathbf{x}_P)}^\mathrm{T} \mathbf{x}_P,
\end{align}
where the gain matrix $\mathbf{K}_{\mathbf{A}_P(\mathbf{x}_P), \mathbf{B}_P(\mathbf{x}_P)}$ is recomputed according to $\mathbf{A}_P(\mathbf{x}_P)$ and $\mathbf{B}_P(\mathbf{x}_P)$ regularly. Matlab simulation code for demonstrating adaptive full-state feedback control of the partial state $\mathbf{x}_P$ is given as follows. 

\begin{framed} 
\noindent \textbf{SingleInvertedPendulumAL.m} \\
\noindent \%\% Single inverted pendulum parameters \\
m1 = 1; L1 = 1; g = 10; \\
\%\% Simulation preliminary configuration \\
dt = 0.001; \% Numerical computation step \\
tSpan = 0:dt:3; \% Simulation time span \\
x = 0.2; \% Cart position  \\
dx = 0; \% Cart velocity \\
y = 0.4*pi; \% Inverted pendulum angle theta  \\
dy = 0;  \% Inverted pendulum angular velocity \\
stt = [y; dy; x; dx]; \% Single inverted pendulum state \\
sttAll = zeros(length(stt), length(tSpan)); k = 0; \% Record states \\
xExpected = 0; yExpected = 0; \% Expected equilibrium status \\
SimConfig = [m1, L1, g, dt]; \\
\%\% Control initialization \\
lambdaE = [-4;-4;-4;-4]; \% Expected eigenvalues \\
iStt = [1 2 4]; \% Indices of partial state control \\
fprintf('Start adaptive control$\backslash$n'); \\
 \\
\%\% Simulation of single inverted pendulum control \\
for t = tSpan \\
$~~~~$ \%\% Control method \\
$~~~~$ if (abs(y)$<$0.1) A21 = g/L1; else A21 = (g/L1)*sin(y)/y; end \\
$~~~~$ A = [0, 1, 0, 0; A21, 0, 0, 0; 0, 0, 0, 1; 0, 0, 0, 0]; \\
$~~~~$ B = [0; -cos(y)/L1; 0; 1]; \% Adaptive linear state-space modelling \\
$~~~~$ sttK = DesignGainMatrix(A(iStt,iStt), B(iStt), lambdaE(iStt)); \\
$~~~~$ acc = -sttK'*stt(iStt); \\
 \\
$~~~~$ \%\% Single inverted pendulum dynamics \\
$~~~~$ stt = DynamicsSIP(SimConfig, stt, acc); \\
$~~~~$ sttC = num2cell(stt); [y, dy, x, dx] = sttC\{:\}; \\
$~~~~$ if (abs(y)$>$=pi/2) fprintf('Control failure!$\backslash$n'); break; end \\
$~~~~$ k = k+1; sttAll(:,k) = stt; \\
$~~~~$ \%\% Single inverted pendulum visualization \\
$~~~~$ if (rem(k,20) == 0) \\
$~~~~$ $~~~~$ DisplaySIP(x, y, L1); pause(dt); \\
$~~~~$ end \\
end
\end{framed}

The visualization code \textbf{DisplaySIP.m} and the single inverted pendulum dynamics code \textbf{DynamicsSIP.m} are given in Section 2.2.3 in Chapter 2. The gain matrix designing code \textbf{DesignGainMatrix.m} is given in Section 2.3.2 in Chapter 2.

\begin{figure}[h!]
\begin{center}
\includegraphics[width=1.0\columnwidth]{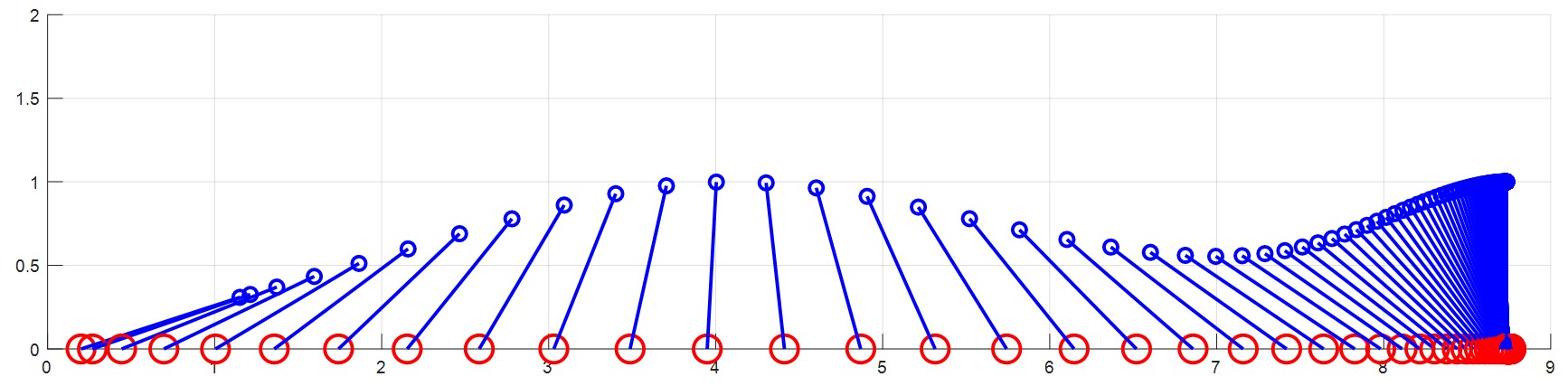}
\end{center}
\caption{Adaptive full-state feedback control of partial inverted pendulum state}
\label{fig:AdaptiveControlSIP}
\end{figure}

After trials with the Matlab simulation code, readers will find that the adaptive linear system model based full-state feedback control method can effectively stabilize the partial state $\mathbf{x}_P$, as demonstrated in Figure \ref{fig:AdaptiveControlSIP}. Readers can see the obvious control convenience brought by adaptive linear state-space modelling based control.

In above simulation, the gain matrix $\mathbf{K}_{\mathbf{A}_P(\mathbf{x}_P), \mathbf{B}_P(\mathbf{x}_P)}$ is recomputed according to $\mathbf{A}_P(\mathbf{x}_P)$ and $\mathbf{B}_P(\mathbf{x}_P)$ in each control period. In fact, a more convenient and efficient realization of the adaptive control method is to establish a look-up table of gain matrices by dividing the state space into \textit{piecewise linear operation subspaces} and pre-computing a gain matrix for each linear operation subspace. 

For example, divide the single inverted pendulum state space into three piecewise linear operation subspaces according to the inverted pendulum angle $\theta$, i.e.
\begin{align*}  
|\theta| \leq \pi/6,  \\
\pi/6 < |\theta| \leq \pi/3,  \\
\pi/3 < |\theta| \leq \theta_{\max}
\end{align*}
respectively. For the first linear operation subspace, set 
\begin{align*}  
\theta = 0
\end{align*}
in the adaptive linear system model described by (\ref{eq:uncertain_SIP_state_DE_linear_partial}) and compute its associated gain matrix $\mathbf{K}_1$. For the second linear operation subspace, set 
\begin{align*}  
\theta = \pi/4
\end{align*}
in the adaptive linear system model described by (\ref{eq:uncertain_SIP_state_DE_linear_partial}) and compute its associated gain matrix $\mathbf{K}_2$. For the third linear operation subspace, set 
\begin{align*}  
\theta = \theta_{\max}
\end{align*}
in the adaptive linear system model described by (\ref{eq:uncertain_SIP_state_DE_linear_partial}) and compute its associated gain matrix $\mathbf{K}_3$. During adaptive control for stabilization of the partial state $\mathbf{x}_P$, the controller can simply check the inverted pendulum angle $\theta$ and select a corresponding gain matrix among 
\begin{align*}  
\{\mathbf{K}_1, \quad \mathbf{K}_2, \quad \mathbf{K}_3\} 
\end{align*}
to use.

Matlab simulation code for demonstrating such convenient and efficient realization of adaptive full-state feedback control of the partial state $\mathbf{x}_P$ is given as follows --- Once the partial state $\mathbf{x}_P$ is stabilized, sliding mode control based on the easier linear system model described by (\ref{eq:SIP_state_DE_linear}) is performed to converge the entire state $\mathbf{x}$ to the expected equilibrium state finally.

\begin{framed} 
\noindent \textbf{SingleInvertedPendulumAL2.m} \\
\noindent \%\% Single inverted pendulum parameters \\
m1 = 1; L1 = 1; g = 10; \\
\%\% Simulation preliminary configuration \\
dt = 0.001; \% Numerical computation step \\
tSpan = 0:dt:10; \% Simulation time span \\
x = 0.2; \% Cart position  \\
dx = 0; \% Cart velocity \\
y = 0.4*pi; \% Inverted pendulum angle theta  \\
dy = 0;  \% Inverted pendulum angular velocity \\
stt = [y; dy; x; dx]; \% Single inverted pendulum state \\
sttAll = zeros(length(stt), length(tSpan)); k = 0; \% Record states \\
xExpected = 0; yExpected = 0; \% Expected equilibrium status \\
SimConfig = [m1, L1, g, dt]; \\
\%\% Control initialization \\
lambdaE = [-4;-4;-4;-4]; \% Expected eigenvalues \\
A = [0, 1, 0, 0; g/L1, 0, 0, 0; 0, 0, 0, 1; 0, 0, 0, 0];  \\
B = [0; -1/L1; 0; 1]; \\
sttK0 = DesignGainMatrix(A, B, lambdaE); \\
iStt = [1 2 4]; \% Indices of partial state control \\
sttK1 = DesignGainMatrix(A(iStt,iStt), B(iStt), lambdaE(iStt)); \\
yA = pi/4; A(2,1) = (g/L1)*sin(yA)/yA; B(2,1) = -cos(yA)/L1; \\
sttK2 = DesignGainMatrix(A(iStt,iStt), B(iStt), lambdaE(iStt)); \\
yA = y; A(2,1) = (g/L1)*sin(yA)/yA; B(2,1) = -cos(yA)/L1; \\
sttK3 = DesignGainMatrix(A(iStt,iStt), B(iStt), lambdaE(iStt)); \\
cAdaptive=1; fprintf('Start adaptive control$\backslash$n'); \\
 \\
\%\% Simulation of single inverted pendulum control \\
for t = tSpan \\
$~~~~$ \%\% Control method \\
$~~~~$ if (y\^{}2+dy\^{}2+dx\^{}2 $>$ 1 \&\& cAdaptive==1) \% Adaptive control \\
$~~~~$ $~~~~$ if (abs(y)$<$pi/6) sttK = sttK1; \\
$~~~~$ $~~~~$ elseif (abs(y)$<$pi/3) sttK = sttK2; \\
$~~~~$ $~~~~$ else sttK = sttK3; end \% Adaptive linear state-space modelling \\
$~~~~$ $~~~~$ acc = -sttK'*stt(iStt); \\
$~~~~$ else \% Switch to sliding mode control \\
$~~~~$ $~~~~$ if (1==cAdaptive) \\
$~~~~$ $~~~~$ $~~~~$ cAdaptive = 0; sttE = [0; 0; x; 0]; sttK = sttK0; \\
$~~~~$ $~~~~$ $~~~~$ fprintf('Switch to sliding mode control$\backslash$n'); \\
$~~~~$ $~~~~$ end \\
$~~~~$ $~~~~$ if (sttE(3)$>$0) sttE(3) = max(sttE(3) - 8*dt, 0); \\
$~~~~$ $~~~~$ else sttE(3) = min(sttE(3) + 8*dt, 0); end \\
$~~~~$ $~~~~$ acc = -sttK'*(stt-sttE); \% Sliding mode FSFC \\
$~~~~$ end \\
 \\
$~~~~$ \%\% Single inverted pendulum dynamics \\
$~~~~$ stt = DynamicsSIP(SimConfig, stt, acc); \\
$~~~~$ sttC = num2cell(stt); [y, dy, x, dx] = sttC\{:\}; \\
$~~~~$ if (x\^{}2+y\^{}2+dy\^{}2+dx\^{}2$<$0.001) fprintf('Control success!$\backslash$n'); break; end \\
$~~~~$ if (abs(y)$>$=pi/2) fprintf('Control failure!$\backslash$n'); break; end \\
$~~~~$ k = k+1; sttAll(:,k) = stt; \\
$~~~~$ \%\% Single inverted pendulum visualization \\
$~~~~$ if (rem(k,20) == 0) \\
$~~~~$ $~~~~$ DisplaySIP(x, y, L1); pause(dt); \\
$~~~~$ end \\
end
\end{framed}

\begin{figure}[h!]
\begin{center}
\includegraphics[width=1.0\columnwidth]{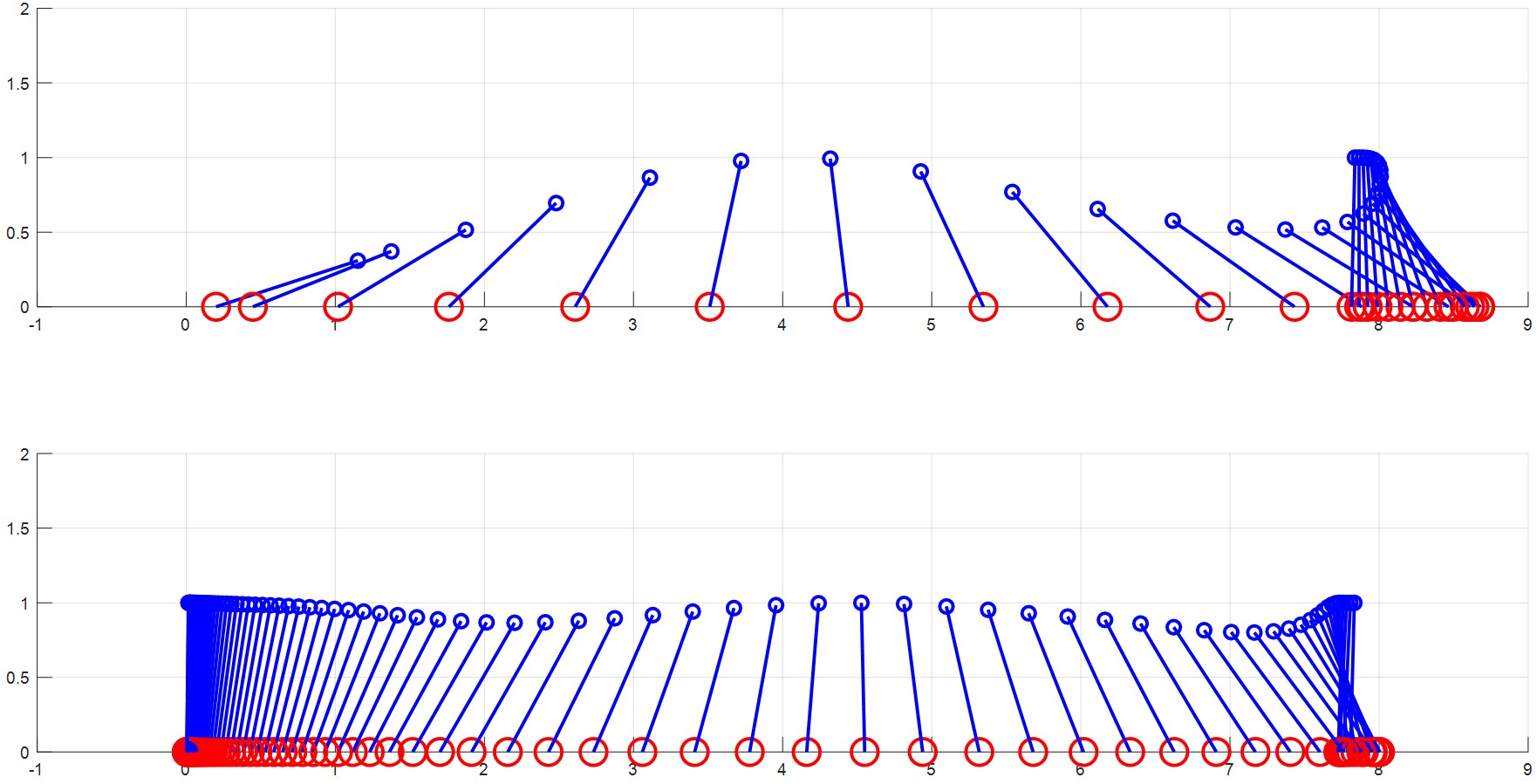}
\end{center}
\caption{Improved adaptive control of inverted pendulum state}
\label{fig:AdaptiveControlSIP2}
\end{figure}

After trials with the Matlab simulation code, readers will also find that the improved version of the adaptive linear system model based full-state feedback control method can effectively stabilize the partial state $\mathbf{x}_P$, as demonstrated in Figure \ref{fig:AdaptiveControlSIP2}. The top sub-figure and bottom sub-figure of Figure \ref{fig:AdaptiveControlSIP2} demonstrate the complete process of single inverted pendulum control. The top sub-figure demonstrates the first sub-process of transforming the single inverted pendulum from the severely inclined initial state (on the left side) to an intermediate vertical state (on the right side) via the improved version of adaptive control, whereas the bottom sub-figure demonstrates the second sub-process of converging the single inverted pendulum from the deviated intermediate vertical state (on the right side) to the expected state (on the left side).

\subsection{System identification}  \label{sec:system_identification}

Revisit the state differential equation (\ref{eq:adaptive_state_DE_linear}) presented in Section \ref{sec:adaptive_linear_SSM_control} 
\begin{align*}  
\frac{\mathrm{d}}{\mathrm{d} t} \mathbf{x} = \mathbf{A}(\mathbf{x}) \mathbf{x} + \mathbf{B}(\mathbf{x}) \mathbf{u}
\end{align*}
which gives a generic formalism of adaptive linear state-space modelling. In fact, in the context of adaptive control, we can also use another formalism instead of the formalism (\ref{eq:adaptive_state_DE_linear}) to highlight existence of certain adaptive mechanism in the system model, namely
\begin{equation}  \label{eq:adaptive_state_DE_linear_Theta}
\frac{\mathrm{d}}{\mathrm{d} t} \mathbf{x} = \mathbf{A}_{\mathbf{\Theta}} \mathbf{x} + \mathbf{B}_{\mathbf{\Theta}} \mathbf{u}.
\end{equation}
Such way of highlighting existence of certain adaptive mechanism can also be extended to generic state-space modelling described by (\ref{eq:state_differential_equation})
\begin{align*}
\frac{\mathrm{d}}{\mathrm{d} t} \mathbf{x} = f(\mathbf{x}, \mathbf{u})
\end{align*}
to obtain a generic formalism of adaptive state-space modelling
\begin{equation}  \label{eq:adaptive_state_differential_equation_Theta}
\frac{\mathrm{d}}{\mathrm{d} t} \mathbf{x} = f_{\mathbf{\Theta}} (\mathbf{x}, \mathbf{u}).
\end{equation}
In (\ref{eq:adaptive_state_DE_linear_Theta}) and (\ref{eq:adaptive_state_differential_equation_Theta}), $\mathbf{\Theta}$ denotes the set of parameters that actually determine how the system model can be adjusted adaptively, though the parameter set $\mathbf{\Theta}$ can always be regarded as a special part of the state $\mathbf{x}$ or certain augmented state version. 

Some explanations hover over the so-called augmented state. Just for thought experiment, suppose the parameter set $\mathbf{\Theta}$ is not part of the state $\mathbf{x}$. As conveyed by (\ref{eq:adaptive_state_differential_equation_Theta}), the parameter set $\mathbf{\Theta}$ indeed matters in determining dynamics of the state $\mathbf{x}$. So we can fairly augment the state $\mathbf{x}$ with the parameter set $\mathbf{\Theta}$ to form a new state
\begin{align*}
\mathbf{x}^A \equiv \begin{bmatrix} \mathbf{x} \\ \mathbf{\Theta} \end{bmatrix}
\end{align*}
and augment (\ref{eq:adaptive_state_differential_equation_Theta}) accordingly to a new system model formalism
\begin{equation}  \label{eq:adaptive_state_differential_equation_Theta_augment}
\frac{\mathrm{d}}{\mathrm{d} t} \mathbf{x}^A \equiv \frac{\mathrm{d}}{\mathrm{d} t} \begin{bmatrix} \mathbf{x} \\ \mathbf{\Theta} \end{bmatrix} = \begin{bmatrix} f_{\mathbf{\Theta}} (\mathbf{x}, \mathbf{u}) \\ g(\mathbf{\Theta}) \end{bmatrix}.
\end{equation}
The new formalism (\ref{eq:adaptive_state_differential_equation_Theta_augment}) can describe system dynamics more completely than the original formalism (\ref{eq:adaptive_state_differential_equation_Theta}) and the augmented state $\mathbf{x}^A$ is more like the ``state'' for the control system than the original vehicle state $\mathbf{x}$. Then we can adopt the augmented state $\mathbf{x}^A$ naturally as the actual state and further treat the parameter set $\mathbf{\Theta}$ as a special part of such state.

Although the system model formalism (\ref{eq:adaptive_state_differential_equation_Theta_augment}) seems to be more natural and more sound, we may still use the original system model formalism (\ref{eq:adaptive_state_differential_equation_Theta}) and treat parameters in $\mathbf{\Theta}$ as unknowns revealed via methods of \textbf{system identification} \cite{Astrom1971, Ljung1987} adaptively.

\subsubsection*{Linear system identification}

For the generic formalism of adaptive linear state-space modelling namely (\ref{eq:adaptive_state_DE_linear_Theta})
\begin{align*}
\dot{\mathbf{x}} \equiv \frac{\mathrm{d}}{\mathrm{d} t} \mathbf{x} = \mathbf{A}_{\mathbf{\Theta}} \mathbf{x} + \mathbf{B}_{\mathbf{\Theta}} \mathbf{u},
\end{align*}
suppose elements of the parametrized state transition matrix $\mathbf{A}_{\mathbf{\Theta}}$ and the parametrized control input matrix $\mathbf{B}_{\mathbf{\Theta}}$ change slowly and may be treated as constant unknowns during a moderate span of control periods. Perform matrix vectorization 
\footnote{Readers can refer to Section 1.4.1 in Chapter 1 for a knowledge of \textit{matrix vectorization} and \textit{Kronecker product} --- Namely Chapter 1 of the author's works \cite{Li2026ACTPA_SJTU_2, Li2026ACTPA_SJTU_1}. Note that this article is Chapter 5 of the works.}
on both sides of (\ref{eq:adaptive_state_DE_linear_Theta}) and resort to Kronecker products as
\begin{align*}
\mbox{vec}(\dot{\mathbf{x}}) &= \mbox{vec}(\mathbf{A}_{\mathbf{\Theta}} \mathbf{x} + \mathbf{B}_{\mathbf{\Theta}} \mathbf{u}) = (\mathbf{x}^\mathrm{T} \otimes \mathbf{I}) \mbox{vec}(\mathbf{A}_{\mathbf{\Theta}}) + (\mathbf{u}^\mathrm{T} \otimes \mathbf{I}) \mbox{vec}(\mathbf{B}_{\mathbf{\Theta}})  \\
  &= \begin{bmatrix} \mathbf{x}^\mathrm{T} \otimes \mathbf{I} & \mathbf{u}^\mathrm{T} \otimes \mathbf{I} \end{bmatrix} \begin{bmatrix} \mbox{vec}(\mathbf{A}_{\mathbf{\Theta}}) \\ \mbox{vec}(\mathbf{B}_{\mathbf{\Theta}}) \end{bmatrix}
\end{align*}
namely
\begin{equation}  \label{eq:linear_sys_identify_vecA+B}
\begin{bmatrix} \mathbf{x}^\mathrm{T} \otimes \mathbf{I} & \mathbf{u}^\mathrm{T} \otimes \mathbf{I} \end{bmatrix} \begin{bmatrix} \mbox{vec}(\mathbf{A}_{\mathbf{\Theta}}) \\ \mbox{vec}(\mathbf{B}_{\mathbf{\Theta}}) \end{bmatrix} = \mbox{vec}(\dot{\mathbf{x}}).
\end{equation}

Given current control period $t$, always consider its $k$ previous control periods, suppose the parametrized state transition matrix $\mathbf{A}_{\mathbf{\Theta}}$ and the parametrized control input matrix $\mathbf{B}_{\mathbf{\Theta}}$ may be treated as constant unknowns during the span of $k+1$ control periods
\begin{align*}
t - k \Delta t, \quad \cdots \quad, \quad t - \Delta t, \quad t,
\end{align*}
and establish equations of the form (\ref{eq:linear_sys_identify_vecA+B}) for the $k+1$ control periods as
\begin{equation}  \label{eq:linear_sys_identify_vecA+B_k+1}
\begin{bmatrix} \mathbf{x}_{t}^\mathrm{T} \otimes \mathbf{I} & \mathbf{u}_{t}^\mathrm{T} \otimes \mathbf{I} \\ \mathbf{x}_{t-1}^\mathrm{T} \otimes \mathbf{I} & \mathbf{u}_{t-1}^\mathrm{T} \otimes \mathbf{I} \\ \vdots & \vdots \\ \mathbf{x}_{t-k}^\mathrm{T} \otimes \mathbf{I} & \mathbf{u}_{t-k}^\mathrm{T} \otimes \mathbf{I} \end{bmatrix} \begin{bmatrix} \mbox{vec}(\mathbf{A}_{\mathbf{\Theta}, t}) \\ \mbox{vec}(\mathbf{B}_{\mathbf{\Theta}, t}) \end{bmatrix} = \begin{bmatrix} \mbox{vec}(\dot{\mathbf{x}}_{t}) \\ \mbox{vec}(\dot{\mathbf{x}}_{t-1}) \\ \vdots \\ \mbox{vec}(\dot{\mathbf{x}}_{t-k}) \end{bmatrix}.
\end{equation}
Solve (\ref{eq:linear_sys_identify_vecA+B_k+1}) to obtain 
\begin{align*}
\begin{bmatrix} \mbox{vec}(\mathbf{A}_{\mathbf{\Theta}, t}) \\ \mbox{vec}(\mathbf{B}_{\mathbf{\Theta}, t}) \end{bmatrix}
\end{align*}
and hence obtain $\mathbf{A}_{\mathbf{\Theta}, t}$ and $\mathbf{B}_{\mathbf{\Theta}, t}$.

\subsubsection*{Application: single inverted pendulum system identification}

Consider single inverted pendulum control and paraphrase (\ref{eq:uncertain_SIP_state_DE_linear}) according to  (\ref{eq:adaptive_state_DE_linear_Theta}) as
\begin{equation}  \label{eq:adaptive_SIP_state_DE_linear_Theta}
\dot{\mathbf{x}} \equiv \frac{\mathrm{d}}{\mathrm{d} t} \mathbf{x} = \begin{bmatrix} 0 & 1 & 0 & 0 \\ \Theta_1 & 0 & 0 & 0 \\ 0 & 0 & 0 & 1  \\ 0 & 0 & 0 & 0 \end{bmatrix} \mathbf{x} + \begin{bmatrix} 0 \\ \Theta_2 \\ 0 \\ 1 \end{bmatrix} a \equiv \mathbf{A}_{\mathbf{\Theta}} \mathbf{x} + \mathbf{B}_{\mathbf{\Theta}} a,
\end{equation}
where the state
\begin{align*}
\mathbf{x} \equiv \begin{bmatrix} \theta & \frac{\mathrm{d} \theta}{\mathrm{d} t} & x & \frac{\mathrm{d} x}{\mathrm{d} t} \end{bmatrix}^\mathrm{T}
\end{align*}
and the parameter set
\begin{align*}
\mathbf{\Theta} \equiv \begin{bmatrix} \Theta_1 \\ \Theta_2 \end{bmatrix} = \begin{bmatrix} \frac{g}{L} \frac{\sin \theta}{\theta} \\ -\frac{\cos \theta}{L} \end{bmatrix}.
\end{align*}
For analysis simplicity, we neglect correlation between $\Theta_1$ and $\Theta_2$, treating them as two independent parameters --- Theoretically this will result in the over-modelling problem
\footnote{Refer to Section 3.3.1 in Chapter 3 --- Namely Chapter 3 of the author's works \cite{Li2026ACTPA_SJTU_2, Li2026ACTPA_SJTU_1}.},
yet which would be largely alleviated or even almost eliminated by establishing redundant equations in terms of $\mathbf{\Theta}$ via (\ref{eq:linear_sys_identify_vecA+B_k+1}).

Extract the sub-model associated with the partial state
\begin{align*}
\mathbf{x}_P \equiv \begin{bmatrix} \theta & \frac{\mathrm{d} \theta}{\mathrm{d} t} & \frac{\mathrm{d} x}{\mathrm{d} t} \end{bmatrix}^\mathrm{T}
\end{align*}
from (\ref{eq:adaptive_SIP_state_DE_linear_Theta}) and obtain
\begin{equation}  \label{eq:adaptive_SIP_state_DE_linear_Theta_partial}
\dot{\mathbf{x}_P} \equiv \frac{\mathrm{d}}{\mathrm{d} t} \mathbf{x}_P = \mathbf{A}_{P, \mathbf{\Theta}} \mathbf{x}_P + \mathbf{B}_{P, \mathbf{\Theta}} a,
\end{equation}
where
\begin{align*}
\mathbf{A}_{P, \mathbf{\Theta}} \equiv \begin{bmatrix} 0 & 1 & 0 \\ \Theta_1 & 0 & 0 \\ 0 & 0 & 0 \end{bmatrix}, \quad \mathbf{B}_{P, \mathbf{\Theta}} \equiv \begin{bmatrix} 0 \\ \Theta_2 \\ 1 \end{bmatrix}.
\end{align*}
Establish a group of equations in terms of $\mathbf{\Theta}$ via (\ref{eq:linear_sys_identify_vecA+B_k+1}) as
\begin{equation}  \label{eq:SIP_identify_vecA+B_k+1}
\begin{bmatrix} \theta_{t} & a_{t} \\ \theta_{t-1} & a_{t-1} \\ \vdots & \vdots \\ \theta_{t-k} & a_{t-k} \end{bmatrix} \begin{bmatrix} \Theta_{1,t} \\ \Theta_{2,t} \end{bmatrix} = \begin{bmatrix} \ddot{\theta}_{t} \\ \ddot{\theta}_{t-1} \\ \vdots \\ \ddot{\theta}_{t-k} \end{bmatrix}.
\end{equation}
Solve (\ref{eq:SIP_identify_vecA+B_k+1}) to obtain $\Theta_{1,t}$, $\Theta_{2,t}$ and further obtain
\begin{align*}
\mathbf{A}_{P, \mathbf{\Theta}, t} \equiv \begin{bmatrix} 0 & 1 & 0 \\ \Theta_{1,t} & 0 & 0 \\ 0 & 0 & 0 \end{bmatrix}, \quad \mathbf{B}_{P, \mathbf{\Theta}, t} \equiv \begin{bmatrix} 0 \\ \Theta_{2,t} \\ 1 \end{bmatrix}.
\end{align*}

Apply the adaptive full-state feedback control method as
\begin{align}  \label{eq:adaptive_SIP_FSFC_a_sys_identify}
a = - \mathbf{K}_{\mathbf{A}_{P, \mathbf{\Theta}, t}, \mathbf{B}_{P, \mathbf{\Theta}, t}}^\mathrm{T} \mathbf{x}_P.
\end{align}
The gain matrix $\mathbf{K}_{\mathbf{A}_{P, \mathbf{\Theta}, t}, \mathbf{B}_{P, \mathbf{\Theta}, t}}$ in (\ref{eq:adaptive_SIP_FSFC_a_sys_identify}) is recomputed regularly according to $\mathbf{A}_{P, \mathbf{\Theta}, t}$ and $\mathbf{B}_{P, \mathbf{\Theta}, t}$ which are revealed via system identification in real time. Matlab simulation code for demonstrating adaptive full-state feedback control of the partial state $\mathbf{x}_P$ with system identification is given as follows. 

\begin{framed} 
\noindent \textbf{SingleInvertedPendulumAdaptiveSI.m} \\
\noindent \%\% Single inverted pendulum parameters \\
m1 = 1; L1 = 1; g = 10; \\
\%\% Simulation preliminary configuration \\
dt = 0.001; \% Numerical computation step \\
tSpan = 0:dt:3; \% Simulation time span \\
x = 0.2; \% Cart position  \\
dx = 0; \% Cart velocity \\
y = 0.4*pi; \% Inverted pendulum angle theta  \\
dy = 0;  \% Inverted pendulum angular velocity \\
stt = [y; dy; x; dx]; \% Single inverted pendulum state \\
sttAll = zeros(length(stt), length(tSpan)); k = 0; \% Record states \\
xExpected = 0; yExpected = 0; \% Expected equilibrium status \\
SimConfig = [m1, L1, g, dt]; \\
\%\% Control initialization \\
lambdaE = [-4;-4;-4;-4]; \% Expected eigenvalues \\
iStt = [1 2 4]; \% Indices of partial state control \\
fprintf('Start adaptive control with system identification$\backslash$n'); \\
kSI = 5; \% System identification span \\
XTheta = zeros(kSI+1,2); YTheta = zeros(kSI+1,1); sttold = stt; \\
 \\
\%\% Simulation of single inverted pendulum control \\
for t = tSpan \\
$~~~~$ \%\% Control method \\
$~~~~$ if (k$<$=kSI) \\
$~~~~$ $~~~~$ acc = 1; \\
$~~~~$ $~~~~$ XTheta(end-k,:) = [stt(1), acc]; \\
$~~~~$ $~~~~$ YTheta(end-k) = (stt(2)-sttold(2))/dt; \\
$~~~~$ else \\
$~~~~$ $~~~~$ XTheta(2:end,:)=XTheta(1:end-1,:); YTheta(2:end)=YTheta(1:end-1); \\
$~~~~$ $~~~~$ XTheta(1,:) = [stt(1), acc]; YTheta(1) = (stt(2)-sttold(2))/dt; \\
$~~~~$ $~~~~$ Theta = XTheta$\backslash$YTheta; \% Linear system identification \\
$~~~~$ $~~~~$ A = [0, 1, 0, 0; Theta(1), 0, 0, 0; 0, 0, 0, 1; 0, 0, 0, 0]; \\
$~~~~$ $~~~~$ B = [0; Theta(2); 0; 1]; \% Adaptive linear state-space modelling \\
$~~~~$ $~~~~$ sttK = DesignGainMatrix(A(iStt,iStt), B(iStt), lambdaE(iStt)); \\
$~~~~$ $~~~~$ acc = -sttK'*stt(iStt); \\
$~~~~$ end \\
$~~~~$ sttold = stt; \\
 \\
$~~~~$ \%\% Single inverted pendulum dynamics \\
$~~~~$ stt = DynamicsSIP(SimConfig, stt, acc); \\
$~~~~$ sttC = num2cell(stt); [y, dy, x, dx] = sttC\{:\}; \\
$~~~~$ if (abs(y)$>$=pi/2) fprintf('Control failure!$\backslash$n'); break; end \\
$~~~~$ k = k+1; sttAll(:,k) = stt; \\
$~~~~$ \%\% Single inverted pendulum visualization \\
$~~~~$ if (rem(k,20) == 0) \\
$~~~~$ $~~~~$ DisplaySIP(x, y, L1); pause(dt); \\
$~~~~$ end \\
end
\end{framed}

The visualization code \textbf{DisplaySIP.m} and the single inverted pendulum dynamics code \textbf{DynamicsSIP.m} are given in Section 2.2.3 in Chapter 2. The gain matrix designing code \textbf{DesignGainMatrix.m} is given in Section 2.3.2 in Chapter 2.

\section{Barrier function control}  \label{sec:control_barrier_function}

Recall the spirit of sliding mode control in handling control system uncertainty, namely to force the state to evolve only in state space regions that tend to be exempt from uncertainty. In fact, whether the state space regions are exempt from uncertainty does not matter. The indeed key point of the spirit is how to guarantee that \textit{the state definitely evolves within certain state space regions}, or stated in a \textit{dual} way, how to guarantee that \textit{the state by no chance evolves into certain state space regions}. Besides sliding mode control, another method also embodies such important spirit. This other method is the method of \textbf{barrier function control} \cite{Prajna2006, Ames2017, Ames2019} which is based on special functions coined as \textbf{barrier functions}
\footnote{The concept \textit{barrier function} may be preceded with the word ``control'' to highlight its control-oriented role, forming the concept \textit{control barrier function}. However, the author prefers to neglect the conceptual nuance between the barrier function and the control barrier function. The author adopts the simple term \textit{barrier function} by default, because normally when people talk about barrier functions, they rarely discuss them purely for conceptual knowledge but naturally tend to bear in mind that the ultimate objective of discussing them consists in control namely to handle control problems in practical applications. So when we use the concept \textit{barrier function}, the control flavour is always there, be we expressing the term ``control'' explicitly or not. It is like when we talk about Lyapunov functions in the control context, we may add ``control'' to the concept \textit{Lyapunov function} to form the concept \textit{control Lyapunov function}, but doing so is unnecessary. People can fairly use the concept \textit{Lyapunov function} instead of the ``control'' highlighted concept version in the control context.}.

\subsection{Safe sets and barrier functions}

The state space regions only within which the state is intended to evolve are called the \textbf{safe state space set} or simply the \textbf{safe set}. In contrast, the state space regions which the state is intended to keep away from are called the \textbf{unsafe state space set} or simply the \textbf{unsafe set}. Denote a generic safe set as $\Omega_S$. The complement of the safe set $\Omega_S$, namely $\overline{\Omega_S}$, denotes the unsafe set.

\subsubsection*{Class $\mathcal{K}$ functions}

Before barrier functions associated with the safe set $\Omega_S$ are discussed, a basic knowledge of the \textit{class $\mathcal{K}$ function} \cite{Khalil2002} is needed. A continuous function
\begin{equation}  \label{eq:class_K_func}
\alpha \mbox{ } : \mbox{ } [0, a) \to [0, \infty)
\end{equation}
is said to belong to the \textit{class $\mathcal{K}$} if
\begin{itemize}
\item it is strictly increasing;
\item it satisfies $\alpha (0) = 0$.
\end{itemize}
It is worth noting that the domain of a class $\mathcal{K}$ function can be infinite (i.e. $a \to \infty$), yet if so, the function value limit at infinity namely $\lim_{r \to \infty} \alpha (r)$ is not necessarily infinite. A continuous function
\begin{equation}  \label{eq:class_K_inf_func}
\alpha \mbox{ } : \mbox{ } [0, \infty) \to [0, \infty)
\end{equation}
is said to belong to the \textit{class $\mathcal{K}_{\infty}$} if
\begin{itemize}
\item it belongs to the \textit{class $\mathcal{K}$}, namely it is a class $\mathcal{K}$ function with $a \to \infty$;
\item it further satisfies $\lim_{r \to \infty} \alpha (r) = \infty$.
\end{itemize}
In other words, the class $\mathcal{K}_{\infty}$ function is a special case of the class $\mathcal{K}$ function whose domain and image are both infinite.

As we can see from definitions of the class $\mathcal{K}$ and the class $\mathcal{K}_{\infty}$ given respectively in (\ref{eq:class_K_func}) and (\ref{eq:class_K_inf_func}), the function domain $\mbox{dom} (\alpha)$, either $[0, a)$ or $[0, \infty)$, has no influence on properties of the two kinds of functions. So definitions of the class $\mathcal{K}$ and the class $\mathcal{K}_{\infty}$ can be extended from the right half of the real axis to the entire real axis. A continuous function
\begin{equation}  \label{eq:extended_class_K_func}
\alpha \mbox{ } : \mbox{ } (-\infty, \infty) \to (-\infty, \infty)
\end{equation}
is said to belong to the \textit{extended class $\mathcal{K}$} if
\begin{itemize}
\item it is strictly increasing;
\item it satisfies $\alpha (0) = 0$.
\end{itemize}
An extended class $\mathcal{K}$ function is said to belong to the \textit{extended class $\mathcal{K}_{\infty}$} if it further satisfies $\lim_{r \to -\infty} \alpha (r) = -\infty$ and $\lim_{r \to \infty} \alpha (r) = \infty$.

\subsubsection*{Barrier functions}

According to analysis presented in Section 1.3.3 in Chapter 1, 
\footnote{Namely Chapter 1 of the author's works \cite{Li2026ACTPA_SJTU_2, Li2026ACTPA_SJTU_1}.}
dynamics of the closed-loop feedback version of a control system is equivalent to dynamics of a self-evolutionary system. For relevant analysis concerning barrier functions, we similarly consider a generic nonlinear self-evolutionary system that adopts state-space modelling described by
\begin{align}  \label{eq:closed-loop_feedback_SDE}
\frac{\mathrm{d}}{\mathrm{d} t} \mathbf{x} = f_c(\mathbf{x}).
\end{align}
Given an intended safe set $\Omega_S$ for the nonlinear self-evolutionary system, a \textbf{barrier function} is a scalar function $h(\mathbf{x})$ in terms of the state $\mathbf{x}$ such that
\begin{itemize}
\item The scalar function $h(\mathbf{x})$ is positive semi-definite over $\Omega_S$, i.e.
\begin{equation}  \label{eq:barrier_function_positive}
h(\mathbf{x}) \geq 0
\end{equation}
for $\mathbf{x} \in \Omega_S$.
\item The scalar function $h(\mathbf{x})$ is zero at the boundary of $\Omega_S$, i.e.
\begin{equation}  \label{eq:barrier_function_zero_boundary}
h(\mathbf{x}) = 0
\end{equation}
for $\mathbf{x} \in \partial \Omega_S$.
\item The scalar function $h(\mathbf{x})$ is strictly positive inside $\Omega_S$, i.e.
\begin{equation}  \label{eq:barrier_function_strict_positive}
h(\mathbf{x}) > 0
\end{equation}
for $\mathbf{x} \in \mbox{Int}(\Omega_S)$.
\item The derivative $\frac{\mathrm{d}}{\mathrm{d} t} h(\mathbf{x})$ plus an extended class $\mathcal{K}$ function $\alpha (h(\mathbf{x}))$ in terms of $h(\mathbf{x})$ is positive semi-definite, i.e.
\begin{equation}  \label{eq:barrier_function_derivative}
\frac{\mathrm{d}}{\mathrm{d} t} h(\mathbf{x}) + \alpha (h(\mathbf{x})) \geq 0 \iff \frac{\mathrm{d}}{\mathrm{d} t} h(\mathbf{x}) \geq - \alpha (h(\mathbf{x})).
\end{equation}
\end{itemize}
According to \cite{Ames2017, Ames2019}, \textit{if there is a barrier function for the nonlinear self-evolutionary system, then we can conclude that the nonlinear self-evolutionary system is safe} namely the state is guaranteed to evolve within the safe set $\Omega_S$.

\begin{framed} 
\noindent \textbf{Barrier function safety criterion}: \textit{If there is a barrier function for the nonlinear self-evolutionary system, then the nonlinear self-evolutionary system is safe}.
\end{framed}

\subsection{Safety guaranteed control}  \label{sec:safety_guaranteed_control}

Like the Lyapunov control strategy presented in Section 4.1.1 in Chapter 4, 
\footnote{Namely Chapter 4 of the author's works \cite{Li2026ACTPA_SJTU_2, Li2026ACTPA_SJTU_1}. Note that this article is Chapter 5 of the works.}
given a nonlinear control system that adopts generic state-space modelling described by (\ref{eq:state_differential_equation})
\begin{align*}
\frac{\mathrm{d}}{\mathrm{d} t} \mathbf{x} = f(\mathbf{x}, \mathbf{u}),
\end{align*}
the strategy of safety guaranteed control is to design a barrier function $h(\mathbf{x})$ on one hand and design a feedback control law of $\mathbf{u}$ in terms of $\mathbf{x}$ on the other hand such that
\begin{equation}  \label{eq:barrier_function_control_derivative}
\frac{\mathrm{d}}{\mathrm{d} t} h(\mathbf{x}) = \nabla h(\mathbf{x})^\mathrm{T} \frac{\mathrm{d}}{\mathrm{d} t} \mathbf{x} = \nabla h(\mathbf{x})^\mathrm{T} f(\mathbf{x}, \mathbf{u}) \geq - \alpha (h(\mathbf{x})).
\end{equation}
Recall definition of the \textit{Lie derivative} given in (4.60), the expression
\footnote{Namely (4.60) in the author's works \cite{Li2026ACTPA_SJTU_2, Li2026ACTPA_SJTU_1}.}
\begin{align*}
\nabla h(\mathbf{x})^\mathrm{T} f(\mathbf{x}, \mathbf{u})
\end{align*}
in (\ref{eq:barrier_function_control_derivative}) is right the Lie derivative of $h(\mathbf{x})$ with respect to $f(\mathbf{x}, \mathbf{u})$ namely 
\begin{align*}
L_{f(\mathbf{x}, \mathbf{u})} h(\mathbf{x}) \equiv \nabla h(\mathbf{x})^\mathrm{T} f(\mathbf{x}, \mathbf{u})
\end{align*}
which is the derivative of $h(\mathbf{x})$ along the direction of the vector $f(\mathbf{x}, \mathbf{u})$.

\begin{framed} 
\noindent \textbf{Barrier function control strategy}: \textit{Design a barrier function $h(\mathbf{x})$ and a feedback control law of $\mathbf{u}$ in terms of $\mathbf{x}$ such that $\frac{\mathrm{d}}{\mathrm{d} t} h(\mathbf{x})$ satisfies (\ref{eq:barrier_function_control_derivative}).}
\end{framed}

The generic nonlinear control system formalism (\ref{eq:state_differential_equation}) and its \textit{single-input-multiple-output} version (4.44) may be too general. As mentioned in Section 4.2.3 in Chapter 4, we withdraw from too general formalisms and instead consider the generic \textit{affine control system} formalism
\begin{equation}  \label{eq:state_differential_equation_MIMO_affine}
\frac{\mathrm{d}}{\mathrm{d} t} \mathbf{x} = \mathbf{f} (\mathbf{x}) + \mathbf{g} (\mathbf{x}) \mathbf{u},
\end{equation}
which is extended from (4.66) by replacing the single-input $u$ with the more general multiple-input $\mathbf{u}$ (including the single-input case as well). Substitute (\ref{eq:state_differential_equation_MIMO_affine}) into (\ref{eq:barrier_function_control_derivative}) and obtain
\begin{equation}  \label{eq:affine_CBF_derivative}
\frac{\mathrm{d}}{\mathrm{d} t} h(\mathbf{x}) = \nabla h(\mathbf{x})^\mathrm{T} \mathbf{f} (\mathbf{x}) + \nabla h(\mathbf{x})^\mathrm{T} \mathbf{g} (\mathbf{x}) \mathbf{u} \geq - \alpha (h(\mathbf{x}))
\end{equation}
namely 
\begin{align*}
L_{\mathbf{f}} h + (L_{\mathbf{g}} h) \mathbf{u} \geq - \alpha (h).
\end{align*}
The inequality given in (\ref{eq:affine_CBF_derivative}) formalizes the affine version of the barrier function control strategy.

\begin{framed} 
\noindent \textbf{Affine control system barrier function control strategy}: \textit{For an affine control system described by (\ref{eq:state_differential_equation_MIMO_affine}), design a barrier function $h(\mathbf{x})$ and a feedback control law of $\mathbf{u}$ in terms of $\mathbf{x}$ such that $\frac{\mathrm{d}}{\mathrm{d} t} h(\mathbf{x})$ satisfies (\ref{eq:affine_CBF_derivative}).}
\end{framed}

A representative method of instantiating the affine control system barrier function control strategy, which is proposed in \cite{Ames2019}, is to modify an existing controller or control law reference $\hat{\mathbf{u}} (\mathbf{x})$ in a minimal way for sake of guaranteeing safety. The control method is formalized as solving of a \textit{quadratic programming} problem
\begin{align}  \label{eq:affine_CBF_QP}
\mathbf{u} = &\arg \min_{\mathbf{u}} \frac{1}{2} \| \mathbf{u} - \hat{\mathbf{u}} \|_2^2  \\
  &\mbox{s.t.  } L_{\mathbf{f}} h + (L_{\mathbf{g}} h) \mathbf{u} \geq - \alpha (h).  \nonumber
\end{align}
The optimization constraint involved in quadratic programming (\ref{eq:affine_CBF_QP}) is right the inequality given in (\ref{eq:affine_CBF_derivative}). In many practical applications, the control input $u$ is single-input, then the quadratic programming formalism (\ref{eq:affine_CBF_QP}) is reduced from multiple-input control to single-input control as
\begin{align}  \label{eq:affine_CBF_QP_single_input}
u = &\arg \min_{u} \frac{1}{2} (u - \hat{u})^2 = \arg \min_{u} | u - \hat{u} |  \\
  &\mbox{s.t.  } L_{\mathbf{f}} h + (L_{\mathbf{g}} h) u \geq - \alpha (h).  \nonumber
\end{align}
There is no essential difference between the formalisms (\ref{eq:affine_CBF_QP}) and (\ref{eq:affine_CBF_QP_single_input}), except that the control input is reduced from the vector version $\mathbf{u}$ to the scalar version $u$.

\subsubsection*{Application: two-dimensional point nonlinear motion barrier function control}

Consider the two-dimensional point nonlinear motion control system described by
\begin{align}  \label{eq:Lyapunov_control_example1}
\frac{\mathrm{d}}{\mathrm{d} t} \begin{bmatrix} x \\ y \end{bmatrix} = \begin{bmatrix} x \sin y \\ y \end{bmatrix} + \begin{bmatrix} 0 \\ 1 \end{bmatrix} u,
\end{align}
where 
\begin{align*}
\mathbf{x} \equiv \begin{bmatrix} x & y \end{bmatrix}^\mathrm{T}
\end{align*}
denotes the system state and $u$ denotes the control input.

As presented in Section 4.1.3 in Chapter 4, derived from the Lyapunov function given in (4.12)
\begin{align*}
V(\mathbf{x}) = \frac{\mathbf{x}^\mathrm{T} \mathbf{x}}{2} = \frac{x^2 + y^2}{2},
\end{align*}
the Lyapunov control law of $u$ specified in (4.13)
\begin{align*}
u = -x^2 \frac{\sin y}{y} - 2 y
\end{align*}
can stabilize the two-dimensional point nonlinear motion control system.

\begin{figure}[h!]
\begin{center}
\includegraphics[width=0.9\columnwidth]{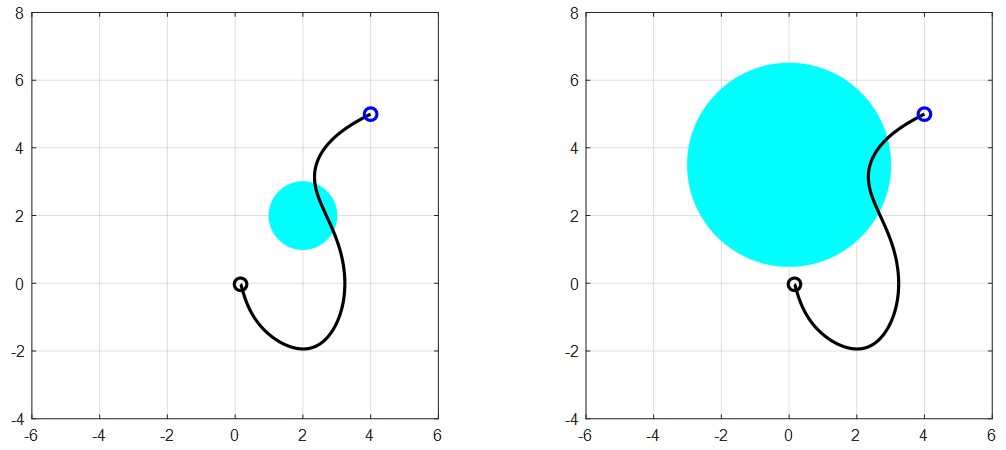}
\end{center}
\caption{Two-dimensional point nonlinear motion control that cannot guarantee safety: (left) unsafe set case 1; (right) unsafe set case 2}
\label{fig:2DMotionUnsafe}
\end{figure}

However, when there are unsafe sets in state space, the Lyapunov control law of $u$ specified in (4.13) may not guarantee safety. For example, given the state space configuration illustrated by the left sub-figure of Figure \ref{fig:2DMotionUnsafe}. The disk area represents the unsafe set, whereas the other part represents the safe set. Although the Lyapunov control law of $u$ specified in (4.13) enables the state to converge (as demonstrated by the black curve), it causes the state to evolve into the unsafe set (as demonstrated by the part of the convergence trajectory that is located inside the disk area).

Apply the affine control system barrier function control method to handle the two-dimensional point nonlinear motion control system. Set the Lyapunov control law of $u$ specified in (4.13) as control law reference $\hat{u} (\mathbf{x})$. Set the barrier function as
\begin{equation}  \label{eq:2D_Pt_NM_barreir_func}
h(\mathbf{x}) = \frac{(x - x_c)^2 + (y - y_c)^2 - r_c^2}{2}
\end{equation}
and set the extended class $\mathcal{K}$ function as
\begin{equation}  \label{eq:2D_Pt_NM_alpha_func}
\alpha (h) = 10 h.
\end{equation}
In (\ref{eq:2D_Pt_NM_barreir_func}), the unsafe set (i.e. disk area) parameters are
\begin{align*}
x_c = 2, \quad y_c = 2, \quad r_c = 1.
\end{align*}
Compute
\begin{align*}
\nabla h(\mathbf{x}) &= \begin{bmatrix} x - x_c \\ y - y_c \end{bmatrix}, \quad \mathbf{f}(\mathbf{x}) = \begin{bmatrix} x \sin y \\ y \end{bmatrix}, \quad \mathbf{g}(\mathbf{x}) = \begin{bmatrix} 0 \\ 1 \end{bmatrix},  \\
L_{\mathbf{f}} h &= \nabla h^\mathrm{T} \mathbf{f} = (x - x_c) x \sin y + (y - y_c) y,  \\
L_{\mathbf{g}} h &= \nabla h^\mathrm{T} \mathbf{g} = y - y_c.
\end{align*}
Solve the quadratic programming problem (\ref{eq:affine_CBF_QP_single_input})
\begin{align*}
u = \arg \min_{u} | u - \hat{u} |  \quad \mbox{s.t.  } L_{\mathbf{f}} h + (L_{\mathbf{g}} h) u \geq - \alpha (h)
\end{align*}
to obtain the barrier function control law. Matlab simulation code for demonstration of two-dimensional point nonlinear motion barrier function control is given as follows. 

\begin{framed} 
\noindent \textbf{TwoDPointNonlinearMotionCBF.m} \\
\noindent \%\% Simulation preliminary configuration \\
dt = 0.001; \% Numerical computation step \\
tSpan = 0:dt:10; \% Simulation time span \\
x = 4; y = 5; \% Nonlinear system state \\
sttAll = zeros(2, length(tSpan)); k = 0; \% Record states in simulation  \\
xExpected = 0; yExpected = 0; \% Expected equilibrium status \\
flgUnsafeSet = 1; \% Unsafe set parameters flag \\
if (flgUnsafeSet) cx = 2; cy = 2; cr = 1; \\
else cx = 0; cy = 3.5; cr = 3; end \\
ca=-pi:pi/90:pi; \% For safe set boundary plotting \\
 \\
\%\% Simulation of nonlinear system control \\
for t = tSpan \\
$~~~~$ \%\% Control method  \\
$~~~~$ \% Lyapunov control as reference \\
$~~~~$ if (abs(y)$>$0.0001) u = -x\^{}2*sin(y)/y - 2*y;  \\
$~~~~$ else u = -x\^{}2 - 2*y; end \\
$~~~~$ \% Control barrier function constraint: Gh'*f + Gh'*g*u + alfh $>$= 0 \\
$~~~~$ h = ((x-cx)\^{}2+(y-cy)\^{}2-cr\^{}2)/2; alfh = 10*h; \\
$~~~~$ Gh = [x-cx; y-cy]; f = [x*sin(y); y]; g = [0; 1]; \\
$~~~~$ Lfh = Gh'*f; Lgh = Gh'*g; \\
$~~~~$ if (Lgh$>$0.0001)  \\
$~~~~$ $~~~~$ umin = -(Lfh+alfh)/Lgh; if (umin$>$u) u = umin; end \\
$~~~~$ elseif (Lgh$<$-0.0001) \\
$~~~~$ $~~~~$ umax = -(Lfh+alfh)/Lgh; if (umax$<$u) u = umax; end \\
$~~~~$ end \\
 \\
$~~~~$ \%\% Nonlinear system dynamics \\
$~~~~$ x = x + x*sin(y)*dt; \\
$~~~~$ y = y + (y+u)*dt; \\
$~~~~$ k = k+1; sttAll(:,k) = [x; y]; \\
$~~~~$ \%\% Nonlinear system visualization \\
$~~~~$ if (rem(k,50) == 0) \\
$~~~~$ $~~~~$ fill(cx+cr*cos(ca),cy+cr*sin(ca),'c','EdgeColor','c'); hold on;  \\
$~~~~$ $~~~~$ plot(x, y, 'ob', 'MarkerSize', 8, 'LineWidth', 2); hold off; \\
$~~~~$ $~~~~$ axis equal; grid on; xlim([-6, 6]); ylim([-4, 8]); pause(10*dt); \\
$~~~~$ end \\
end \\
figure(1), fill(cx+cr*cos(ca),cy+cr*sin(ca),'c','EdgeColor','c'); \\
hold on; plot(sttAll(1,:), sttAll(2,:), 'k', 'LineWidth', 2);  \\
plot(sttAll(1,1), sttAll(2,1), 'ob', 'MarkerSize', 8, 'LineWidth', 2); \\
plot(sttAll(1,end), sttAll(2,end), 'ok', 'MarkerSize', 8, 'LineWidth', 2);  \\
axis equal; grid on; xlim([-6, 6]); ylim([-4, 8]); hold off;
\end{framed}

\begin{figure}[h!]
\begin{center}
\includegraphics[width=0.4\columnwidth]{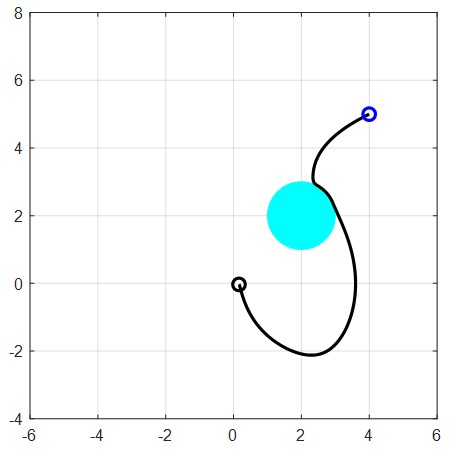}
\end{center}
\caption{Two-dimensional point nonlinear motion barrier function control}
\label{fig:2DMotionCBF}
\end{figure}

The simulation result is demonstrated in Figure \ref{fig:2DMotionCBF}. The two-dimensional point not only can succeed in converging to the expected state, but also can circumvent the unsafe set (or in other words, can guarantee evolving within the safe set) --- Readers may notice a small discrepancy between the expected state namely $(0, 0)$ and the terminal end of the state evolution trajectory. Despite existence of the small discrepancy which is due to the limited time span of simulation, the success of state convergence is well reflected.

It is worth noting that the Lie derivative of $h(\mathbf{x})$ with respect to $\mathbf{g}(\mathbf{x})$ is singular if
\begin{align*}
y = y_c \iff L_{\mathbf{g}} h = 0.
\end{align*}
To handle singularity of $L_{\mathbf{g}} h$, an expedient way is to simply adopt the control law reference $\hat{u} (\mathbf{x})$ without solving the quadratic programming problem (\ref{eq:affine_CBF_QP_single_input}) when $y$ is close to $y_c$. In other words, it is only when $y$ is not so close to $y_c$ that the quadratic programming problem (\ref{eq:affine_CBF_QP_single_input}) is solved to obtain the barrier function control law that replaces the control law reference. Such expedient way of handling singularity of $L_{\mathbf{g}} h$ corresponds to the following code block in the simulation code \textbf{TwoDPointNonlinearMotionCBF.m}.

\begin{framed} 
\noindent $~~~~$ if (Lgh$>$0.0001)  \\
$~~~~$ $~~~~$ umin = -(Lfh+alfh)/Lgh; if (umin$>$u) u = umin; end \\
$~~~~$ elseif (Lgh$<$-0.0001) \\
$~~~~$ $~~~~$ umax = -(Lfh+alfh)/Lgh; if (umax$<$u) u = umax; end \\
$~~~~$ end
\end{framed}

\subsubsection*{Application: single inverted pendulum barrier function control}

Take single inverted pendulum control as example. Consider the affine control system model formalism (4.43) for the single inverted pendulum control system
\begin{align*}
\frac{\mathrm{d}}{\mathrm{d} t} \begin{bmatrix} \theta \\ \frac{\mathrm{d} \theta}{\mathrm{d} t} \\ x \\ \frac{\mathrm{d} x}{\mathrm{d} t} \end{bmatrix} = \begin{bmatrix} \frac{\mathrm{d} \theta}{\mathrm{d} t} \\ \frac{\sin \theta}{L} g \\ \frac{\mathrm{d} x}{\mathrm{d} t} \\ 0 \end{bmatrix} + \begin{bmatrix} 0 \\ - \frac{\cos \theta}{L} \\ 0 \\ 1 \end{bmatrix} a \equiv \mathbf{f}(\mathbf{x}) \mathbf{x} + \mathbf{g}(\mathbf{x}) a
\end{align*}
where the state
\begin{align*}
\mathbf{x} \equiv \begin{bmatrix} \theta & \frac{\mathrm{d} \theta}{\mathrm{d} t} & x & \frac{\mathrm{d} x}{\mathrm{d} t} \end{bmatrix}^\mathrm{T}.
\end{align*}
For concrete configuration of parameters, let 
\begin{align*}  
L = 1, \quad g = 10, 
\end{align*}
then we have
\begin{align*}
\mathbf{f}(\mathbf{x}) = \begin{bmatrix} \frac{\mathrm{d} \theta}{\mathrm{d} t} \\ 10 \sin \theta \\ \frac{\mathrm{d} x}{\mathrm{d} t} \\ 0 \end{bmatrix}, \quad \mathbf{g}(\mathbf{x}) = \begin{bmatrix} 0 \\ - \cos \theta \\ 0 \\ 1 \end{bmatrix}.
\end{align*}

Apply the affine control system barrier function control method to handle the single inverted pendulum control system. Set the single inverted pendulum full-state feedback control law
\begin{align*}
a &= - \begin{bmatrix}  -131.60 & -41.60 & -25.60 & -25.60 \end{bmatrix} \mathbf{x}  \\
  &= 131.60 \theta + 41.60 \dot{\theta} + 25.60 x + 25.60 \dot{x}
\end{align*}
presented in Section 2.2.3 in Chapter 2 as control law reference $\hat{u} (\mathbf{x})$. Set the barrier function as
\begin{equation}  \label{eq:SIP_barreir_func}
h(\mathbf{x}) = \frac{25 (\theta_B^2 - \theta^2) + (\dot{\theta}_B^2 - \dot{\theta}^2)}{2},
\end{equation}
where
\begin{align*}
\theta_B = \frac{\pi}{15}, \quad \dot{\theta}_B = 2.
\end{align*}
Set the extended class $\mathcal{K}$ function as
\begin{equation}  \label{eq:SIP_alpha_func}
\alpha (h) = h.
\end{equation}
Compute
\begin{align*}
\nabla h(\mathbf{x}) &= \begin{bmatrix} -25 \theta \\ -\dot{\theta} \\ 0 \\ 0 \end{bmatrix}, \quad \mathbf{f}(\mathbf{x}) = \begin{bmatrix} \dot{\theta} \\ 10 \sin \theta \\ \dot{x} \\ 0 \end{bmatrix}, \quad \mathbf{g}(\mathbf{x}) = \begin{bmatrix} 0 \\ - \cos \theta \\ 0 \\ 1 \end{bmatrix},  \\
L_{\mathbf{f}} h &= \nabla h^\mathrm{T} \mathbf{f} = -25 \dot{\theta} \theta - 10 \dot{\theta} \sin \theta,  \\
L_{\mathbf{g}} h &= \nabla h^\mathrm{T} \mathbf{g} = \dot{\theta} \cos \theta.
\end{align*}
Solve the quadratic programming problem (\ref{eq:affine_CBF_QP_single_input})
\begin{align*}
u = \arg \min_{u} | u - \hat{u} |  \quad \mbox{s.t.  } L_{\mathbf{f}} h + (L_{\mathbf{g}} h) u \geq - \alpha (h)
\end{align*}
to obtain the barrier function control law. Matlab simulation code for demonstration of single inverted pendulum barrier function control is given as follows.

\begin{framed} 
\noindent \textbf{SingleInvertedPendulumCBF.m} \\
\noindent \%\% Single inverted pendulum parameters \\
m1 = 1; L1 = 1; g = 10; \\
\%\% Simulation preliminary configuration \\
dt = 0.001; \% Numerical computation step \\
tSpan = 0:dt:10; \% Simulation time span \\
x = 20; \% Cart position (far away from expected state) \\
dx = 0; \% Cart velocity \\
y = 0.2; \% Inverted pendulum angle theta  \\
dy = 0;  \% Inverted pendulum angular velocity \\
stt = [y; dy; x; dx]; \% Single inverted pendulum state \\
sttAll = zeros(length(stt), length(tSpan)); k = 0; \% Record states \\
xExpected = 0; yExpected = 0; \% Expected equilibrium status \\
SimConfig = [m1, L1, g, dt]; \\
\%\% Design the gain matrix via the general method \\
A = [0, 1, 0, 0; g/L1, 0, 0, 0; 0, 0, 0, 1; 0, 0, 0, 0]; \\
B = [0; -1/L1; 0; 1]; \\
lambdaE = [-4;-4;-4;-4]; \% Expected eigenvalues \\
sttK = DesignGainMatrix(A, B, lambdaE); \% Obtain the gain matrix \\
yCBF = pi/15; dyCBF = 2; \% Safe set parameters \\
fprintf('Gain matrix K: '); sttK' \\
 \\
\%\% Simulation of single inverted pendulum control \\
for t = tSpan \\
$~~~~$ \%\% Control method \\
$~~~~$ \% Full-state feedback control as reference \\
$~~~~$ acc = -sttK'*stt;  \\
$~~~~$ \% Control barrier function constraint: Gh'*f + Gh'*g*u + alfh $>$= 0 \\
$~~~~$ h = (25*(yCBF\^{}2-y\^{}2)+(dyCBF\^{}2-dy\^{}2))/2;  \\
$~~~~$ Gh = [-25*y; -dy; 0; 0];  \\
$~~~~$ fX = [dy; g*sin(y)/L1; dx; 0]; gX = [0; -cos(y)/L1; 0; 1]; \\
$~~~~$ Lfh = Gh'*fX; Lgh = Gh'*gX; alfh = h;  \\
$~~~~$ if (Lgh$>$0.0001)  \\
$~~~~$ $~~~~$ accmin = -(Lfh+alfh)/Lgh; if (accmin$>$acc) acc = accmin; end \\
$~~~~$ elseif (Lgh$<$-0.0001) \\
$~~~~$ $~~~~$ accmax = -(Lfh+alfh)/Lgh; if (accmax$<$acc) acc = accmax; end \\
$~~~~$ end \\
 \\
$~~~~$ \%\% Single inverted pendulum dynamics \\
$~~~~$ stt = DynamicsSIP(SimConfig, stt, acc); \\
$~~~~$ sttC = num2cell(stt); [y, dy, x, dx] = sttC\{:\}; \\
$~~~~$ if (abs(y)$>$=pi/2) fprintf('Control failure!$\backslash$n'); break; end \\
$~~~~$ k = k+1; sttAll(:,k) = stt; \\
$~~~~$ \%\% Single inverted pendulum visualization \\
$~~~~$ if (rem(k,20) == 0) \\
$~~~~$ $~~~~$ DisplaySIP(x, y, L1); pause(dt); \\
$~~~~$ end \\
end
\end{framed}

If the initial cart position is far away from the expected state (as in above demonstrated simulation), the adopted full-state feedback control law reference by itself will cause the single inverted pendulum to fall down. As already explained in Section \ref{sec:sliding_mode_augmented_FSFC} in this article
\footnote{Namely Chapter 5 of the author's works \cite{Li2026ACTPA_SJTU_2, Li2026ACTPA_SJTU_1}.}, 
the adopted full-state feedback control law reference only focuses on converging the final state to the expected state as soon as possible, without considering intermediate state evolution during the control process. When the initial deviation of the cart position is large, the adopted full-state feedback control law reference tends to generate drastic control input of cart acceleration that causes the single inverted pendulum state to evolve into state space where the essential modelling assumption is violated.

On the other hand, the single inverted pendulum barrier function control law obtained by solving the quadratic programming problem (\ref{eq:affine_CBF_QP_single_input}) enables the single inverted pendulum state to keep away from undesirable part of state space (i.e. the unsafe set) and evolve only within the safe set.

\subsection{Lyapunov-barrier function control}

The barrier function control method is intended for guaranteeing safety of state evolution. However, the barrier function control method itself may not guarantee convergence of state evolution. For example, still consider the two-dimensional point nonlinear motion control system and suppose the unsafe set is the disk area illustrated in the right sub-figure of Figure \ref{fig:2DMotionUnsafe}, with the disk parameters
\begin{align*}
x_c = 0, \quad y_c = 3.5, \quad r_c = 3.
\end{align*}
Apply again the two-dimensional point nonlinear motion barrier function control law presented in Section \ref{sec:safety_guaranteed_control} to see its performance when facing the new unsafe set --- For simulation, readers only need to change the value of the variable ``flgUnsafeSet'' from 1 to 0 in the Matlab simulation code \textbf{TwoDPointNonlinearMotionCBF.m} and run the Matlab simulation code again.

\begin{figure}[h!]
\begin{center}
\includegraphics[width=0.4\columnwidth]{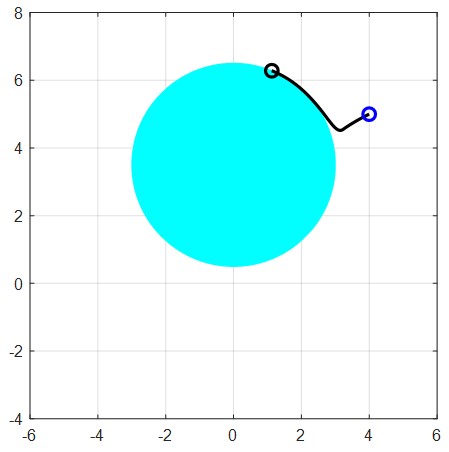}
\end{center}
\caption{Failure of two-dimensional point nonlinear motion barrier function control}
\label{fig:2DMotionCBFfailure}
\end{figure}

Readers will observe that in the presence of the new unsafe set, the two-dimensional point nonlinear motion barrier function control law causes the two-dimensional point to get stuck at a state far away from the expected state, as demonstrated in Figure \ref{fig:2DMotionCBFfailure}. Safety of the two-dimensional point nonlinear motion control system is indeed guaranteed, but the control objective is not achieved.

For sake of achieving the control objective on one hand and guaranteeing control safety on the other hand, a natural idea is to combine the Lyapunov control strategy that aims at the former and the barrier function control strategy that aims at the latter. This leads to an augmented strategy namely \textbf{Lyapunov-barrier function control}. Still withdraw from the too general formalism (\ref{eq:state_differential_equation}) and instead consider the generic affine control system formalism (\ref{eq:state_differential_equation_MIMO_affine})
\begin{align*}
\frac{\mathrm{d}}{\mathrm{d} t} \mathbf{x} = \mathbf{f} (\mathbf{x}) + \mathbf{g} (\mathbf{x}) \mathbf{u}.
\end{align*}
The affine version of the Lyapunov-barrier function control strategy consists in not only the inequality given in (\ref{eq:affine_CBF_derivative})
\begin{align*}
\frac{\mathrm{d}}{\mathrm{d} t} h(\mathbf{x}) = \nabla h(\mathbf{x})^\mathrm{T} \mathbf{f} (\mathbf{x}) + \nabla h(\mathbf{x})^\mathrm{T} \mathbf{g} (\mathbf{x}) \mathbf{u} \geq - \alpha (h(\mathbf{x}))
\end{align*}
i.e.
\begin{align*}
L_{\mathbf{f}} h + (L_{\mathbf{g}} h) \mathbf{u} \geq - \alpha (h)
\end{align*}
but also another inequality
\begin{equation}  \label{eq:affine_Lyapunov_CBF_derivative}
\frac{\mathrm{d}}{\mathrm{d} t} V(\mathbf{x}) = \nabla V(\mathbf{x})^\mathrm{T} \mathbf{f} (\mathbf{x}) + \nabla V(\mathbf{x})^\mathrm{T} \mathbf{g} (\mathbf{x}) \mathbf{u} \leq - \gamma (V(\mathbf{x}))
\end{equation}
i.e.
\begin{align*}
L_{\mathbf{f}} V + (L_{\mathbf{g}} V) \mathbf{u} \leq - \gamma (V),
\end{align*}
where $V(\mathbf{x})$ is a Lyapunov function and $\gamma (V)$ in terms of $V$ is an extended class $\mathcal{K}$ function.

\begin{framed} 
\noindent \textbf{Affine control system Lyapunov-barrier function control strategy}: \textit{For an affine control system described by (\ref{eq:state_differential_equation_MIMO_affine}), design a Lyapunov function $V(\mathbf{x})$, a barrier function $h(\mathbf{x})$, and a feedback control law of $\mathbf{u}$ in terms of $\mathbf{x}$ such that $\frac{\mathrm{d}}{\mathrm{d} t} V(\mathbf{x})$ satisfies (\ref{eq:affine_Lyapunov_CBF_derivative}) and $\frac{\mathrm{d}}{\mathrm{d} t} h(\mathbf{x})$ satisfies (\ref{eq:affine_CBF_derivative}).}
\end{framed}

A method of instantiating the affine control system Lyapunov-barrier function control strategy may be formalized also as solving of a quadratic programming problem
\begin{align}  \label{eq:affine_Lyapunov_CBF_QP_naive}
\mathbf{u} = &\arg \min_{\mathbf{u}} \frac{1}{2} \mathbf{u}^\mathrm{T} \mathbf{H}(\mathbf{x}) \mathbf{u}  \\
  &\mbox{s.t.  } L_{\mathbf{f}} V + (L_{\mathbf{g}} V) \mathbf{u} \leq - \gamma (V),  \nonumber  \\  
  &\qquad L_{\mathbf{f}} h + (L_{\mathbf{g}} h) \mathbf{u} \geq - \alpha (h).  \nonumber
\end{align}
However, the Lyapunov function related constraint in the quadratic programming problem (\ref{eq:affine_Lyapunov_CBF_QP_naive}) is too restrictive and may cause unsolvability of (\ref{eq:affine_Lyapunov_CBF_QP_naive}).

A better method of instantiating the affine control system Lyapunov-barrier function control strategy, which is proposed in \cite{Ames2019}, can be formalized as a relaxed version of (\ref{eq:affine_Lyapunov_CBF_QP_naive}) namely
\begin{align}  \label{eq:affine_Lyapunov_CBF_QP}
\mathbf{u} = &\arg \min_{\mathbf{u}} \frac{1}{2} \mathbf{u}^\mathrm{T} \mathbf{H}(\mathbf{x}) \mathbf{u} + \frac{\lambda}{2} \delta^2 \\
  &\mbox{s.t.  } L_{\mathbf{f}} V + (L_{\mathbf{g}} V) \mathbf{u} \leq - \gamma (V) + \delta,  \nonumber  \\  
  &\qquad L_{\mathbf{f}} h + (L_{\mathbf{g}} h) \mathbf{u} \geq - \alpha (h),  \nonumber
\end{align}
where $\mathbf{H}(\mathbf{x})$ denotes a positive definite matrix and $\delta$ denotes the relaxation variable. We may also set a control law reference $\hat{\mathbf{u}} (\mathbf{x})$ and incorporate it into (\ref{eq:affine_Lyapunov_CBF_QP}) as
\begin{align}  \label{eq:affine_Lyapunov_CBF_QP_with_ref}
\mathbf{u} = &\arg \min_{\mathbf{u}} \frac{1}{2} (\mathbf{u} - \hat{\mathbf{u}})^\mathrm{T} \mathbf{H}(\mathbf{x}) (\mathbf{u} - \hat{\mathbf{u}}) + \frac{\lambda}{2} \delta^2 \\
  &\mbox{s.t.  } L_{\mathbf{f}} V + (L_{\mathbf{g}} V) \mathbf{u} \leq - \gamma (V) + \delta,  \nonumber  \\  
  &\qquad L_{\mathbf{f}} h + (L_{\mathbf{g}} h) \mathbf{u} \geq - \alpha (h).  \nonumber
\end{align}

\subsubsection*{Application: two-dimensional point nonlinear motion Lyapunov-barrier function control}

Consider the two-dimensional point nonlinear motion control system described by (\ref{eq:Lyapunov_control_example1})
\begin{align*}
\frac{\mathrm{d}}{\mathrm{d} t} \begin{bmatrix} x \\ y \end{bmatrix} = \begin{bmatrix} x \sin y \\ y \end{bmatrix} + \begin{bmatrix} 0 \\ 1 \end{bmatrix} u.
\end{align*}
Apply the affine control system Lyapunov-barrier function control method to handle the two-dimensional point nonlinear motion control system. Set the Lyapunov function as that in (4.12)
\begin{align*}
V(\mathbf{x}) = \frac{\mathbf{x}^\mathrm{T} \mathbf{x}}{2} = \frac{x^2 + y^2}{2}
\end{align*}
and set the corresponding extended class $\mathcal{K}$ function as
\begin{equation}  \label{eq:2D_Pt_NM_gamma_func}
\gamma(V) = V.
\end{equation}
Set the barrier function as that in (\ref{eq:2D_Pt_NM_barreir_func})
\begin{align*}
h(\mathbf{x}) = \frac{(x - x_c)^2 + (y - y_c)^2 - r_c^2}{2}
\end{align*}
and set the corresponding extended class $\mathcal{K}$ function as that in (\ref{eq:2D_Pt_NM_alpha_func})
\begin{align*}
\alpha (h) = 10 h.
\end{align*}

Set the Lyapunov control law of $u$ specified in (4.13) 
\begin{align*}
u = -x^2 \frac{\sin y}{y} - 2 y
\end{align*}
as control law reference $\hat{u} (\mathbf{x})$. Compute
\begin{align*}
\nabla V(\mathbf{x}) &= \begin{bmatrix} x \\ y \end{bmatrix}, \quad \nabla h(\mathbf{x}) = \begin{bmatrix} x - x_c \\ y - y_c \end{bmatrix}, \quad \mathbf{f}(\mathbf{x}) = \begin{bmatrix} x \sin y \\ y \end{bmatrix}, \quad \mathbf{g}(\mathbf{x}) = \begin{bmatrix} 0 \\ 1 \end{bmatrix},  \\
L_{\mathbf{f}} V &= \nabla V^\mathrm{T} \mathbf{f} = x^2 \sin y + y^2, \quad L_{\mathbf{g}} V = \nabla V^\mathrm{T} \mathbf{g} = y,  \\
L_{\mathbf{f}} h &= \nabla h^\mathrm{T} \mathbf{f} = (x - x_c) x \sin y + (y - y_c) y, \quad L_{\mathbf{g}} h = \nabla h^\mathrm{T} \mathbf{g} = y - y_c.
\end{align*}
Set
\begin{align*}
\mathbf{H}(\mathbf{x}) = 1, \quad \lambda = 0.5^2
\end{align*}
and solve the quadratic programming problem (\ref{eq:affine_Lyapunov_CBF_QP_with_ref})
\begin{align*}
u = &\arg \min_{u} \frac{1}{2} (u - \hat{u})^2 + \frac{0.5^2}{2} \delta^2 \\
  &\mbox{s.t.  } L_{\mathbf{f}} V + (L_{\mathbf{g}} V) \mathbf{u} \leq - \gamma (V) + \delta,  \quad L_{\mathbf{f}} h + (L_{\mathbf{g}} h) \mathbf{u} \geq - \alpha (h)  \nonumber
\end{align*}
to obtain the Lyapunov-barrier function control law. Matlab simulation code for demonstration of two-dimensional point nonlinear motion Lyapunov-barrier function control is given as follows.

\begin{framed} 
\noindent \textbf{TwoDPointNonlinearMotionLyapunovCBF.m} \\
\noindent \%\% Simulation preliminary configuration \\
dt = 0.001; \% Numerical computation step \\
tSpan = 0:dt:10; \% Simulation time span \\
x = 4; y = 5; \% Nonlinear system state \\
sttAll = zeros(2, length(tSpan)); k = 0; \% Record states in simulation  \\
xExpected = 0; yExpected = 0; \% Expected equilibrium status \\
flgUnsafeSet = 0; \% Unsafe set parameters flag \\
if (flgUnsafeSet) cx = 2; cy = 2; cr = 1; \\
else cx = 0; cy = 3.5; cr = 3; end \\
ca=-pi:pi/90:pi; \% For safe set boundary plotting \\
 \\
\%\% Simulation of nonlinear system control \\
for t = tSpan \\
$~~~~$ \%\% Control method \\
$~~~~$ f = [x*sin(y); y]; g = [0; 1]; \\
$~~~~$ \% Lyapunov control as reference \\
$~~~~$ if (abs(y)$>$0.0001) u = -x\^{}2*sin(y)/y - 2*y;  \\
$~~~~$ else u = -x\^{}2 - 2*y; end \\
$~~~~$ \% Lyapunov function constraint: GV'*f + GV'*g*u + gammaV - delta $<$= 0 \\
$~~~~$ V = (x\^{}2+y\^{}2)/2; gammaV = V; \\
$~~~~$ GV = [x; y]; LfV = GV'*f; LgV = GV'*g; \\
$~~~~$ \% Control barrier function constraint: Gh'*f + Gh'*g*u + alfh $>$= 0 \\
$~~~~$ h = ((x-cx)\^{}2+(y-cy)\^{}2-cr\^{}2)/2; alfh = 10*h; \\
$~~~~$ Gh = [x-cx; y-cy]; Lfh = Gh'*f; Lgh = Gh'*g; \\
$~~~~$ \% Quadratic programming \\
$~~~~$ if (abs(y)$>$0.0001) \\
$~~~~$ $~~~~$ H = diag([1/2, 0.5\^{}2/2]); c = [-u; 0]; \\
$~~~~$ $~~~~$ A = [LgV, -1; -Lgh, 0]; b = [-LfV-gammaV; Lfh+alfh]; \\
$~~~~$ $~~~~$ udelta = quadprog(H,c,A,b); u = udelta(1); \\
$~~~~$ end \\
 \\
$~~~~$ \%\% Nonlinear system dynamics \\
$~~~~$ x = x + x*sin(y)*dt; \\
$~~~~$ y = y + (y+u)*dt; \\
$~~~~$ k = k+1; sttAll(:,k) = [x; y]; \\
$~~~~$ \%\% Nonlinear system visualization \\
$~~~~$ if (rem(k,50) == 0) \\
$~~~~$ $~~~~$ fill(cx+cr*cos(ca),cy+cr*sin(ca),'c','EdgeColor','c'); hold on;  \\
$~~~~$ $~~~~$ plot(x, y, 'ob', 'MarkerSize', 8, 'LineWidth', 2); hold off; \\
$~~~~$ $~~~~$ axis equal; grid on; xlim([-6, 6]); ylim([-4, 8]); pause(10*dt); \\
$~~~~$ end \\
end \\
figure(1), fill(cx+cr*cos(ca),cy+cr*sin(ca),'c','EdgeColor','c'); \\
hold on; plot(sttAll(1,:), sttAll(2,:), 'k', 'LineWidth', 2);  \\
plot(sttAll(1,1), sttAll(2,1), 'ob', 'MarkerSize', 8, 'LineWidth', 2); \\
plot(sttAll(1,end), sttAll(2,end), 'ok', 'MarkerSize', 8, 'LineWidth', 2);  \\
axis equal; grid on; xlim([-6, 6]); ylim([-4, 8]); hold off;
\end{framed}

\begin{figure}[h!]
\begin{center}
\includegraphics[width=0.9\columnwidth]{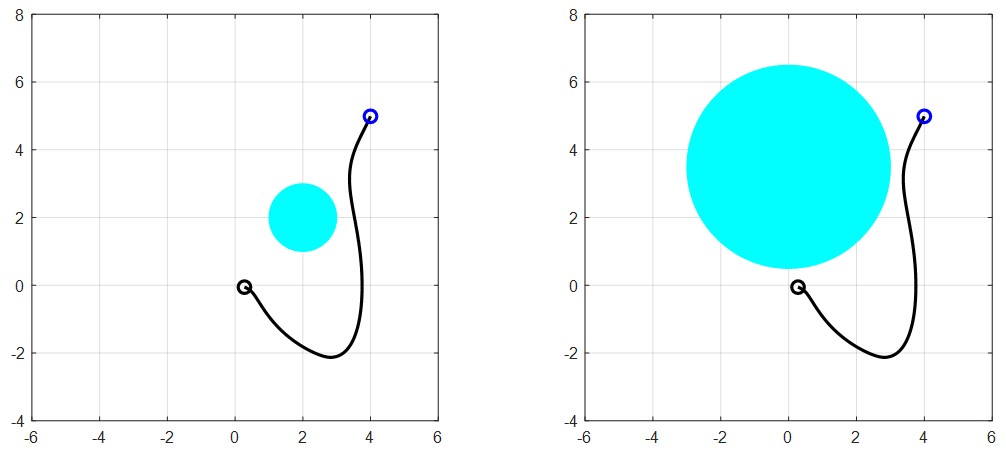}
\end{center}
\caption{Two-dimensional point nonlinear motion Lyapunov-barrier function control}
\label{fig:2DMotionLyapunovCBF}
\end{figure}

The simulation result is demonstrated in Figure \ref{fig:2DMotionLyapunovCBF}. No matter for the old unsafe set illustrated in the left sub-figure of Figure \ref{fig:2DMotionUnsafe} or the new unsafe set illustrated in the right sub-figure of Figure \ref{fig:2DMotionUnsafe}, the Lyapunov-barrier function control method always enables the two-dimensional point to converge to the expected state while guaranteeing safety as well.



\newpage
\addcontentsline{toc}{chapter}{Bibliography}

\fancyhf{} 

\bibliographystyle{unsrt}
\bibliography{LI_Hao_Refs_ACTPA}

\fancyhead[LE,RO]{\thepage}
\fancyhead[RE]{\textit{ \nouppercase{\leftmark}} }
\fancyhead[LO]{\textit{ \nouppercase{\rightmark}} }

\end{document}